\documentclass[epj]{svjour}
\usepackage{graphics}
\usepackage{axodraw}
\usepackage{epsfig}
\usepackage[latin1]{inputenc}
\usepackage{amsmath}
\usepackage{amssymb}
\usepackage{cite}

\newcommand{\lam}{\Lambda}

\newcommand{\kn}{K^0_S} 
\newcommand{\mgevsq}{{\rm GeV}^{2} } 
\newcommand{\mgev}{{\rm GeV} }

\newcommand{\dg}{\ensuremath{^{\circ}}}
\def\xbj{\ensuremath{x_{Bj}}}
\newcommand{\gev}{\,\mbox{GeV}}

\newcommand{\ptpair}{p_{T c\bar{c}}}
\def\question#1{}
\def\answer#1{}
\def\prp{\perp}

\def\kt{\ensuremath{k_\prp}}

\def\pt{\ensuremath{p_\prp}}
\def\pt{\ensuremath{p_\prp}}

\def\cascade{{\sc Cascade}}

\def\as{\ensuremath{\alpha_s}}

\def\bb{{\bf b}}

\def\xb{{\bf x}}

\def\gs{{\gamma^*}}
\def\br{{\bf r}}
\def\QQ{Q\bar{Q}}
\def\bas{{\bar\alpha}_s}
\def\ee{e^+e^-}
\def\cM{{\cal{M}}}
\def\Om{\Omega}
\def\half{\mbox{\small $\frac{1}{2}$}}
\def\out{{\rm out}}
\def\Eout{E_{\out}}
\newcommand{\dif}{\mathrm{d}}
\newcommand{\diff}[1]{\frac{\mathrm{d}#1}{#1}}
\newcommand{\CCFM}{Ciafaloni:1988ur,Catani:1990yc,Catani:1990sg,Marchesini:1994wr}
\newcommand{\BFKL}{Kuraev:1976ge,Kuraev:1977fs,Balitsky:1978ic}

\newcommand{\CASCADEMC}{Jung:2000hk,Jung:2001hx}

\newcommand{\Pmax}{\bar{q}}

\begin{document}
\title{ Small \boldmath$x$ Phenomenology - summary 
of the 3rd Lund Small \boldmath$x$  Workshop in 2004}
\subtitle{{\sc The Small $x$ Collaboration } }
\author{Jeppe~R.~Andersen\inst{1}
 \and Serguei~Baranov \inst{2} 
 \and Jochen~Bartels\inst{3} 
 \and Gergely~G.~Barnafoldi\inst{4}
 \and Grigorios~Chachamis\inst{3}
 \and John~Collins\inst{5}
 \and Guenter~Grindhammer\inst{6}
 \and Gösta~Gustafson\inst{7}
 \and Magnus~Hansson\inst{8}
 \and Gunnar~Ingelman\inst{9}
 \and Hannes~Jung\inst{10}
 \and Leif~Jönsson\inst{8}
 \and Albert~Knutsson\inst{8}
 \and Henri~Kowalski\inst{10}
 \and Krzysztof~Kutak\inst{3}
 \and Albrecht~Kyrieleis\inst{12}
 \and Peter~Levai\inst{4}
 \and Artem~Lipatov\inst{11}
 \and Leif~Lönnblad\inst{7}
 \and Michael~Lublinsky\inst{13}
 \and Giuseppe~Marchesini\inst{14}
 \and Izabela~Milcewicz\inst{15}
 \and Christiane~Risler\inst{10}
 \and Agustin~Sabio-Vera\inst{16}
 \and Malin~Sjödahl\inst{7}
 \and Anna~Stasto\inst{17}
 \and Jacek~Turnau\inst{15}
 \and Graeme~Watt\inst{18}
 \and Nikolai~Zotov\inst{11} \\ \\
{Edited by: Gösta~Gustafson, Hannes~Jung, Leif~Jönsson, Leif~Lönnblad}
}       
%
%
\institute{
  Cavendish Laboratory, University of Cambridge, UK  
  \and Lebedev~Institute~of~Physics, Moscow, Russia            
  \and Hamburg University, FRG                                 
  \and KFKI RMKI, Budapest, Hungary                                  
  \and Penn State Univ., 104 Davey Lab., University Park PA 16802, USA 
  \and Max Planck Institut, Munich, FRG,                       
  \and Department of Theoretical Physics, Lund University, Sweden 
  \and Department of Physics, Lund University, Sweden          
  \and University of Uppsala, Sweden                           
  \and DESY, Hamburg, FRG                                      
  \and Skobeltsyn~Institute~of~Nuclear~Physics, Moscow~State~University, Moscow,
 Russia                                                        
  \and University of Manchester, UK                             
  \and University of Connecticut, USA                          
  \and University of Milan-Bicocca and INFN, Sezione di Milano-Bicocca, Italy 
  \and H. Niewodniczanski Institute of Nuclear Physics, Cracow, Poland  
  \and CERN, Geneva, Switzerland                               
  \and DESY, Hamburg, FRG and
       H. Niewodniczanski Institute of Nuclear Physics, Cracow, Poland 
  \and IPPP, Durham,UK  and DESY, Hamburg, FRG                       
}
%
%
%
\abstract{ A third workshop on small-$x$ physics, within the Small-x
  Collaboration, was held in Hamburg in May 2004 with the aim of
  overviewing recent theoretical progress in this area and
  summarizing the experimental status.
%
} 
\maketitle
%
\section{Introduction}
\label{sec:intro}

In this report we summarize some of the recent developments in
small-$x$ physics, based on presentations and discussions during the
\textit{Lund Small-x} workshop held in DESY, Hamburg in May 2004.

Although accepted as an integral part of the Standard Model, QCD is
still not a completely understood theory. The qualitative aspects of
asymptotic freedom and confinement are under control, but the
quantitative predictive power of the theory is not at a satisfactory
level. In particular this is true for the non-perturbative regime,
where most of our understanding comes from phenomenological models,
such as the Lund string fragmentation model, and also from lattice
gauge calculations and effective theories, such as chiral perturbation
theory. For the perturbative aspects of QCD, the situation is more
satisfactory. In the weak coupling limit, the collinear factorization
theorem with so-called DGLAP evolution\cite{Gribov:1972ri,
  Lipatov:1975qm, Altarelli:1977zs, Dokshitzer:1977sg} is working well
and is under good theoretical control. Many cross sections have been
calculated to next-to-leading order (NLO), several even to
next-to-next-to-leading order, and some calculations involving
(next-to)$^3$-leading order have begun (see e.g.\ \cite{Moch:2004sf} and
references therein). The quantitative precision in this regime is
approaching the per-mille level, which is very encouraging although
still very far from the precision in QED.

However, there is a domain, still in the perturbative regime, where
our understanding is lacking. This is the region of high energy and
moderate momentum transfer, such as small-$x$ Deeply Inelastic
Scattering (DIS) as measured at HERA and low to medium $E_\perp$ jet
production at the Tevatron. In this region, the collinear
factorization must break down as the perturbative expansion becomes
plagued by large logarithms of the ratio between the total collision
energy and the momentum transfer of the hard sub-process, which needs
to be resummed to all orders to obtain precision predictions from QCD.
These logarithms arise from the large increase of the phase space
available for additional gluon emissions, resulting in a rapid rise of
the gluon density in hadrons with increasing collision energy or,
equivalently, decreasing momentum fraction, $x$.

In this high energy limit, QCD is believed to be correctly
approximated by the BFKL
evolution\cite{\BFKL}, and cross sections
should be possible to predict using
\kt-factorization~\cite{Gribov:1984tu,Levin:1991ry,Catani:1991eg,Collins:1991ty} 
where \textit{off-shell} matrix
elements are convoluted with \textit{unintegrated} parton densities
obeying BFKL evolution. However, so far the precision in the
predictions from \kt-factorization has been very poor. Although BFKL
evolution correctly predicted the strong rise of the $F_2$ structure
function with decreasing $x$ at HERA on a qualitative level, it turned
out that the next-to-leading order corrections to BFKL are
huge\cite{Fadin:1998py,Ciafaloni:1998gs}, basically making any
calculation with leading-logarithmic accuracy in \kt-factorization
useless.  

Several attempts have been made to tame the NLO corrections to BFKL by
e.g.\ matching to the collinear limit\cite{Ciafaloni:1999yw} and
matching this with off-shell matrix elements or impact factors
calculated to NLO. Another strategy is based on the fact that a large
part of the NLO corrections to BFKL can be traced to the lack of
energy and momentum conservation in the LO
evolution\cite{Salam:1999cn}. Although energy and momentum is still
not conserved in NLO evolution, the contributions from ladders which
violates energy--momentum conservations are reduced. Amending the
leading-logarithmic evolution with kinematical constraints, either
approximately in analytical calculations\cite{Kwiecinski:1996td} or
exactly in Monte-Carlo
programs\cite{Orr:1997im,Kharraziha:1998dn,Jung:2001hx,Andersen:2003gs},
should possibly lead to more reasonable QCD predictions, although
still formally only to leading logarithmic accuracy. However, so far
none of these strategies have been able to fulfill their ambitions,
and the reproduction of available data is still not satisfactory.

The plot thickens further when considering the increase in gluon
density at small $x$. At high enough energy the density of gluons
becomes so high that they must start to overlap and recombine, and we
will encounter the phenomena of multiple interactions, saturation and
rapidity gaps. In the non-perturbative region these phenomena have
already been established, but there is currently no consensus on
whether effects of recombination of perturbative gluons have been seen
at e.g.\ HERA. Perturbative recombination would require non-linear
evolution equations, which then also could break \kt-factorization.

In our first review \cite{Andersson:2002cf} we focused on the
theoretical and phenomenological aspects of \kt-factorization, while
in the second \cite{Andersen:2003xj} we also gave an overview of
experimental results in the small-$x$ region. In this third review we
will continue to present recent developments in these areas, but also
give an overview and introduction to saturation effects and non-linear
evolution.

The layout of this report is as follows. First we discuss some recent
developments of \kt-factorization in section~\ref{sec:ktfac}, starting
with the unintegrated parton densities (section~\ref{sec:updf}) and
\textit{doubly} unintegrated parton densities (\ref{sec:duPFD}) and
continuing with recent advances in NLO calculations (\ref{sec:nlobfkl}
and \ref{sec:nloimpact}). Then, in section \ref{sec:ktapp} we describe
some phenomenological applications of \kt-factorization, looking at
how to use them to obtain QCD predictions for heavy quark
(\ref{sec:heavytev}) and quarkonium (\ref{sec:onium}) production. In
section~\ref{sec:jetphysics} we present the recent investigations by
Marchesini and Mueller relating some aspects of jet physics to BFKL
dynamics, which could make it possible to study this kind of evolution
also in other environments. In section \ref{sec:saturation} we give an
introduction and overview of saturation phenomena and non-linear
evolution. Section \ref{sec:agk} also deals with saturation, but in
the context of the so-called AGK cutting rules which enables us to
relate saturation with multiple scatterings and diffraction. In
section \ref{sec:exp} we review some recent experimental results
relating to the issues in the previous sections, beginning with
multiple interactions and underlying events in section
\ref{sec:expmi}, followed by rapidity gaps between jets in
\ref{sec:jetgapjet}, jet-production at small-$x$ in \ref{sec:xjets}
and production of strange particles in DIS in section
\ref{sec:strange}.  Finally we present a brief summary and outlook in
section~\ref{sec:conclusions}.

\section{The \kt-factorization formalism}
\label{sec:ktfac}


\textit{Main author H.~Jung}\\

In the high energy limit, 
cross sections can be calculated using
$\kt$ -factorization~\cite{Gribov:1984tu,Levin:1991ry,Catani:1991eg,Collins:1991ty}
with convolution of a off-shell ($\kt$ dependent) partonic cross section
$\hat{\sigma}(\frac{x}{z},\kt^2) $ and an $\kt$ - unintegrated parton
density function ${\cal F}(z,\kt^2)$:
\begin{equation}
 \sigma  = \int 
\frac{dz}{z} d^2 \kt \hat{\sigma}(\frac{x}{z},\kt^2) {\cal F}(z,\kt^2)
\label{kt-factorisation}
\end{equation}
The unintegrated gluon density ${\cal F}(z,\kt^2)$ is described by the
BFKL~\cite{\BFKL} evolution equation in the region of asymptotically
large energies (small $x$).  An appropriate description valid for both
small and large $x$ is given by the CCFM evolution
equation~\cite{\CCFM}, resulting in an unintegrated gluon density,
${\cal A} (x,\kt^2,\Pmax^2 ) $, which is a function also of the
additional scale, $\Pmax $.  Here and in the following
we use the following classification scheme: $x{\cal G}(x,\kt^2)$
describes DGLAP type unintegrated gluon distributions, $x{\cal
  F}(x,\kt^2)$ is used for pure BFKL and $x{\cal A}(x,\kt^2,\Pmax^2)$
stands for a CCFM type or any other type having two scales involved.
Different approaches to the unintegrated parton density functions have 
been discussed in detail in \cite{Andersson:2002cf,Andersen:2003xj}.
\par
\begin{figure}[h]
\begin{center}
\includegraphics[width=0.53\textwidth]{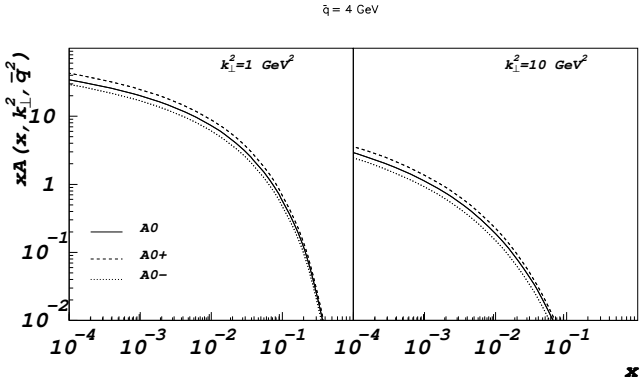}
\caption{\label{seta+-comparison}
Comparison of the CCFM uPDF obtained after changing the renormalization scale
in the off-shell matrix element by
a factor 2 up and down.
}
\end{center}
\end{figure}

While still being formally at leading order, the unintegrated gluon
densities incorporate effects from the next-to-leading order in the
collinear approach~\cite{Ryskin:2000bz}. This is 
discussed in more detail in the next
subsections. To further connect to the uncertainty estimates of cross
section calculated in the collinear approach, the change of the
renormalization and factorization scales are used to estimate the
influence and size of higher order corrections. In \cite{Jung:2004gs}
the CCFM unintegrated PDFs are determined such that the structure
function $F_2$ as measured at H1~\cite{Aid:1996au,Adloff:2000qk} and
ZEUS~\cite{Derrick:1996hn,Chekanov:2001qu} can be described after
convolution with the off-shell matrix element. This fit is repeated
for the renormalization scale in the off-shell matrix element varied
by a factor of 2 up and down, resulting in new sets of
PDFs\cite{Jung:2004gs}, {\it set A0+} and {\it set A0-}. These PDFs
are compared with the central set {\it set A0} in
Fig.~\ref{seta+-comparison}.


\subsection{Future fits of uPDF parameterizations}
\label{sec:param}

\textit{Main author M.~Hansson}\\
 
There are a number of possible measurements sensitive to the
transverse momentum of the propagating gluons in the gluon ladder, and
thereby suitable for investigations concerning the unintegrated gluon
density of the proton. One possible observable is the difference in
azimuthal angle, $\Delta \phi ^{\star}$, of a dijet system in the
hadronic center of mass frame. The differential cross section
$\frac{d\sigma}{d\Delta \phi ^{\star}}$ has been measured at the 
Tevatron~\cite{Acosta:2004nj,Abazov:2004hm,Abbott:1999se,Abe:1995mj,%
Abe:1996zt,Acosta:2003nn}  and 
only recently 
at HERA \cite{Chekanov:2005zg,Flucke:2005ux}. 
The quantity
\begin{equation}
  S=\frac{\int_0^{\alpha}N_{dijet}(\Delta\phi^{\star},x,Q^2)
  d\Delta\phi^{\star}}{\int_0^{\pi}N_{dijet}(\Delta\phi^{\star},x,Q^2)
  d\Delta\phi^{\star}},
\end{equation}
first proposed in \cite{Szczurek:2000pj}, has been measured
\cite{Aktas:2003ja} and showed a large sensitivity to the unintegrated
gluon density. Another measurement, proposed in
\cite{Luszczak:2004ag}, would be to measure
$\frac{d\sigma}{dp_{1,t}^2dp_{2,t}^2}$ where $dp_{i,t}^2$ are the
transverse momenta of a charm anti-charm pair. In
\cite{Luszczak:2004ag}, also an alternative to this was discussed,
namely to measure the quantity
\begin{equation}
f(p_{\max}^2>kp_{\min}^2;W)\equiv\frac{\sigma(p_{\max}^2>kp_{\min}^2;W)}{\sigma(W)}
\end{equation}
where $p_{\max}^2=\max(dp_{1,t}^2,dp_{2,t}^2)$,
$p_{\min}^2=\min(dp_{1,t}^2,dp_{2,t}^2)$ and $k$ is a constant. This
quantity would be a measure of the spread in the $p_{1,t}^2\times
p_{2,t}^2$ plane. Yet another possibility would be a direct
reconstruction of $x_g$ and $k_{g,t}^2$ from (DIS) multijet events,
thereby mapping the unintegrated gluon density directly.

The unintegrated gluon density could also be constrained from global
fits. So far, only fits to $F_2$ have been made \cite{Hansson:2003xz},
and a global fit using various data such as forward jets, 2+n jets,
heavy quarks and azimuthal jet-jet correlations would further
constrain the unintegrated gluon density.


\subsection{The need for doubly unintegrated parton density functions } 
\label{sec:updf}

\textit{Main author J.~Collins}\\

Conventional parton densities are defined in terms of an integral over
all transverse momentum and virtuality for a parton that initiates a
hard scattering.  While such a definition of an integrated parton
density is appropriate for very inclusive quantities, such as the
ordinary structure functions $F_1$ and $F_2$ in DIS, the definition
becomes increasingly unsuitable as one studies less inclusive cross
sections.  Associated with the use of integrated parton densities are
approximations on parton kinematics that can readily lead to
unphysical cross sections when enough details of the final state are
investigated.  

We propose that it is important to the future use of pQCD that a
systematic program be undertaken to reformulate factorization results
in terms of fully unintegrated densities, which are differential in
both transverse momentum and virtuality.  These densities are called
``doubly unintegrated parton densities'' by Watt, Martin and Ryskin
\cite{Watt:2003mx,Watt:2003vf} (discussed in the next section), and ``parton
correlation functions'' by Collins and Zu \cite{Collins:2004vq}; these
authors have presented the reasoning for the inadequacy, in different
contexts, of the more conventional approach.  The new methods have
their motivation in contexts such as Monte-Carlo event generators
where final-state kinematics are studied in detail.  Even so, a
systematic reformulation for other processes to use unintegrated
densities would present a unified methodology.

\begin{figure}
  \centering
  \includegraphics[scale=0.4]{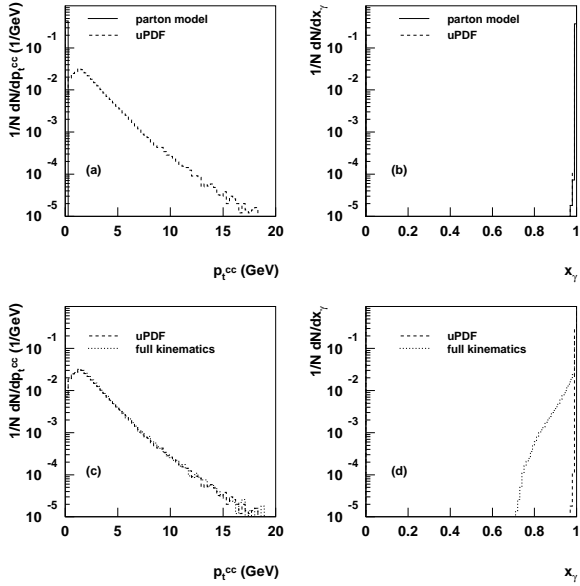}
  \caption{
    (a) and (b): Comparison between use of simple LO parton model
    approximation and of the use of $\kt$ densities for the $\pt$ of
    $c\bar{c}$ pairs in photoproduction, and for the $x_\gamma$.  (c) and
    (d): Comparison of use of $\kt$ densities and full simulation.  }
  \label{fig:ccbar.1}
\end{figure}

These methods form an extension of $\kt$-factorization, which has so
far been applied in small-$x$ processes and, as the CSS
formalism\cite{Collins:1984kg}, in the transverse-momentum
distribution of the Drell-Yan and related processes.

The problem that is addressed is nicely illustrated by considering
photoproduction of $c\bar{c}$ pairs.  In Figs.~\ref{fig:ccbar.1}, we
compare three methods of calculation carried out 
within the \cascade\ event generator \cite{\CASCADEMC}:
\begin{itemize}
\item Use of a conventional gluon density that is a function of parton
  $x$ alone.
\item Use of a $\kt$ density that is a function of parton $x$ and
  $\kt$.  These are the objects usually called ``unintegrated parton
  densities''.
\item Use of a ``doubly unintegrated density'' that is a function of parton
  $x$, $\kt$ and virtuality, that is, of the complete parton 4-momentum,
  in \cascade\ taken after the full simulation of the initial state parton
  showering.
\end{itemize}
The partonic subprocess in all cases is the lowest order
photon-gluon-fusion process $\gamma+g\longrightarrow c+\bar{c}$.
  Two differential cross sections are
plotted: one as a function of the transverse momentum of the $c\bar{c}$
pair, and the other as a function of the $x_\gamma$ of the pair.
By $x_\gamma$ is meant the fractional momentum of the photon carried by the
$c\bar{c}$ pair, calculated in the light-front sense as
\begin{displaymath}
 x_\gamma =  \frac{\sum_{i=c,\bar{c}} (E_i - p_{z\;i}) }{2 y E_e} 
    =  \frac{p_{c\bar{c}}^-}{q^-}.
\end{displaymath}
Here $E_e$ is the electron beam energy and the coordinates are
oriented so that the electron and proton beams are in the $-z$ and
$+z$ directions respectively.

In the normal parton model approximation for the hard scattering, the
gluon is assigned zero transverse momentum and virtuality, so that the
cross section is restricted to $\ptpair=0$ and $x_\gamma=1$, as
shown by the solid lines in Fig.\ \ref{fig:ccbar.1}(a,b).  When a
$\kt$ dependent gluon density is used, quite large gluonic $\kt$ can be
generated, so that the $\ptpair$ distribution is spread out in
a much more physical way, as given by the dashed line in Fig.\ 
\ref{fig:ccbar.1}(a).  But as shown in plot (b), $x_\gamma$ stays close to
unity. 
Neglecting the full recoil mass $m$ 
is equivalent of taking $k^2= \frac{-\kt^2}{1-x}$ with 
$k^2 $ being the virtuality of the gluon,
$\kt^2$ its transverse momentum and $x$ its light cone energy
fraction.  This gives a particular value to the gluon's $k^-$. 
When we also take into account the correct virtuality of the gluon, there
is no noticeable change in the $\ptpair$ distribution --- see Fig.\ 
\ref{fig:ccbar.1}(c) (dotted line) --- since that is already made broad by the
transverse momentum of the gluon.  But the gluon's $k^-$ is able to
spread out the $x_\gamma$ distribution, as in Fig.\ \ref{fig:ccbar.1}(d) with
the dotted line. This is equivalent with a proper treatment of the kinematics
and  results in   $k^2= \frac{-\kt^2 - x m^2 }{1-x}$, which can be significant
for finite $x$.
Clearly, the use of
the simple parton-model kinematic approximation gives unphysically
narrow distributions.  The correct physical situation is that the
gluon surely has a distribution in transverse momentum and virtuality,
and for the considered cross sections neglect of parton transverse
momentum and virtuality leads to wrong results.  It is clearly better
to have a correct starting point even at LO, 
for differential cross sections such as we have plotted.

Therefore it is
highly desirable to reformulate perturbative QCD methods in terms of
doubly unintegrated parton densities from the beginning.  A full
implementation will be able to use the full power of calculations at
NLO and beyond.  


\subsection{Doubly unintegrated PDFs}
\label{sec:duPFD}

\textit{Main author G.~Watt}\\

The notation for the two-scale unintegrated gluon distribution, $x\mathcal{A}(x,k_\perp^2,\bar{q}^2)$, used 
in \cite{Andersson:2002cf,Andersen:2003xj} and elsewhere in this report, is related to that used in this 
section by
\begin{equation}
  \label{eq:notation}
  x\mathcal{A}(x,k_\perp^2,\bar{q}^2) \leftrightarrow f_g(x,k_t^2,\mu^2)/k_t^2.
\end{equation}

\subsubsection{Unintegrated PDFs from integrated ones} \label{sec:pdf2updf}
Existing analyses of the CCFM equation are based on numerical solution via Monte Carlo methods.  
Kimber, Martin and Ryskin \cite{Kimber:2001sc} showed that, in a certain approximation,
it is possible to obtain two-scale UPDFs, $f_a(x,k_t^2,\mu^2)$, from single-scale distributions, 
with the dependence on the second scale $\mu$ introduced only in the \emph{last} step of the evolution.  
It was found that this ``last-step'' prescription gave similar results whether the single-scale 
distributions were evolved with a unified BFKL-DGLAP equation \cite{Kwiecinski:1997ee} or purely 
with the DGLAP equation, indicating that angular ordering is more important than small-$x$ effects.  
Here, we summarize the procedure \cite{Kimber:2001sc,Watt:2003mx} for obtaining UPDFs from the 
conventional DGLAP-evolved integrated PDFs, $a(x,\mu^2) = xg(x,\mu^2)$ or $xq(x,\mu^2)$.

The UPDFs are constructed to satisfy the normalization conditions
\begin{equation} \label{eq:updfnorm}
  \int_0^{\mu^2}\!\diff{k_t^2}\,f_a(x,k_t^2,\mu^2) = a(x,\mu^2),
\end{equation}
which are ensured by defining the UPDFs to be \cite{Kimber:2001sc,Watt:2003mx}
\begin{align}
  f_a(x,k_t^2,\mu^2) \equiv &\frac{\partial}{\partial \ln
    k_t^2}\left[\,a(x,k_t^2)\,
    T_a(k_t^2,\mu^2)\,\right]\nonumber\\
  =& T_a(k_t^2,\mu^2)\,\frac{\alpha_S(k_t^2)}{2\pi}\nonumber\\
  &\times\sum_{b=g,q}\,\int_x^1\!\dif{z}\,P_{ab }(z)\,b \left
  (\frac{x}{z}, k_t^2 \right), \label{eq:updf}
\end{align}
where the Sudakov form factors are
\begin{align}
  &T_a(k_t^2,\mu^2) \equiv\nonumber\\
  &\exp\left(-\int_{k_t^2}^{\mu^2}\!\diff{\kappa_t^2}\,
    \frac{\alpha_S(\kappa_t^2)}{2\pi}\,
    \sum_{b=g,q}\,\int_0^1\!\dif{\zeta}\;\zeta \,P_{ba}(\zeta)\right),
  \label{eq:sudakov}
\end{align}
and $P_{ba}$ are the unregulated LO DGLAP splitting kernels.

In addition, it is necessary to apply angular-ordering constraints due to color coherence, which 
regulate the singularities in \eqref{eq:updf} and \eqref{eq:sudakov} arising from soft gluon emission.  
These constraints are not applied for quark emission where there is no ``coherence'' effect.  The 
explicit expressions for the unintegrated gluon and quark distributions are given in \cite{Watt:2003mx}.

This approach to UPDFs amounts to relaxing the DGLAP approximation of strongly-ordered 
transverse momenta along the evolution chain only in the \emph{last} evolution step.  
If we consider DIS in the Breit frame, where the proton has 4-momentum $p$ and the 
virtual photon has 4-momentum $q$, then the penultimate parton in the evolution chain, 
with 4-momentum $k_{n-1}=(x/z)\,p$, splits to a final parton with 4-momentum
\begin{equation} \label{eq:k}
  k_n \equiv k \equiv (k^+,k^-,\boldsymbol{k_t}) = x\,p - \beta\,q^\prime + k_\perp,
\end{equation}
where the plus and minus components are $k^\pm\equiv k^0\pm k^3$.  In the Breit frame:
\begin{gather}
  p=(Q/\xbj,0,\boldsymbol{0}),\\
  q^\prime\equiv q+\xbj\,p=(0,Q,\boldsymbol{0}),\\
  k_\perp=(0,0,\boldsymbol{k_t}), \label{eq:kperpdef}
\end{gather}
so that $p^2=0={q^\prime}^2$, $q^2 = -Q^2$ and $k_\perp^2 = -k_t^2$.
The condition that the parton emitted in the last evolution step is on-shell, $(k_{n-1}-k_n)^2=0$, gives
\begin{equation} \label{eq:beta}
  \beta = \frac{\xbj}{x}\frac{z}{(1-z)}\frac{k_t^2}{Q^2},
\end{equation}
so $k^2=-k_t^2/(1-z)$.  In the high-energy (small-$x$) limit, where gluons 
dominate, we have $z\to0$, so $k\simeq x\,p+k_\perp$ and $k^2\simeq -k_t^2$.  
Cross sections can then be calculated using the $k_t$-factorization formalism,
\begin{equation} \label{eq:ktfact}
  \sigma^{\gamma^*p} = \int_{\xbj}^1\!\diff{x}\,\int_0^{\infty}\!\diff{k_t^2}\;
  f_g(x,k_t^2,\mu^2)\;\hat{\sigma}^{\gamma^*g^*},
\end{equation}
where the partonic cross section $\hat{\sigma}^{\gamma^*g^*}$ is calculated 
with an off-shell incoming gluon.

\subsubsection{Doubly-unintegrated PDFs}

\begin{figure}
  \begin{center}
    \includegraphics[width=0.5\textwidth]{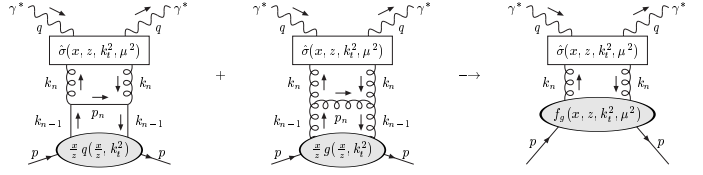}
    \caption{Illustration of $(z,k_t)$-factorization for the 
    doubly-unintegrated gluon distribution, $f_g(x,z,k_t^2,\mu^2)$, 
    shown in the final diagram.  In the first two diagrams the 
    penultimate parton in the DGLAP evolution chain, with 4-momentum 
    $k_{n-1}=(x/z)\,p$, splits into a gluon with 4-momentum 
    $k_n\equiv k=x\,p-\beta\,q^\prime+k_\perp$. \label{fig:zktfactg}}
  \end{center}
\end{figure}

Away from the high-energy limit, where we have finite $z$, the partonic 
cross section of \eqref{eq:ktfact} will necessarily have some $z$ dependence 
through the $q^\prime$ component, i.e.~the minus component, of the 4-momentum 
$k$ \eqref{eq:k}.  Therefore, we should consider \emph{doubly}-unintegrated 
PDFs (DUPDFs), $f_a(x,z,k_t^2,\mu^2)$, which satisfy
\begin{equation}
  \int_x^1\!\dif{z}\,f_a(x,z,k_t^2,\mu^2)=f_a(x,k_t^2,\mu^2).
\end{equation}
From \eqref{eq:updf}, the DUPDFs are
\begin{align}
  f_a(x,z,k_t^2,\mu^2) = & T_a(k_t^2,\mu^2)\,\frac{\alpha_S(k_t^2)}{2\pi}
  \nonumber\\
  &\times\sum_{b=g,q}P_{ab }(z)\,b\left(\frac{x}{z},k_t^2\right),
\end{align}
apart from the angular-ordering constraints.  The explicit expressions for 
the doubly-unintegrated gluon and quark distributions are given 
in \cite{Watt:2003mx}.  The $k_t$-factorization formula \eqref{eq:ktfact} 
is then generalized to the ``$(z,k_t)$-factorization'' formula \cite{Watt:2003mx}
\begin{multline} \label{eq:zktfact}
  \sigma^{\gamma^*p} = \\
  \sum_{a=g,q} \int_{\xbj}^1\!\diff{x}\,\int_x^1\!\dif{z}\,
  \int_0^{\infty}\!\diff{k_t^2}\;f_a(x,z,k_t^2,\mu^2)\;
  \hat{\sigma}^{\gamma^*a^*}.
\end{multline}
Note that $f_a(x,z,k_t^2,\mu^2)$ are linear densities in $z$, but 
logarithmic in $x$ and $k_t^2$.  This idea is illustrated in 
Fig.~\ref{fig:zktfactg} for the case $a=g$.
It is not immediately obvious how the partonic cross sections 
$\hat{\sigma}^{\gamma^*a^*}$ in \eqref{eq:zktfact} should be 
calculated.  Recall that they can be written
\begin{equation} \label{eq:sigmahat}
  \hat{\sigma} = \int\!\dif\Phi\,\lvert\mathcal{M}\rvert^2 \,/\, F,
\end{equation}
where $\dif\Phi$ is the phase space element, 
$\lvert\mathcal{M}\rvert^2$ is the squared matrix element, 
and $F$ is the flux factor.  The phase space element 
$\dif\Phi$ can be calculated with the full kinematics, 
that is, with $k = x\,p-\beta\,q^\prime+k_\perp$.  The flux 
factor $F$ is taken to be the same as in collinear factorization 
(and in $k_t$-factorization), that is, $F=4x\,p\cdot q$.  The 
last evolution steps in Fig.~\ref{fig:zktfactg} only factorize from 
the rest of the diagram, to give the LO DGLAP splitting kernels, in 
the leading logarithmic approximation (LLA), that is, in either the 
collinear ($k_t\to 0$) or high-energy ($z\to0$) limits.  Therefore, 
$\lvert\mathcal{M}\rvert^2$ should be evaluated with either 
$k=x\,p$ or $k=x\,p+k_\perp$, in order to provide the factorization 
between the DUPDF and the subprocess labeled $\hat{\sigma}$ in Fig.~\ref{fig:zktfactg}. 
For the specific case of inclusive jet production in DIS and 
working in an axial gluon gauge, it was observed in \cite{Watt:2003mx} 
that the main effect of the ``beyond LLA'' terms (proportional to 
$\beta$ \eqref{eq:beta}) was to suppress soft gluon emission, and 
that these terms made a negligible difference to the cross section 
when the angular-ordering constraints were applied.

\begin{figure}
  \begin{center}
    \includegraphics[clip,width=0.5\textwidth]{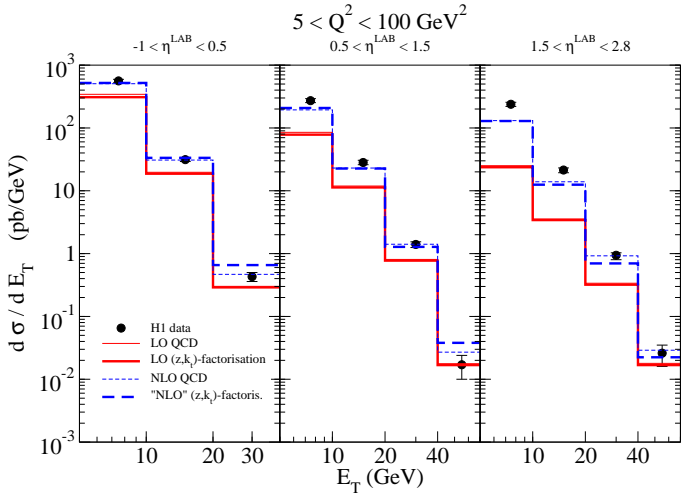}
    \caption{Comparison with H1 inclusive jet production data 
    \cite{Adloff:2002ew} in three pseudorapidity ($\eta^\textsc{lab}$) bins. 
    The predictions of the $(z,k_t)$-factorization approach based on 
    DUPDFs \cite{Watt:2003mx} (which is much simpler to implement) are 
    in good agreement with the conventional QCD approach.  In some 
    bins the predictions of the latter approach are hidden beneath the 
    bold lines of the $(z,k_t)$-factorization approach, at the respective order.}
    \label{fig:H1plot}
  \end{center}
\end{figure}

The prescription adopted in \cite{Watt:2003mx} was to evaluate 
$\lvert\mathcal{M}\rvert^2$ in the collinear approximation 
($k=x\,p$), so that a $(z,k_t)$-factorization calculation 
approximately reproduces the collinear factorization 
calculation starting one rung down as in the first two diagrams 
of Fig.~\ref{fig:zktfactg}, that is, where the subprocess is 
evaluated at one order higher in $\alpha_S$.  This was demonstrated 
for inclusive jet production in DIS, where the LO subprocess is 
simply $\gamma^*q^*\to q$.  Similarly, a ``NLO'' calculation, 
where the subprocesses are $\gamma^*g^*\to q\bar{q}$ and 
$\gamma^*q^*\to qg$, was found to give results close to the 
conventional NLO QCD calculation, where the subprocesses are 
$\mathcal{O}(\alpha_S^2)$; see Fig.~\ref{fig:H1plot}.

\begin{figure}
  \centering
  \includegraphics[width=0.5\textwidth,clip]{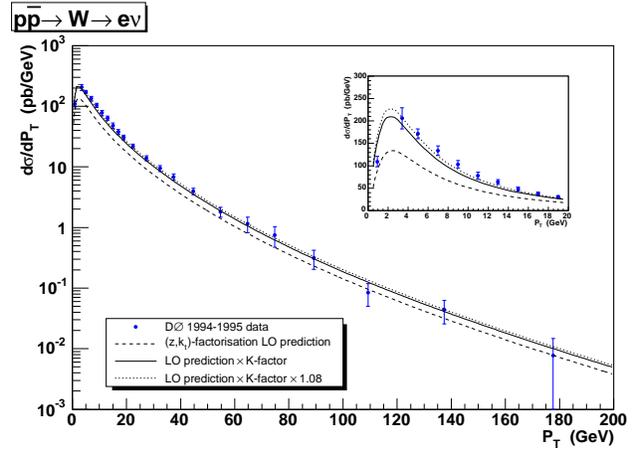}
  \caption{$p_T$ distribution of W bosons produced at the 
  Tevatron calculated using $(z,k_t)$-factorization \cite{Watt:2003vf}, 
  compared to D{\O} data \cite{Abbott:2000xv}.}
  \label{fig:zktwprod}
\end{figure}

\begin{figure}
  \centering
  \includegraphics[width=0.5\textwidth,clip]{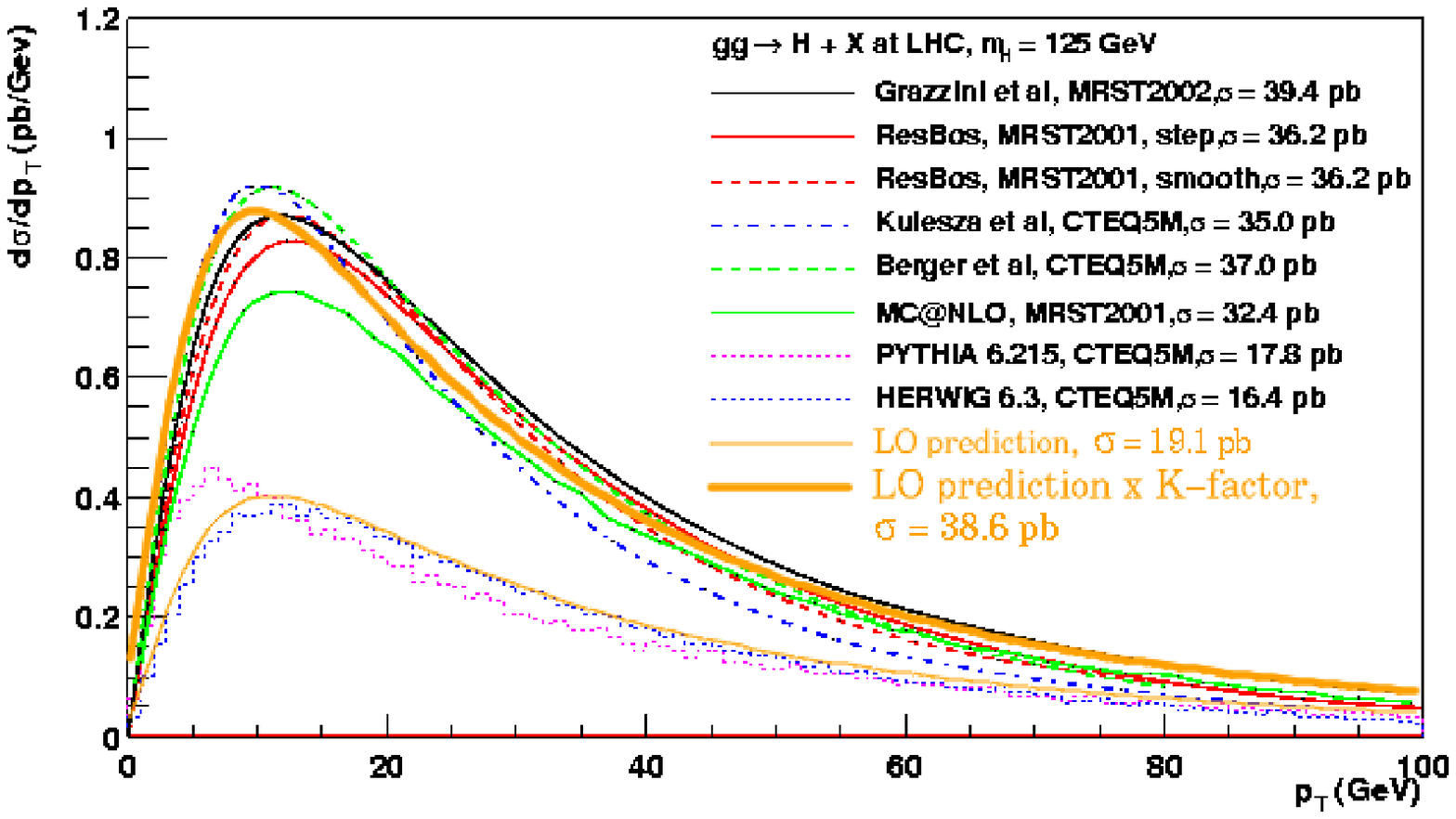}\\
  \includegraphics[width=0.5\textwidth,clip]{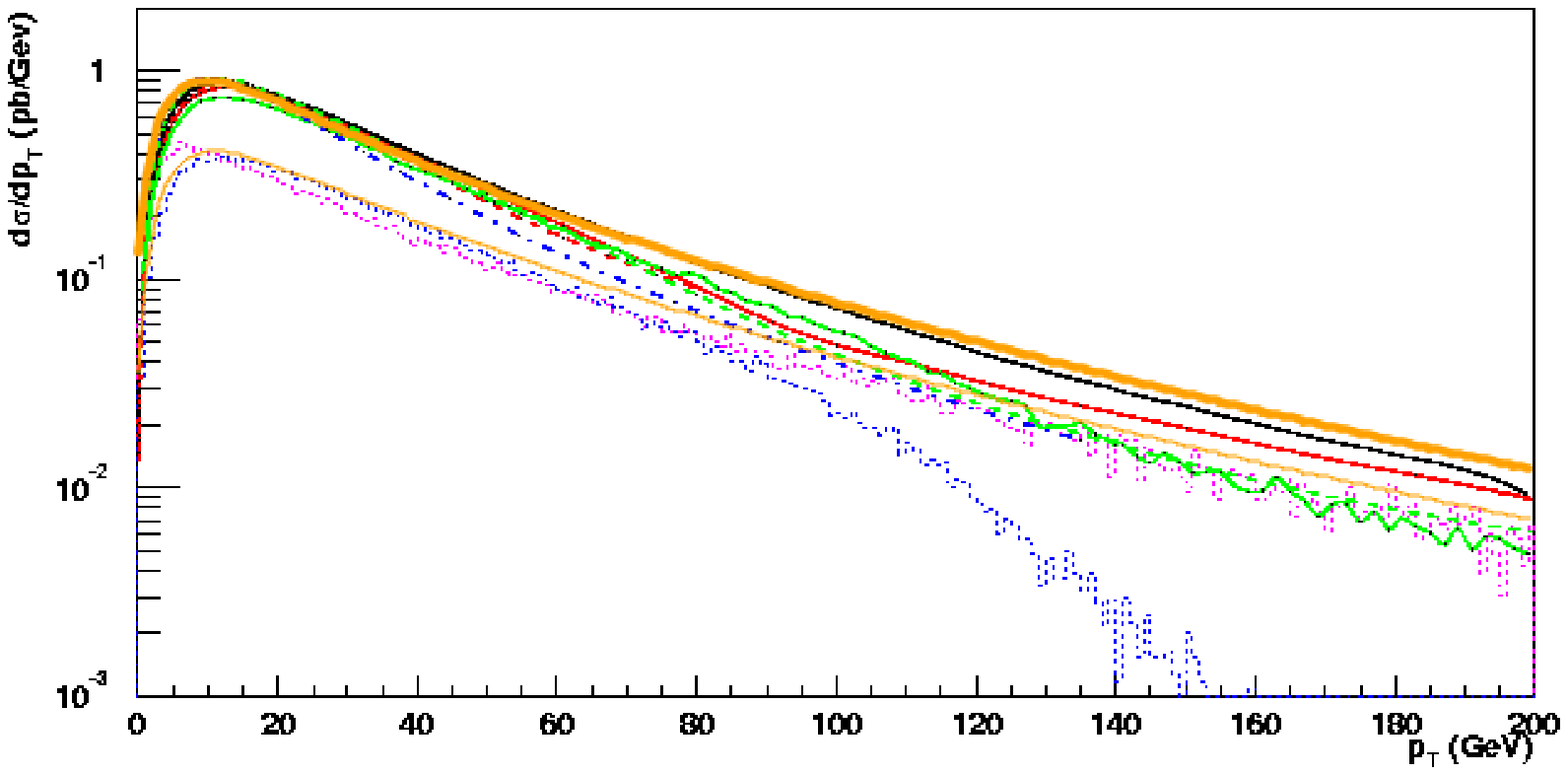}
  \caption{$p_T$ distribution of SM Higgs bosons produced at the LHC with 
  mass 125 GeV calculated using $(z,k_t)$-factorization \cite{Watt:2003vf}, 
  compared to various resummed and parton shower predictions which are all
   matched to fixed-order calculations at large $p_T$ (apart from \textsc{herwig}) \cite{Balazs:2004rd}.}
  \label{fig:zkthiggs}
\end{figure}

In \cite{Watt:2003vf}, the $(z,k_t)$-factorization formalism was extended 
to hadron--hadron collisions and applied to predict the $p_T$ distributions 
of vector bosons ($V=W,Z$) and Standard Model Higgs bosons ($H$).  
For $p_T\ll M_{V,H}$, fixed-order collinear factorization calculations 
diverge, with $\ln(M_{V,H}/p_T)$ terms appearing in the perturbation 
series due to soft and collinear gluon emission.  Traditional calculations 
combine fixed-order perturbation theory at high $p_T$ with either analytic 
resummation or numerical DGLAP-based parton shower formalisms at low $p_T$, 
with some matching criterion to decide when to switch between the two.  
It has been shown in \cite{Gawron:2003np,Kwiecinski:2003fu} that UPDFs 
obtained from an approximate solution of the CCFM evolution equation embody 
the conventional soft gluon resummation formulae.  In the framework of 
$(z,k_t)$-factorization, the lowest order subprocesses are simply 
$q_1^*\,q_2^*\to V$ and $g_1^*\,g_2^*\to H$.  A good description was obtained 
in \cite{Watt:2003vf} of the $p_T$ distributions of $W$ and $Z$ bosons 
produced at the Tevatron Run 1 over the whole $p_T$ range; 
see Fig.~\ref{fig:zktwprod}.
The predicted Higgs $p_T$ distribution at the LHC was found to reproduce, 
to a fair degree, the predictions of more elaborate theoretical studies 
\cite{Balazs:2004rd}, in particular the NNLL+NLO resummation approach of 
Grazzini \emph{et al.} \cite{Bozzi:2003jy}; see Fig.~\ref{fig:zkthiggs}.
Alternative predictions for Higgs production at the LHC using the 
$k_t$-factorization approach have been made in 
\cite{Gawron:2003np,Jung:2003wu,Lipatov:2005at,Luszczak:2005xs}.

Note that matrix-element corrections are necessary in DGLAP-based 
parton shower simulations at large $p_T$.  Without such corrections, 
the \textsc{herwig} parton shower prediction falls off dramatically at 
large $p_T\gtrsim M_H$ \cite{Corcella:2004fr}; see Fig.~\ref{fig:zkthiggs}.  
The same effect is observed in \textsc{herwig} predictions for the 
$p_T$ distributions of $W$ and $Z$ bosons \cite{Corcella:1999gs}, 
whereas in Fig.~\ref{fig:zktwprod} the Tevatron data at large 
$p_T\gtrsim M_W$ are well-described \emph{without} explicit 
matrix-element corrections.  Also, the $(z,k_t)$-factorization 
prediction for Higgs production is found to be close to the NLO
 fixed-order result at large $p_T$, see Fig.~\ref{fig:zkthiggs}, 
 suggesting that a large part of the subleading terms are included 
 by accounting for the precise kinematics in the $g_1^*\,g_2^*\to H$ subprocess.

The integrated PDFs used as input in \cite{Watt:2003mx,Watt:2003vf} were 
determined from a global fit to data using the conventional collinear 
approximation \cite{Martin:2002dr}.  A more precise treatment would 
determine the integrated PDFs, used as input to the last evolution step, 
from a new global fit to data using the $(z,k_t)$-factorization formalism.


\subsection{NLO BFKL}
\label{sec:nlobfkl}

\textit{Main author J.~Andersen and A.~Sabio-Vera}\\

Since the completion of the calculation of the next--to--leading (NLL)
corrections to the BFKL equation~\cite{Fadin:1998py,Ciafaloni:1998gs}
for the forward kernel there has been a large activity focused on the
study of the fundamental properties of the NLL gluon Green's function
in the Regge limit of QCD at high energies~\cite{Ross:1998xw,Salam:1998tj,Ciafaloni:2003kd,Ciafaloni:2003rd,Ciafaloni:2003ek,Ciafaloni:2002xk,Ciafaloni:2002xf,Ciafaloni:2000cb,Ciafaloni:1999au,Ciafaloni:1999yw,Ciafaloni:1998iv,Forshaw:1999xm,Forshaw:2000hv,Kovchegov:1998ae,Schmidt:1999mz,Altarelli:2003hk,Altarelli:2001ji,Altarelli:2000mh,Altarelli:1999vw}.  Recently, a powerful approach has been developed
which allows for the complete and exact analysis of the solution at
NLL. In Ref.~\cite{Andersen:2003an} it was demonstrated how it is possible to use
$D = 4 + 2 \epsilon$ dimensional regularization together with an
effective gluon mass ($\lambda$) to explicitly show the cancellation
of simple and double poles in $\epsilon$. This procedure carries a
logarithmic dependence in $\lambda$ which numerically cancels out when
the full NLL BFKL evolution is taken into account for a given
center--of--mass energy, this being a natural consequence of the
infrared finiteness of the full kernel. The basis of this approach is
the iterated form of the solution for the NLL BFKL equation, {\it
  i.e.}
\begin{equation}
  \label{solution1}
  \begin{split}
    &f({\bf k}_a ,{\bf k}_b, {\rm Y}) = e^{\omega_0^\lambda \left({\bf
          k}_a\right) {\rm Y}}
    \left\{\frac{}{}\delta^{(2)} ({\bf k}_a - {\bf k}_b) \right. \\
    &\hspace{5mm}+ \sum_{n=1}^{\infty} \prod_{i=1}^{n} \int d^2 {\bf k}_i
    \left[\frac{\theta\left({\bf k}_i^2 - \lambda^2\right)}{\pi {\bf
          k}_i^2} \xi\left({\bf k}_i\right) +\right.\\
&\hspace{35mm}\left.\widetilde{\mathcal{K}}_r
      \left({\bf k}_a+\sum_{l=0}^{i-1}{\bf k}_l,
        {\bf k}_a+\sum_{l=1}^{i}{\bf k}_l\right)\frac{}{}\right]\\
     &\hspace{10mm}\left. \times \int_0^{y_{i-1}} d y_i ~ e^{\left(
          \omega_0^\lambda\left({\bf k}_a+\sum_{l=1}^i {\bf
              k}_l\right)-\omega_0^\lambda\left({\bf
              k}_a+\sum_{l=1}^{i-1} {\bf k}_l\right) \right)y_i}\right.\\
      &\hspace{30mm}\left.\delta^{(2)} \left(\sum_{l=1}^{n}{\bf k}_l + {\bf k}_a - {\bf k}_b
      \right)\right\},
  \end{split}
\end{equation}
where the strong ordering in longitudinal components of the parton emission is 
encoded in the nested integrals in rapidity with an upper limit set by the 
logarithm of the total energy in the process, $y_0 = {\rm Y}$. The 
Reggeized form of the gluon propagators in the $t$--channel,  
$\omega_0^\lambda \left({\bf q}\right)$, in this approach reads
\begin{equation}
\label{trajectory}
\begin{split}
  \omega_0^\lambda \left({\bf q}\right)=
  -\bar{\alpha}_s& \ln{\frac{{\bf q}^2}{\lambda^2}} +
  \frac{\bar{\alpha}_s^2}{4}\left[\frac{\beta_0}{2 N_c}\ln{\frac{{\bf
          q}^2}{\lambda^2}}\ln{\frac{{\bf q}^2
        \lambda^2}{\mu^4}}\right.\\
  &+\left.\left(\frac{\pi^2}{3}-\frac{4}{3}-
      \frac{5}{3}\frac{\beta_0}{N_c}\right)\ln{\frac{{\bf
          q}^2}{\lambda^2}}+6 \zeta(3)\right]
\end{split}
\end{equation}
with 
\begin{eqnarray}
\xi \left({\rm X}\right) &\equiv& \bar{\alpha}_s +  
\frac{{\bar{\alpha}_s}^2}{4}\left(\frac{4}{3}-\frac{\pi^2}{3}+\frac{5}{3}\frac{\beta_0}{N_c}-\frac{\beta_0}{N_c}\ln{\frac{{\rm X}}{\mu^2}}\right) 
\end{eqnarray}
being the corresponding part in the real emission kernel. To complete the 
real part of the NLL kernel there are other more complicated terms in 
$\tilde{\cal K}_r$ which do not generate $\epsilon$ singularities when 
integrated over the full phase space of the emissions, for details see 
Ref.~\cite{Andersen:2003an}.

\begin{figure}[tbp]
  \centering
  \epsfig{width=0.5\textwidth,file=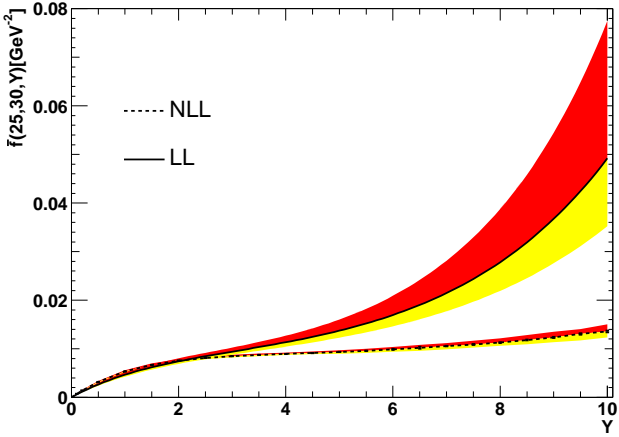,angle=0}
  \epsfig{width=0.5\textwidth,file=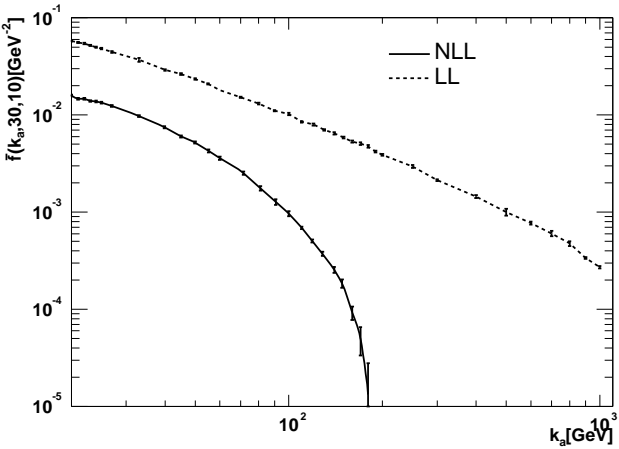,angle=0}
  \caption{Analysis of the gluon Green's function as obtained from the NLL 
BFKL equation.}
\label{JAfigure1}
\end{figure}

The numerical implementation and analysis of the form of 
solution as in Eq.~(\ref{solution1}) was carried out in Ref.~\cite{Andersen:2003wy}. At 
the light of this study the known feature of a lower intercept 
at NLL with respect to leading--order (LL) was confirmed. As in this approach 
it is not needed to expand on any eigenfunctions there are no instabilities 
in the energy growth. This is highlighted at the left hand side of 
Fig.~\ref{JAfigure1} where 
the bands correspond to uncertainties in the choice of renormalization scale.

\question{This figure needs larger labels... is hardly readable. In this figure,
the function f of eq(13) is plotted as a function of Y and k. We definitely need
a better description of this figure, and what exactly is plotted. What are the
two scales ? Maybe a small sketch could help ?}

However, the space where the convergence of the perturbative expansion
is poor is not in energy but in transverse momenta. In particular,
when the two transverse scales entering the forward gluon Green's
function are of comparable magnitude then the NLL corrections are
smaller when compared to LL, this can be seen in the bottom plot of
Fig~\ref{JAfigure1}. However when the ratio between these scales
largely departs from unity then the $|{\rm NLL-LL}|$ difference
becomes large, driving, as it is well--known, the gluon Green's
function into an oscillatory behavior with negative values.

\begin{figure}[tbp]
  \centering
  \epsfig{width=0.5\textwidth,file=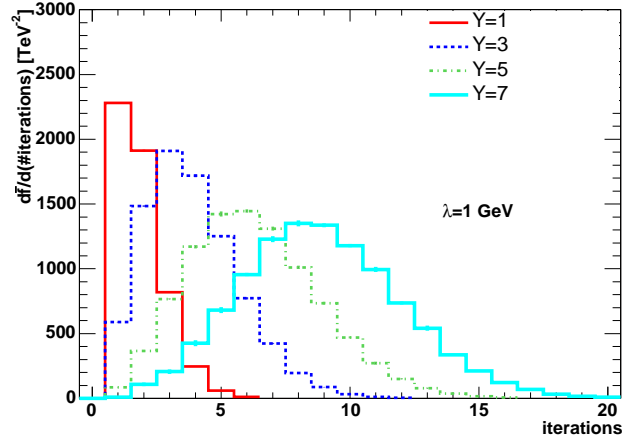,angle=0}
  \epsfig{width=0.5\textwidth,file=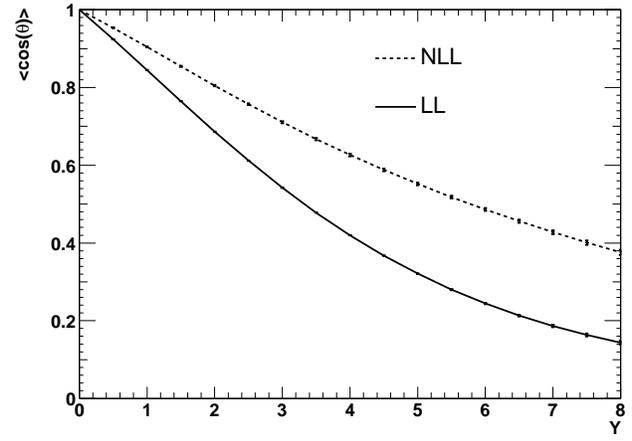,angle=0}
  \caption{Distribution in the number of iterations and angular dependence 
of the NLL gluon Green's function.}
\label{JAfigure2}
\end{figure}

The main advantage of the method here described is that the Green's
function is generated integrating the phase space using a Monte Carlo
sampling of the different parton configurations. This feature allows
for a full control of the average multiplicities and angular
dependences. The former can be extracted from the Poisson--like
distribution in the number of rungs, or iterations of the kernel,
needed to reach a convergent solution. This is obtained numerically in
the upper part of Fig.~\ref{JAfigure2}, where we see e.g.\ that for
$Y=5$ it is should be enough to include $\sim15$ rungs/iterations.
\question{For this figure also the values should be explained: what is
  plotted against what ?, What is actually obtained {\it numerically
    at the upper part of Fig.} ?}  At the lower part of the same
figure the angular correlations in the azimuthal angle of dijets with
similar and large transverse energy, and low hadronic activity in
between, is studied in a toy cross--section with simplified impact
factors. The increase of the angular correlation when the NLL terms
are included in such observable is a characteristic feature of these
corrections.  \question{So, what does that mean? Does it mean, that
  the decorrelation, which we had thought is a feature of the
  unordered cascade, decreases if one goes to NLO? If that is the
  case, this should be more stressed, as it could change our
  picture..}  This study is possible within this approach in an
immediate manner because the NLL kernel is treated in full, without
angular averaging, so there is no need to use a Fourier expansion in
angular variables via the introduction of conformal spins.

\begin{figure}[tbp]
  \centering
  \epsfig{height=0.5\textwidth,file=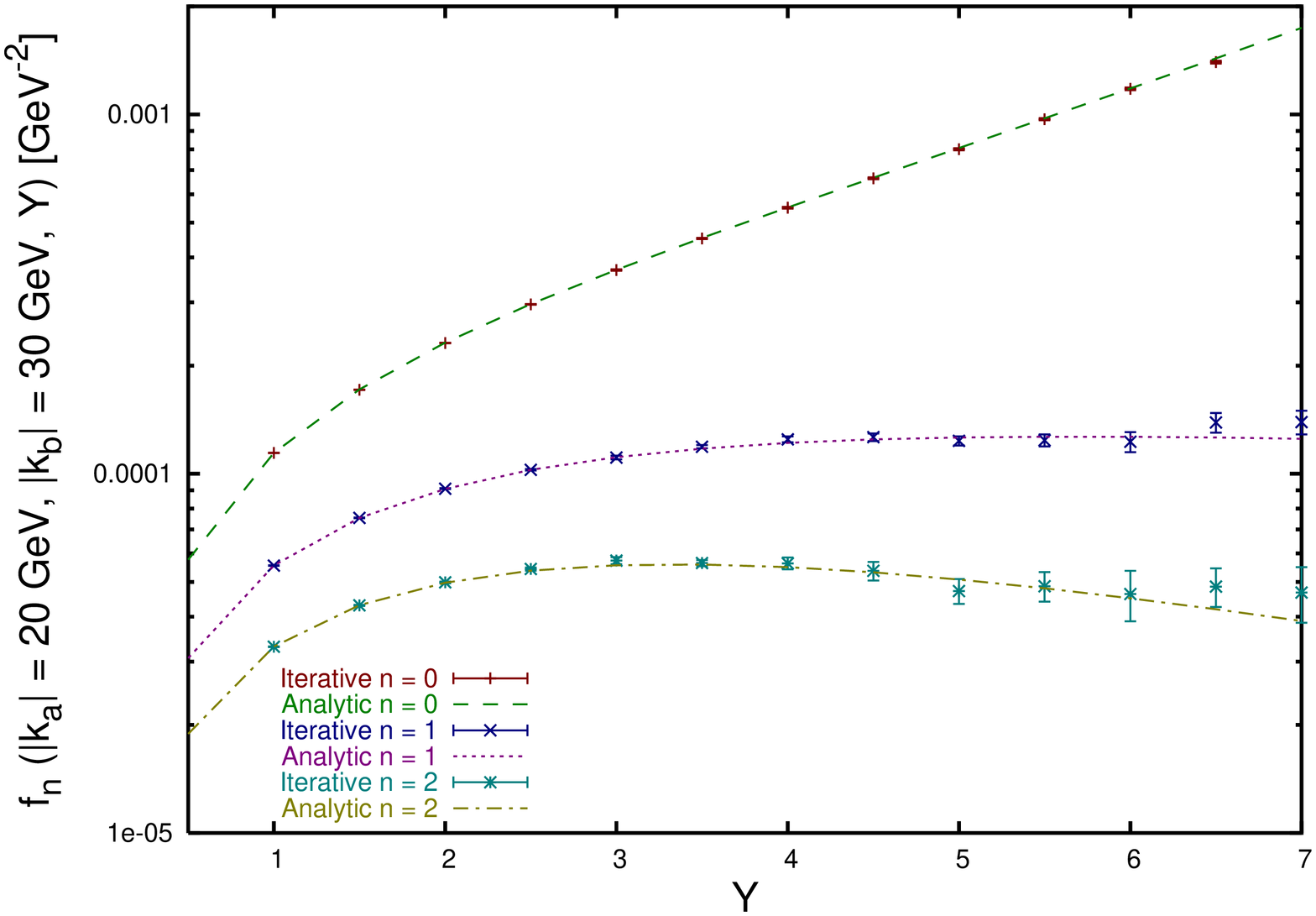,angle=-90}
  \epsfig{height=0.5\textwidth,file=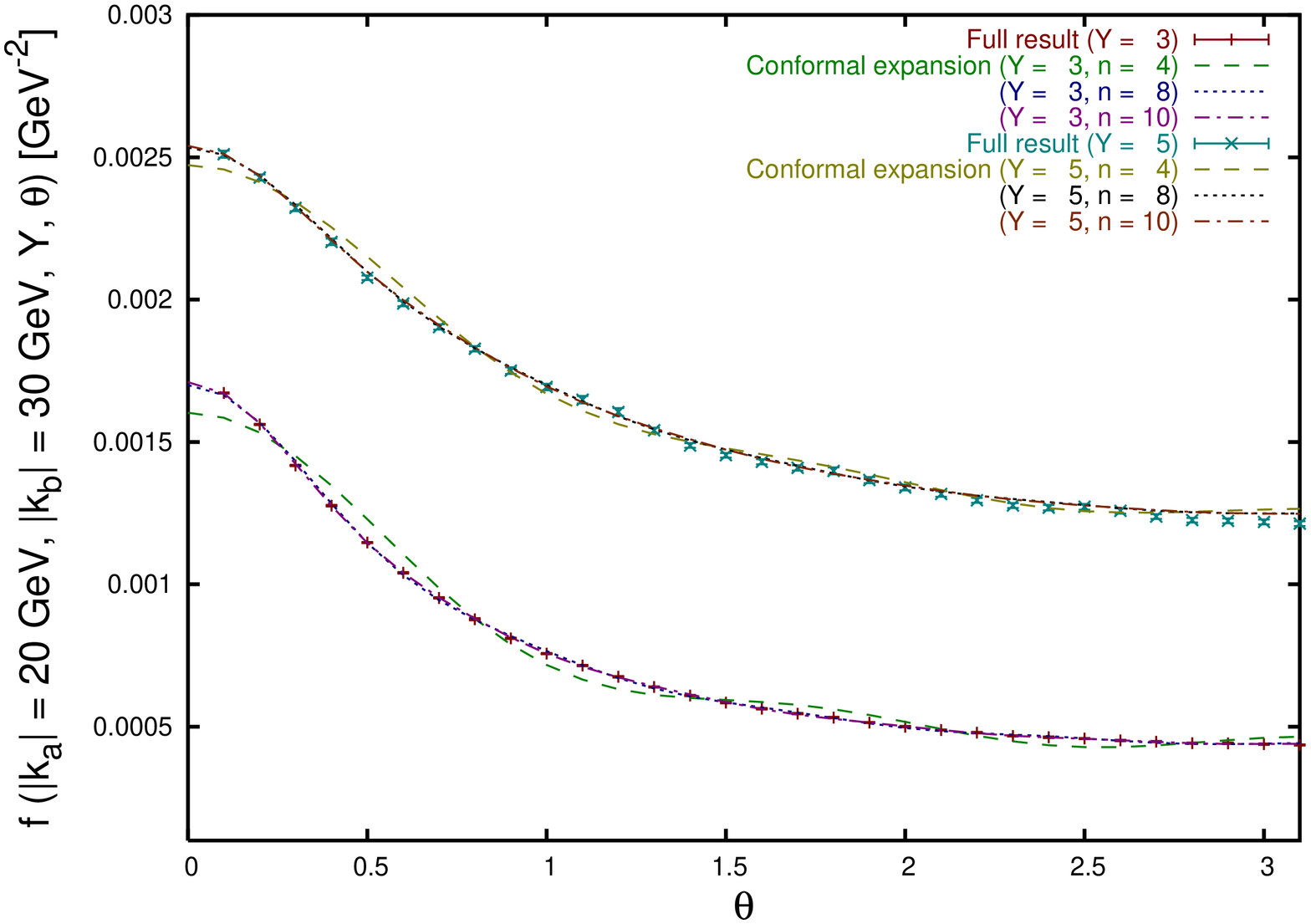,angle=-90}
  \caption{Projections on conformal spins of the N=4 SUSY NLL BFKL
    Green's function.}
\label{JAfigure3}
\end{figure}

An interesting theoretical development in the context of NLL BFKL was the 
calculation of the forward NLL kernel in the conformally invariant 
$N=4$ super Yang--Mills theory~\cite{Kotikov:2000pm,Kotikov:2002ab}. In such field theory the 
coupling remains a constant even at NLL, opening the possibility of finding 
the solution of the BFKL equation in a straightforward way because the 
LL eigenfunctions are also so at NLL. In particular, the kernel was calculated 
for all conformal spins in Ref.~\cite{Kotikov:2000pm,Kotikov:2002ab} allowing for the direct test 
of the angular structure of the solution as obtained from the method here 
described. This comparison between both approaches was performed in 
Ref.~\cite{Andersen:2004uj}. In this case the gluon Regge trajectory reads (with $a$ denoting the coupling constant)

\begin{equation}
\omega_0^\lambda \left({\bf{q}}\right) =  - a \ln{\frac{\bf{q}^2}{\lambda^2}}+ \frac{a^2}{4} \left[\left(\frac{\pi^2}{3}-\frac{1}{3}\right)\ln{\frac{\bf{q}^2}{\lambda^2}}+ 6 \, \zeta(3) \right]
\label{trajsusy}
\end{equation}
and $\xi =  a + a^2 \left(\frac{1}{12}-\frac{\pi^2}{12} \right)$ is 
a constant without logarithmic dependence. For a precise determination of 
the contribution to the gluon Green's function stemming from the different 
Fourier components in the azimuthal angle, {\it i.e.}
\begin{equation}
f\left(\bf{k}_a,\bf{k}_b, {\rm Y}\right)=
\sum_{n=-\infty}^{\infty} f_n\left(|\bf{k}_a|,|\bf{k}_b|, {\rm Y}\right) e^{i n \theta},
\end{equation}
it is enough to extract the coefficients of the expansion, either using 
the kernel calculated in~\cite{Kotikov:2000pm,Kotikov:2002ab}
\begin{equation}
f_n\left(|\bf{k}_a|,|\bf{k}_b|, {\rm Y}\right) =
\frac{1}{\pi |\bf{k}_a| |\bf{k}_b|} 
\int \frac{d \gamma}{2 \pi i} 
\left(\frac{\bf{k}_a^2}{\bf{k}_b^2}\right)^{\gamma-\frac{1}{2}}
e^{\omega_n (a,\gamma) {\rm Y}},
\end{equation}
or making use of the iterative solution explained in this section~\cite{Andersen:2004uj}:
\begin{equation}
f_n\left(|\bf{k}_a|,|\bf{k}_b|, {\rm Y}\right) =
\int_0^{2 \pi} \frac{d\theta}{2 \pi} \, 
f\left(\bf{k}_a,\bf{k}_b, {\rm Y}\right) \cos{\left(n \theta\right)}. 
\end{equation}
The results from these two independent alternatives are shown to
coincide in Fig.~\ref{JAfigure3}.  \question{Again this figure needs
  larger labels, and an explanation what is plotted against what.}  In
the upper part the $n=0$ Fourier component clearly dominates at large
energies, decreasing the angular correlations as the energy increases.
In the lower part it is shown how the convergence in the angular
variable on the transverse plane is achieved after only a few terms in
the Fourier expansion for different values of the available energy in
the scattering process.

In this section a new analysis of the gluon Green's function as obtained from 
the NLL BFKL kernel has been presented. The method of solution is based on 
the Monte Carlo integration of the phase space of different partonic 
configurations in the multi--Regge and quasi--multi--Regge kinematics. This 
method has many advantages with respect to previous analysis of the same 
problem. It allows for a reliable study of angular dependences in a 
straightforward manner, the multiplicities in the evolution are under control, 
and it provides an exact solution even with running coupling terms which 
break the scale invariance in the kernel. Many other studies are on their 
way using this procedure, as for example, deep inelastic scattering, the 
non--forward case and the matching of this solution to different impact 
factors for the final calculation of cross--sections at NLL where the BFKL 
approach will be relevant at present and planned colliders.


\subsection{Resummation at small $x$}

\textit{Main author A.~Stasto}\\

\begin{figure}
\centerline{\epsfig{file=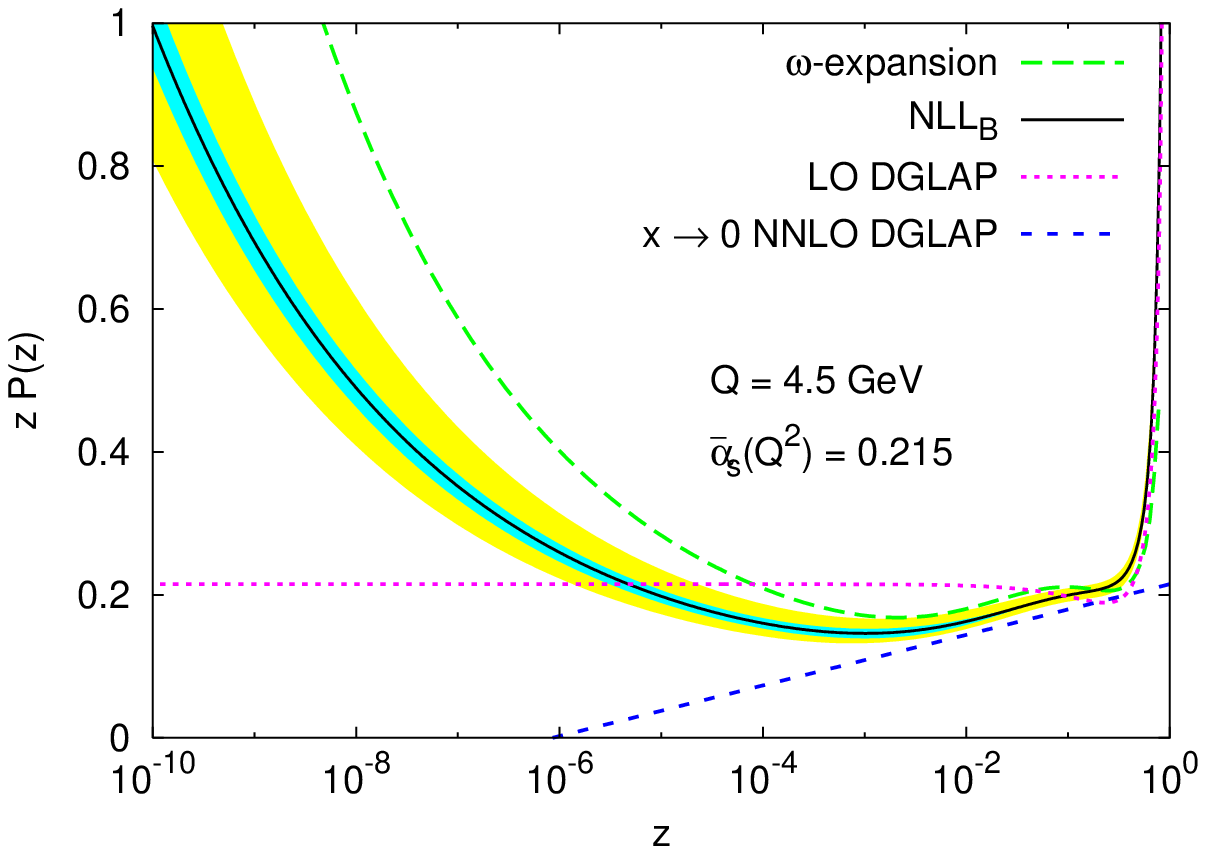,width=7cm}}
\label{fig:pggresum}
\end{figure}

The large magnitude of the NLLx correction in the high energy limit, as well 
as the instabilities
associated with it, motivate the study of the resummation  procedure in the 
limit of small $x$.
In particular it has been observed that, by taking into account collinear 
limits correctly in the NLLx
equation, as it is required by the DGLAP dynamics, stabilizes the high energy 
expansion.
To understand this in more detail let us recall the structure of the LLx BFKL 
equation in the 
Mellin space where the Mellin variable $\gamma $ is conjugated to the logarithm 
of the transverse momentum $\ln k_T^2/\Lambda^2$
\begin{equation}
\chi^{(0)}(\gamma) \; = \; 2 \psi(1) -\psi(\gamma)-\psi(1-\gamma) \; 
\sim \;\frac{1}{\gamma}+\frac{1}{1-\gamma}
\end{equation}
where in the pole expansion of the kernel eigenvalue we have retained only
 leading
collinear and anticollinear poles. These correspond exactly to the DGLAP 
strong
ordering of transverse momenta along the gluon ladder. In the NLLx case the
 eigenvalue
function takes on a complicated functional form which in the collinear limit is
\begin{equation}
\chi^{(1)}(\gamma) \simeq \frac{A_1(0)}{\gamma^2}+\frac{A_1(0)}{(1-\gamma)^2}
-\frac{1}{2\gamma^3}
-\frac{1}{2 (1-\gamma)^3)}+{\cal O}(\frac{1}{\gamma},\frac{1}{1-\gamma})
\label{eq:nllx}
\end{equation}
with $A_1(0)=-11/12$. Note the negative sign of the NLLx contribution.
It turns out that the collinear approximation above reproduces the exact 
result within  $\sim 7\%$ of accuracy.
The terms proportional to $A_1(0)$ are related to the non-singular in $x$ 
part of the LO DGLAP splitting function, whereas the cubic poles come from 
the energy scale choice.
The highly singular form of the NLLx correction as it is seen from 
eq.(\ref{eq:nllx}) is the source
of the large correction and potentially unstable behavior.
The resummation procedure presented in \cite{Ciafaloni:1999yw} is  based
 on four key ingredients:
\begin{itemize}
\item Taking into account the full splitting function at LO in the DGLAP 
approximation.
\item Incorporating the energy scale change in the form of the kinematical 
constraint.
\item Running of the coupling constant $\as$
\item Subtraction of the double and single poles in order to avoid double 
counting.
\end{itemize}
In \cite{Ciafaloni:2003rd} a procedure based on the numerical solution of the
BFKL equation in momentum
space was presented. It takes into account all of the above-mentioned 
ingredients and yields
stable result for the intercept and the gluon Green's function. Furthermore, 
the procedure for extraction the resummed splitting function was also 
presented, which is more relevant for application to the
deep inelastic scattering processes such as measured at HERA. 
In Fig.\ref{fig:pggresum} we show the
resummed splitting function obtained  in the resummed scheme
\cite{Ciafaloni:2003rd},
 together with the renormalization scale variation and the singular 
 in $x$ part of the NNLO DGLAP splitting function. 
  The characteristic 
 feature of the resummed splitting function $P_{gg}$ is the 
 strong preasymptotic behavior at intermediate values of 
 $x\simeq 10^{-3}-10^{-4}$ which 
 manifests itself in the dip of the splitting function, only later 
 followed by the increase at very small $x$.
 Also interesting is the fact that the small $x$ part of the  
 NNLO DGLAP splitting function matches
 nearly exactly  with the initial decrease of the resummed splitting 
 function.
 The existence of the dip rather than an 
 increase at values of $x\sim 10^{-4}$ 
 can have an interesting impact on the phenomenology.

\subsection{The NLO $\gs$ impact factor}
\label{sec:nloimpact}

\textit{Main author A.~Kyrieleis}\\

\begin{center}
  \begin{figure}[h]
    \epsfig{file=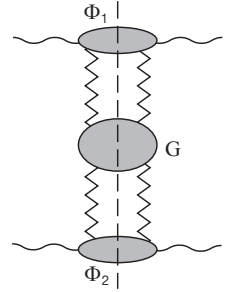,width=5.2cm}
    \caption{\label{BFKL_fac} $\sigma_{tot}^{\gamma^*\gamma^*}$ in the
      framework of BFKL}
  \end{figure}
\end{center}

One of the most attractive observables to test the BFKL approach is
the total cross section for $\gs\gs$ scattering.  To calculate this
observable in the framework of NLO BFKL the $\gs$ impact factor
($\Phi$) at NLO is needed in addition to the universal BFKL Green
function (G), see Fig.\ref{BFKL_fac}.
If the NLO BFKL equation is solved in the momentum space the numerical
value of the $\gs$ impact factor has to be known as a function of the
Reggeon momentum and of the energy scale.

Besides this, the NLO $\gs$ impact factor allows to approach the
resummation of the next-to-leading logs(1/x) in the quark anomalous
dimensions.  It also provides the full information necessary to
investigate the color dipole picture at NLO which, at LO, is one of
the important ingredients to the QCD evolution based upon the
Balitsky-Kovchegov equation (see section 5 below).  At the first
small-x workshop \cite{Andersson:2002cf} first steps in the
calculation of this impact factor have been presented.

The virtual and the real corrections of the $\gs$ impact factor are
calculated from the photon-Reggeon vertices for $q\bar q$ and $q\bar
qg$ production, respectively.  Both vertices are known
\cite{Bartels:2000gt,Fadin:2001ap,Bartels:2001mv,Bartels:2002uz}.
What remains to complete the calculation of the NLO photon impact
factor after the infrared divergences of the virtual and of the real
parts have been combined \cite{Bartels:2002uz} are the integrations over the
$q\bar q$ and $q\bar qg$ phase space, respectively.
 
Recently, the phase space integration in the real corrections have
been performed for the case of longitudinal photon polarization,
\cite{Bartels:2004bi}.  The integration over the transverse momenta
have been carried out analytically.
To this end the Feynman diagrams were treated separately giving rise
to additional divergences that have been regularized.
As the result, a convergent Feynman parameter integral has been
obtained for each Feynman diagram (or small groups of them). These
results can serve as a starting point for further analytical
investigations, in particular because the Mellin transform of the real
corrections w.r.t the Reggeon momentum can be easily obtained.

The remaining integrations in the real corrections (longitudinal $\gs$
polarization) have been carried out numerically \cite{Bartels:2004bi}.
The result is a function $\Phi^{real}$ of two dimensionless (scaled by
the photon virtuality) variables: the Reggeon momentum $\br^2$ and the
energy scale $s_0$.  A physical scattering amplitude (e.g. for the
$\gs\gs$ scattering process) involving the BFKL Green's function and
the impact factors has to be invariant under changes of $s_0$. The
$s_0$ dependence of the $\gs$ impact factor therefore represents an
important issue.  $s_0$ enters the NLO $\gs$ impact factor as a cutoff
to exclude that region of the $q{\bar q}g$ phase space where the gluon
is separated in rapidity from the $q{\bar q}$ pair (LLA).  The virtual
corrections are therefore independent of $s_0$ and the integration of
the real corrections alone already allows to study the $s_0$
dependence of the NLO $\gs$ impact factor. Let us define, as part of
the full NLO impact factor:
\begin{displaymath}
  \Phi' = g^2\Phi^{(0)} + g^4\Phi^{\mathrm real}
\end{displaymath}
where $g^2 \Phi^{(0)}$ denotes the LO $\gs$ impact factor and
$g^2=4\pi \alpha_s$.
\begin{figure}[h]
  \begin{center}
    \vspace{-.8cm}
    \epsfig{file=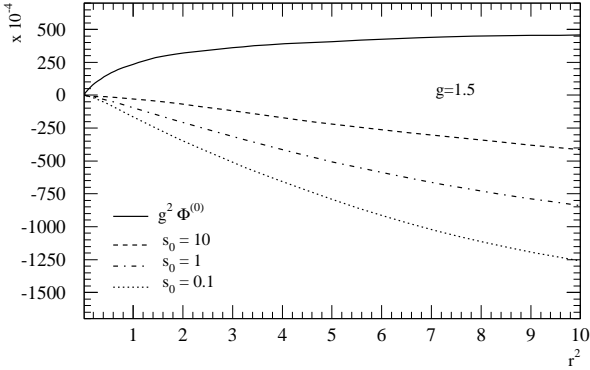,width=9cm}
    \caption{\label{ifac} $\Phi' $ at different different values of
      $s_0$}
  \end{center}
\end{figure}
Choosing $Q^2=15$ GeV$^2$ for the photon virtuality leads
$\alpha_s(Q^2)=0.18$ or $g=1.5$.  Fig.\ref{ifac} compares $\Phi'$ to
the LO impact factor as function of $\br^2$ at different values of
$s_0$.  The real corrections are negative and rather large. More
important, $\Phi'$ becomes, in absolute terms, more significant for
smaller values of $s_0$.  This implies that the $\gs$ impact factor
tends to become smaller with decreasing $s_0$. Since a decrease of
$s_0$ in the energy dependence $(\frac{s}{s_0})^\omega$ will enhance
the scattering amplitude, the combined $s_0$ dependence of the impact
factors and the BFKL Green's function has to compensate this growth.
The result for the $s_0$ behavior of the $\gs$ impact factor is
therefore, at least, consistent with the general expectation. To check
the $s_0$ (in)dependence of the full scattering amplitude and to
compute $\sigma^{\gs\gs}$, at least for longitudinal $\gs$
polarization, the phase space integration in the virtual corrections
is the only piece missing.


\section{Applications of \kt-factorization}
\label{sec:ktapp}

In collinear factorisation the transverse momenta of the incoming partons are
neglected whereas they are included in \kt-factorization if the 
same order in \as\  of the calculation is considered. Thus in collinear
factorsiation these transverse momentum effects
come in as a next-to-leading order level. 

In the following sections we discuss some applications of
\kt-factorization to describe heavy quark production in $p\bar{p}$
collisions.

\subsection{Heavy quark production at the Tevatron}
\label{sec:heavytev}

\textit{Main author N.~Zotov}\\

Heavy quark production in hard collisions of hadrons has been
considered as a clear test of perturbative QCD. Such processes provide
also some of the most important backgrounds to new physics phenomena
at high energies.
  
Bottom production at the Tevatron in the $\kt$-factorization approach
was considered earlier
in~\cite{Levin:1991ry,Ryskin:1995sj,Ryskin:2000bz,Ryskin:1999yq,Hagler:2000dd,Jung:2001rp,Baranov:2000gv,Lipatov:2001ny}.
Here we use the $\kt$-factorization approach for a more detailed
analysis of the experimental
data~\cite{Abbott:1999se,Abe:1996zt,Acosta:2002qk,Acosta:2001rz,Abbott:1999wu}.
The analysis also covers the azimuthal correlations between $b$ and
$\bar b$ quarks and their decay muons.  Some of these results have
been presented earlier in
Refs.~\cite{Lipatov:2001ny,Baranov:1999ja,Baranov:2000uv,Zotov:2003gc,Lipatov:2002tc,Kotikov:2001ct,Kotikov:2002nh}
(see also \cite{Andersson:2002cf,Andersen:2003xj}).

\subsection{Theoretical framework} 

In the $\kt$-factorization approach, the differential cross section
for inclusive heavy quark production may be written as
(see~~\cite{Baranov:2004eu})
\begin{align}
  d\sigma(p\bar p\to& Q\bar Q\,X) =\nonumber\\
  &{1\over 16\pi (x_1\,x_2\,s)^2}\,\mathcal{A}(x_1,{\mathbf q}_{1T}^2,\mu^2)\,
  \mathcal{A}(x_2,{\mathbf q}_{2T}^2,\mu^2)\nonumber\\
  \times&
  \sum {|M|^2_{{\rm SHA}}(g^*g^*\to Q\bar Q)}\nonumber\\
  \times& dy_1\,dy_2\,d{\mathbf p}_{2T}^2\,d{\mathbf q}_{1T}^2\,
  d{\mathbf q}_{2T}^2{d\phi_1\over 2\pi}\,{d\phi_2\over 2\pi}\,
  {d\phi_Q\over 2\pi},
  \label{nikolai-eq1}
\end{align}

\noindent where $\mathcal{A}(x_1,{\mathbf q}_{1T}^2,\mu^2)$ and 
$\mathcal{A}(x_2,{\mathbf q}_{2T}^2,\mu^2)$ are unintegrated gluon
distributions in the proton, ${\mathbf q}_{1T}$, ${\mathbf q}_{2T}$,
${\mathbf p}_{2T}$ and $\phi_1$, $\phi_2$, $\phi_Q$ are transverse
momenta and azimuthal angles of the initial BFKL gluons and final
heavy quark respectively, $y_1$ and $y_2$ are the rapidities of heavy
quarks in the $p\bar p$ center of mass frame.  $\sum {|M|^2_{{\rm
      SHA}}(g^*g^*\to Q\bar Q)}$ is the off mass shell matrix element,
where the symbol $\sum$ in (\ref{nikolai-eq1}) indicates an averaging
over initial and a summation over the final polarization states. The
expression for $\sum {|M|^2_{{\rm SHA}}(g^*g^*\to Q\bar Q)}$ coincides
with the one presented in~\cite{Catani:1991eg}.

In the numerical analysis, we have used the KMS
parameterization~\cite{Kwiecinski:1997ee} for the $\kt$-dependent
gluon density.  It was obtained from a unified BFKL and DGLAP
description of $F_2$ data and includes the so called consistency
constraint~\cite{Kwiecinski:1996td}.  The consistency constraint
introduces a large correction to the LO BFKL equation; about 70\% of
the full NLO corrections to the BFKL exponent $\lambda$ are
effectively included in this constraint, as is shown
in~\cite{Kwiecinski:1996td,Kwiecinski:1999hv}.

\subsection{Numerical results} 

In this section we present the numerical results of our calculations 
and compare them with $B$-meson production at 
D0~\cite{Abbott:1999se,Abbott:1999wu},
CDF~\cite{Abe:1996zt,Acosta:2002qk,Acosta:2001rz} and
UA1~\cite{Albajar:1993be}.

Besides the choice of the unintegrated
gluon distribution, the results depend on the 
bottom quark mass, the factorization scale $\mu^2$ and the 
$b$ quark fragmentation function. As an example, Ref.~\cite{Cacciari:2002pa}
used a special choice of the $b$-quark fragmentation function, as a
way to increase the $B$ meson cross section in the observable range
of transverse momenta. In the present paper we convert $b$ quarks 
into $B$ mesons using the standard Peterson fragmentation function
~\cite{Peterson:1982ak} 
with $\epsilon = 0.006$. Regarding the other parameters, we use  
$m_b = 4.75\,{\rm GeV}$ and $\mu^2 = {\mathbf q}_{T}^2$ as 
in~\cite{Levin:1991ry,Hagler:2000dd}.

\begin{figure}
  \begin{center}
    \includegraphics[clip,width=0.35\textwidth]{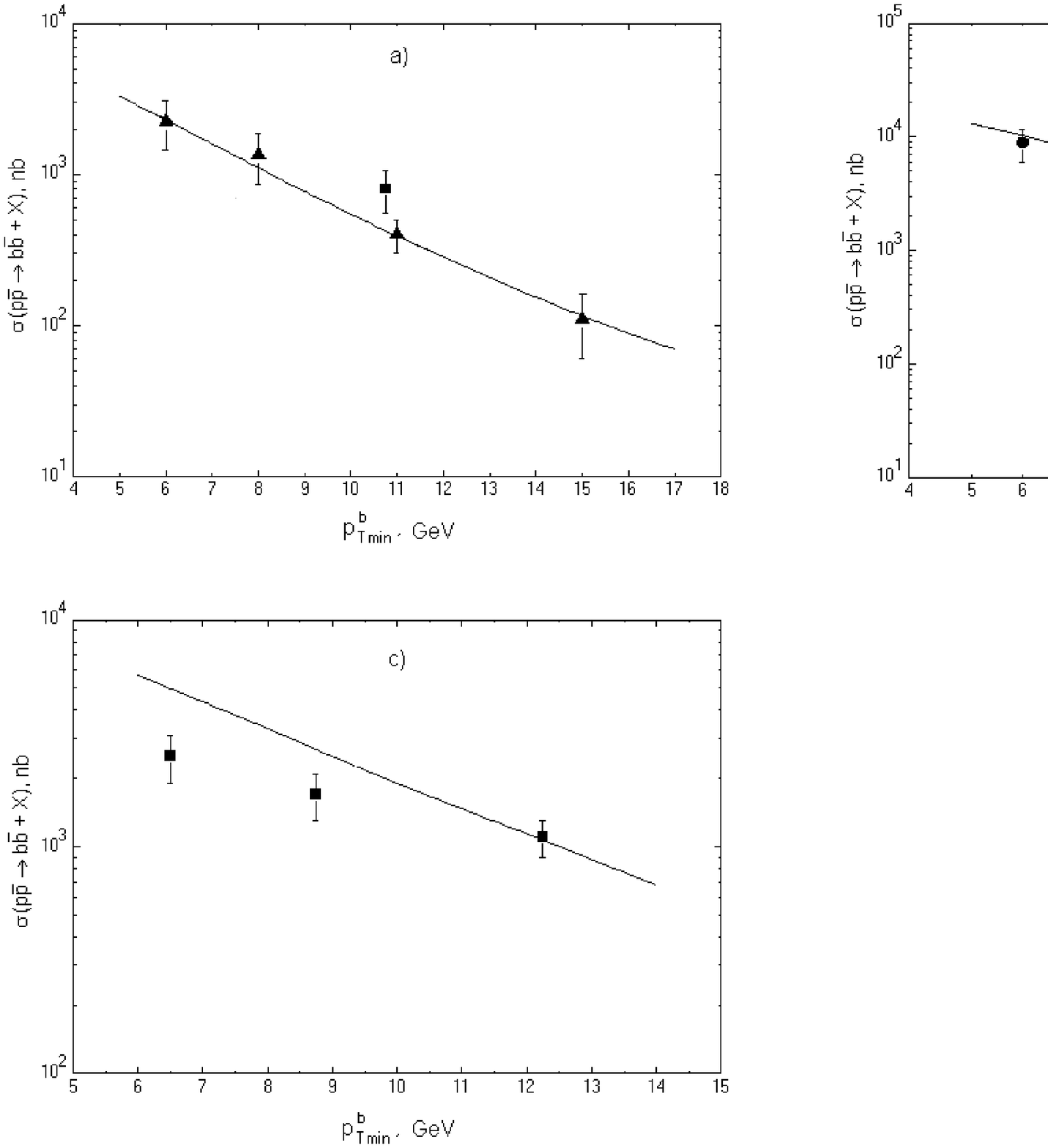}
    \includegraphics[clip,width=0.35\textwidth]{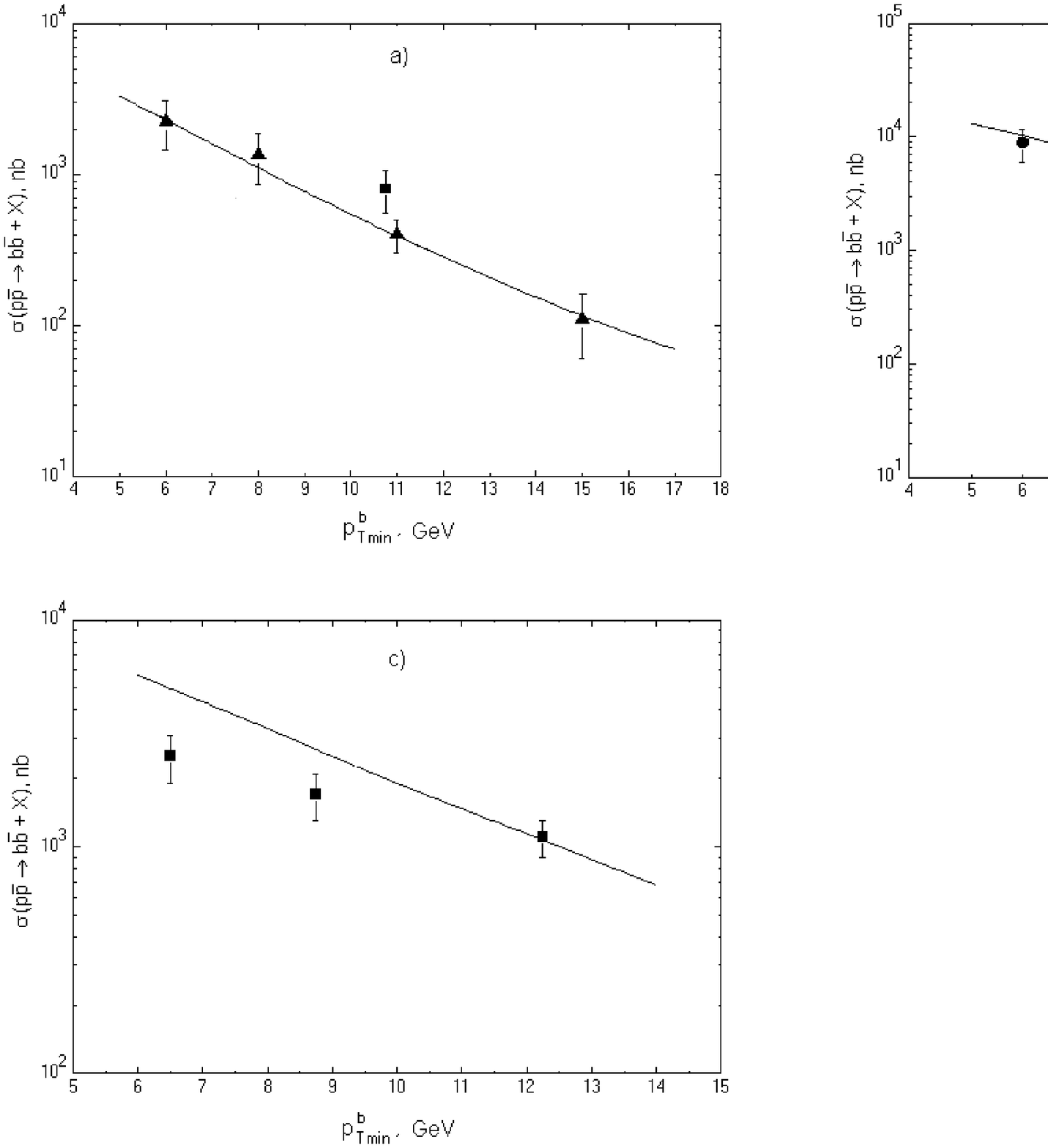}
    \includegraphics[clip,width=0.35\textwidth]{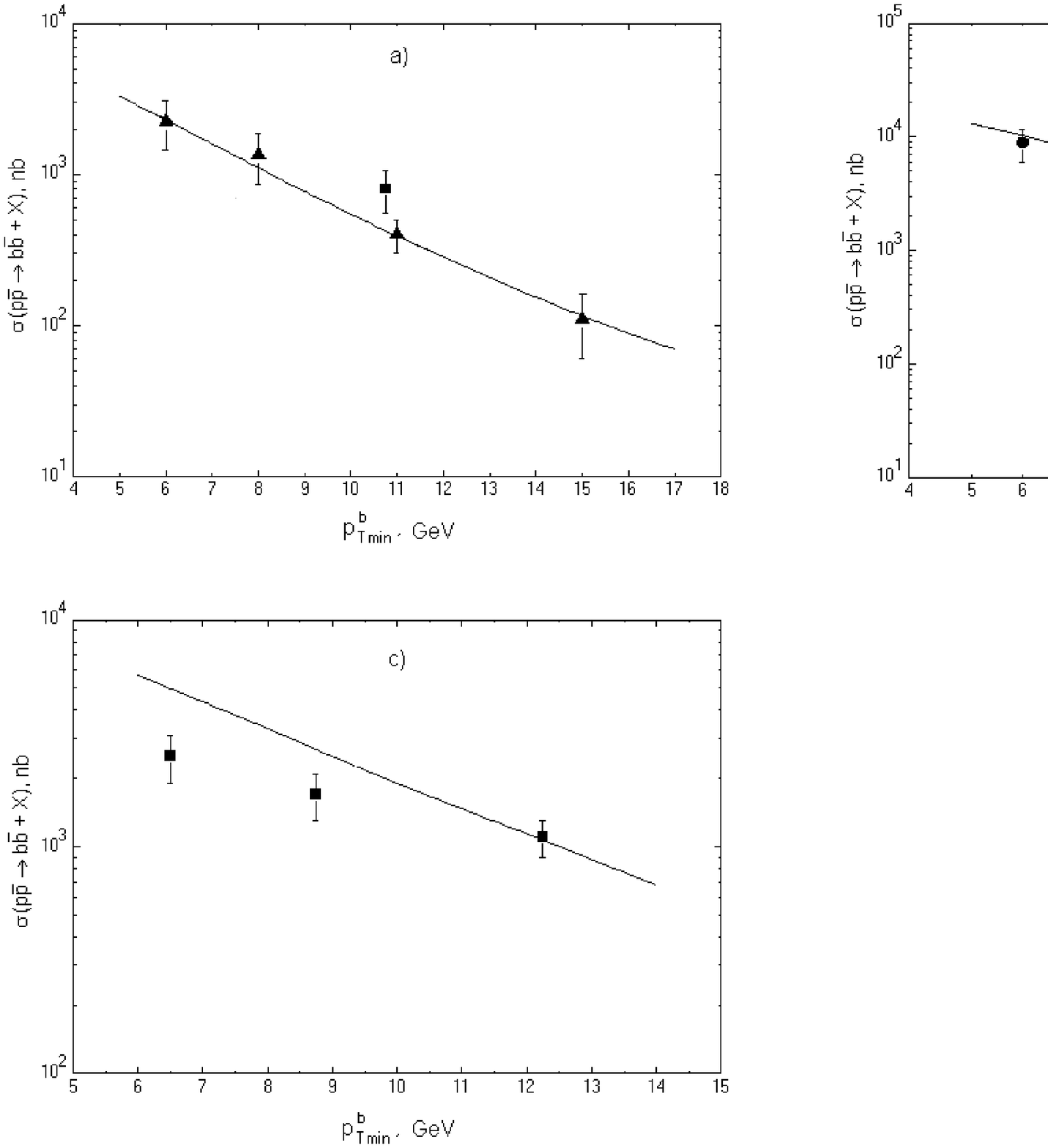}
  \end{center}
  \caption{The $b$ quark transverse momentum distribution
    (integrated from $p_{T\min}^b$) at Tevatron conditions presented
    in the form of integrated cross sections.  The curves correspond
    to the $\kt$-factorization results with the KMS unintegrated gluon
    distribution.  Experimental data are from
    UA1~\protect\cite{Albajar:1993be} (Fig. a)),
    D0~\protect\cite{Abbott:1999se} (Fig. b)), and
    CDF~\protect\cite{Abe:1996zt,Acosta:2002qk} (Fig. c)).}
\label{fig6}
\end{figure}

\begin{figure}
  \begin{center}
    \epsfig{figure=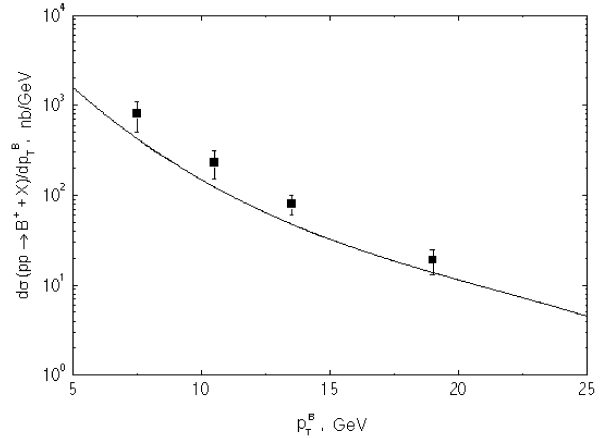,width=0.5\textwidth}
  \end{center}
  \caption{Theoretical predictions for the $B$ meson $p_T$ spectrum compared 
    to the CDF~\protect\cite{Acosta:2001rz} data. Curve is the same as
    in Fig.~\protect\ref{fig6}.}
\label{fig7}
\end{figure}

\begin{figure}
  \begin{center}
    \includegraphics[clip,width=0.35\textwidth]{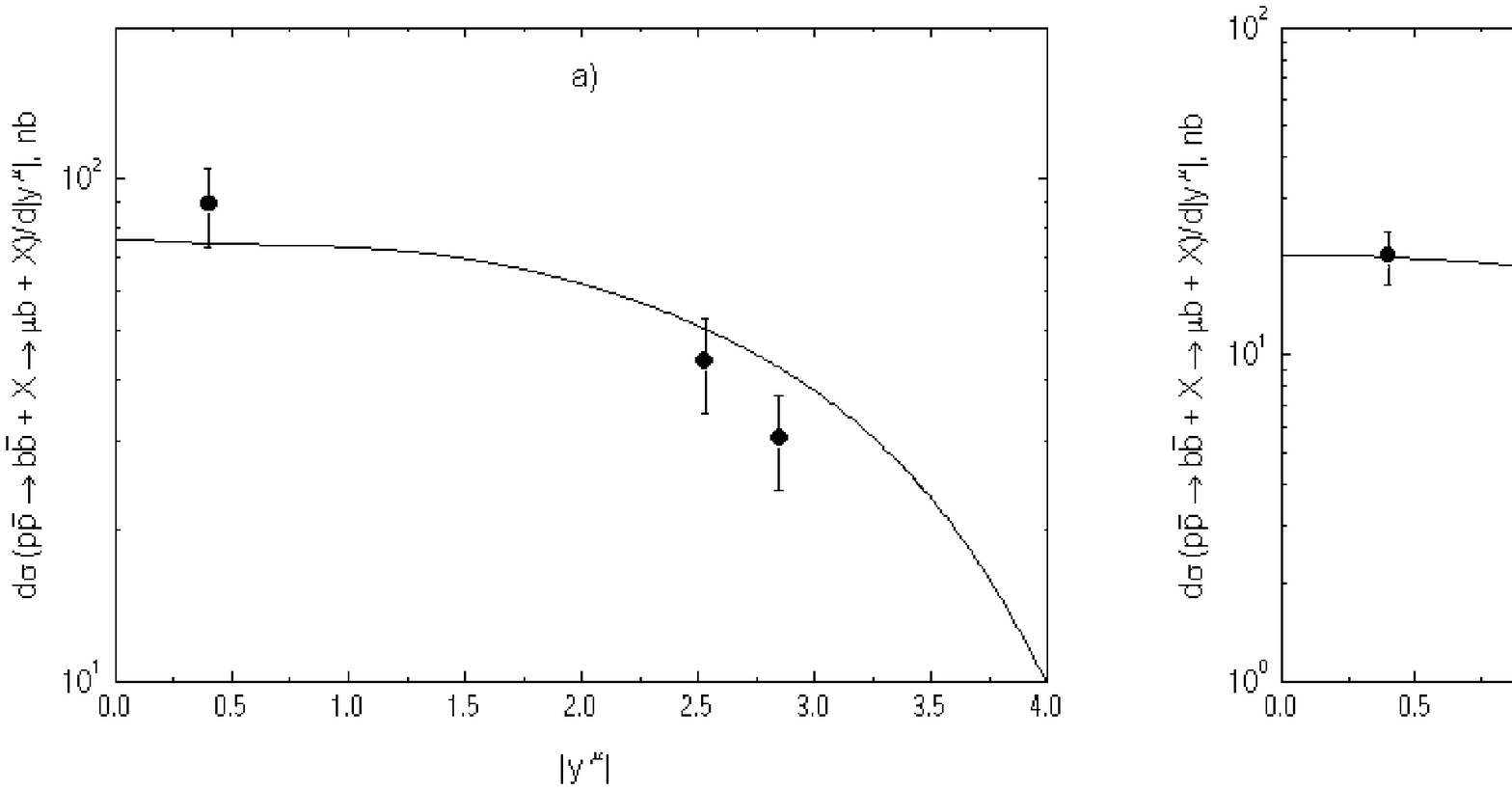}
    \includegraphics[clip,width=0.35\textwidth]{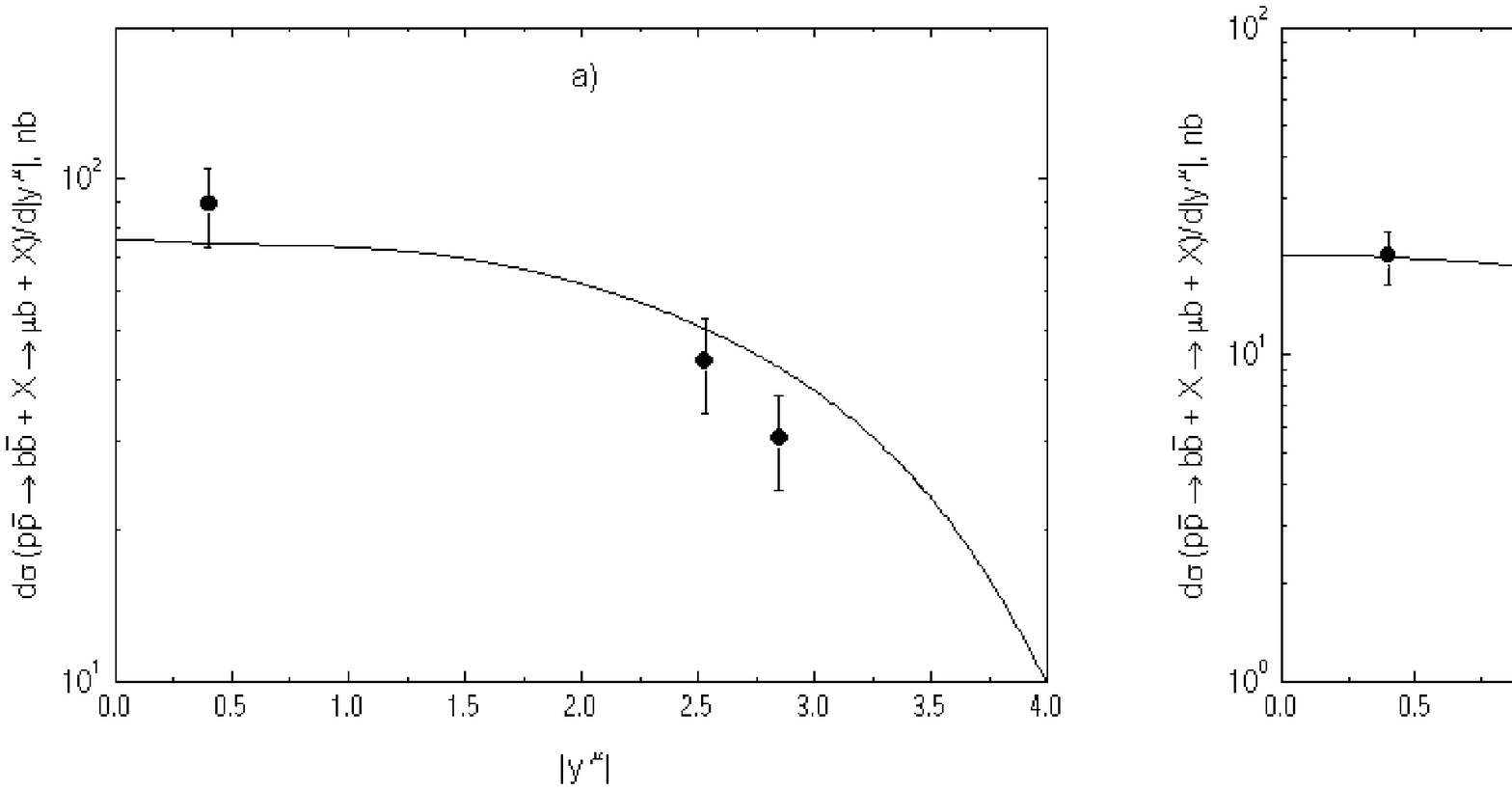}
  \end{center}
  \caption{The cross section for muons from $B$ meson decay as a function of
    rapidity compared to the D0 data~\protect\cite{Abbott:1999wu}. The
    curves are the same as in Fig.~\protect\ref{fig6}.}
\label{fig8}
\end{figure}

\begin{figure}
  \begin{center}
    \epsfig{figure=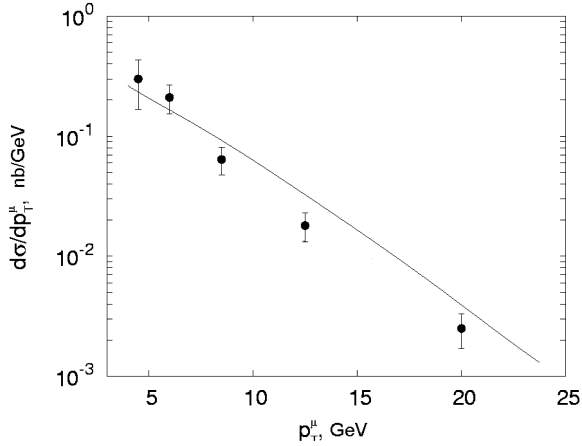,width=8.5cm}
  \end{center}
  \caption{Predictions for the leading muon $p_T$ spectrum in the
    $b\bar b$ production events compared to the D0
    data~\protect\cite{Abbott:1999se}.  The curve is the same as in
    Fig.~\protect\ref{fig6}.}
  \label{fig9}
\end{figure}

\begin{figure}
  \begin{center}
    \epsfig{figure=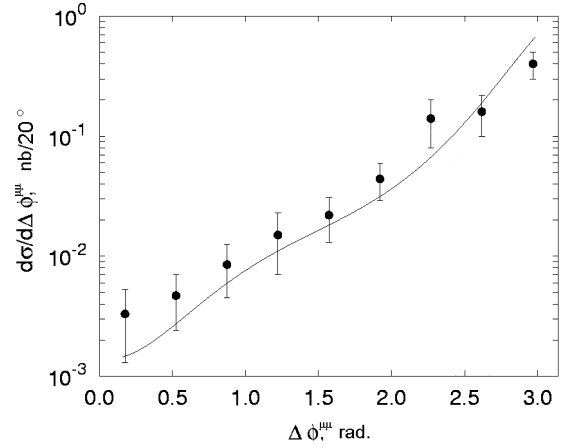,width=8.5cm}
  \end{center}
  \caption{Azimuthal muon-muon correlations at Tevatron conditions.
    The curve is the same as in Fig.~\protect\ref{fig6}. Experimental
    data are from the D0 collaboration~\protect\cite{Abbott:1999se}.}
  \label{fig10}
\end{figure}

The results of the calculations are shown in
Figs.~\ref{fig6}-\ref{fig10}.  Fig.~\ref{fig6} displays the $b$ quark
transverse momentum distribution at Tevatron conditions presented in
the form of integrated cross sections. The following cuts were
applied: (a) $|y_1| < 1.5$, $|y_2| < 1.5$, $\sqrt s = 630 \,{\rm
  GeV}$; (b) $|y_1| < 1$, $\sqrt s = 1800 \,{\rm GeV}$; and (c) $|y_1|
< 1$, $|y_2| < 1$, $\sqrt s = 1800 \,{\rm GeV}$.  One can see
reasonable agreement with the experimental data.

Fig.~\ref{fig7} shows the prediction for the $B$ meson $p_T$ spectrum
at $\sqrt s = 1800 \,{\rm GeV}$ compared to the CDF
data~\cite{Abe:1996zt} within the experimental cuts $|y| < 1$, where
also a fair agreement is found between results obtained in the
$\kt$-factorization approach and experimental data.

The  D0  data include also muons originating from
the semileptonic decays of $B$-mesons. To produce muons from $B$
mesons in theoretical calculations, we simulate their semileptonic
decay according to the standard electroweak theory. In Fig.~\ref{fig8}
we show the rapidity distribution $d\sigma/d|y^{\mu}|$ for decay muons
with $p_T^{\mu} > 5 \,{\rm GeV}$.

Fig.~\ref{fig9} shows the leading muon $p_T$ spectrum for $b\bar b$ 
production events
compared to the D0 data. The cuts applied to both muons are given by
$4 < p_{T}^{\mu} < 25 \,{\rm GeV}$, $|\eta^{\mu}| < 0.8$ and
$6 < m^{\mu\mu} < 35 \,{\rm GeV}$. The leading muon in the event 
is defined as the muon with largest $p_{T}^{\mu}$-value.
In all the above cases a rather good description of the experimental
measurements is achieved.

It has been pointed out that investigations of $b\bar b$ correlations, 
such as the azimuthal opening angle between $b$ and $\bar b$ quarks 
(or between their decay muons), allow additional details of the
$b$ quark production to be tested, since these quantities are
sensitive to the relative contributions of the different production 
mechanisms~\cite{Levin:1991ry,Ryskin:1995sj,Ryskin:2000bz,Ryskin:1999yq,Hagler:2000dd,Baranov:2000gv}.
In the collinear approach at LO the gluon-gluon fusion mechanism gives
simply a delta function, $\delta (\Delta \phi^{b\bar b} - \pi)$,
for the distribution
in the azimuthal angle difference $\Delta \phi^{b\bar b}$.
In the $\kt$-factorization approach the 
non-vanishing initial gluon transverse momenta, 
$q_{1T}$ and $q_{2T}$, implies that this back-to-back 
quark production kinematics is modified.
In the collinear approximation this effect can only be achieved
if NLO contributions are included.

The differential $b\bar b$ cross section $d\sigma/d\Delta \phi^{\mu \mu}$
is shown in Fig.~\ref{fig10} (from~\cite{Baranov:2004eu}). 
The following cuts were
applied to both muons: 
$4 < p_{T}^{\mu} < 25 \,{\rm GeV}$, 
$|\eta^{\mu}| < 0.8$ and $6 < m^{\mu\mu} < 35 \,{\rm GeV}$.
We note a significant deviation from the pure back-to-back production,
corresponding to $\Delta\phi^{\mu\mu}\approx\pi$.
There is good agreement between the KMS prediction 
and the experimental data, which shows that for these correlations 
the $k_\perp$-factorization scheme
with LO matrix elements very well reproduces the NLO effects due to the 
gluon evolution.

\subsection{Quarkonium production}
\label{sec:onium}

\textit{Main author S.~Baranov}\\

\def\kf {$\kt$-factorization }
\def\J {$J/\psi$}
\def\xc {$\chi_c$ }
\def\Y {$\Upsilon$ }
\def\xb {$\chi_b$ }
\def\R {{\cal R}}
\def\O {{\cal O}}
\def\Landau {Landau-Yang }

The \kf approach has rather successfully described the production of
open charm and beauty, as discussed in the previous section, but also
hadroproduction of heavy quarkonium states, \J, \xc and \Y mesons, at
the Tevatron are well
described\cite{Hagler:2000dd,Hagler:2000eu,Yuan:2000cp,Yuan:2000qe}.
In many cases, however,
the data can also be described within the usual collinear parton
model, if the relevant next-to-leading order QCD corrections are taken
into account, or if the so called color-octet mechanism is included.

In this context, the theoretical predictions on \J~spin alignment made
in Ref. \cite{Baranov:1998af} are of particular interest, as the
collinear and \kf approaches show qualitatively different behavior.
Note that the \kf approach provides the only known (up to date)
explanation of the \J~ polarization phenomena observed at the Tevatron
\cite{Baranov:2002cf} and at HERA \cite{Lipatov:2002tc}.

It would be interesting and important to find other examples, where
the difference between the collinear and noncollinear approaches would
be manifested in a clear and unambiguous way. In this section we
suggest such a process. We analyze the production of $P$-wave
quarkonium states (namely the \xc and \xb mesons) in high energy
hadronic collisions and demonstrate the dramatic difference between
the different theoretical calculations.

Naively one could expect a difference from the fact that the
production of $\chi_1$ states in the $2\to 1$ gluon-gluon fusion
process is forbidden, if the initial gluons are on shell, but is
allowed if the gluons are off shell. However, the real situation is
complicated by the necessity to take into account also the $2\to 2$
processes. The results of our analysis are presented in the
next subsection.

We begin our discussion with showing the predictions of the collinear
parton model for the production of $P$-wave charmonia at Tevatron
conditions. The color-singlet production scheme refers to the $2\to
2$ gluon-gluon fusion subprocess
\begin{equation}
  g+g\to\chi +g.
  \label{gg-xg}
\end{equation}
(It would be inadequate to rely upon the $2\to 1$ subprocess
$g+g\to\chi$ in this case, because the final state particle would then
be produced with zero transverse momentum, and thus could not be
detected experimentally.)  The computational technique is explained
in detail elsewhere \cite{Krasemann:1978jc,Guberina:1980dc,Kniehl:2003pc}.

For the sake of definiteness, we only present the parameter setting
used in our calculations. Throughout the paper we use the LO GRV set 
\cite{Gluck:1994uf} for gluon densities in the proton, and
the value for the \xc wave function, $|\R'_{\chi_c}(0)|^2 = 0.075$\ 
GeV$^5$, taken from the potential model of Ref. \cite{Eichten:1995ch}.
The renormalization scale in the strong coupling constant
$\alpha_s(\mu_R^2/\Lambda^2)$ is set to
$\mu_R^2=m_{\chi}^2+p_{T,\chi}^2$ with $\Lambda$=200 MeV.  The
integration over the final state phase space is restricted to the
pseudorapidity interval $-0.6<\eta(\chi_c)<0.6$, in accord with the
experimental cuts used by the CDF collaboration
\cite{Abe:1992ww,Abe:1993vw,Abe:1995dv,Abe:1997yz,Affolder:2001ij,Affolder:1999wm}.

Since in the collinear formalism the predictions based on the color-singlet mechanism alone are
known to be inconsistent with the data
\cite{Abe:1992ww,Abe:1993vw,Abe:1995dv,Abe:1997yz,Affolder:2001ij,Affolder:1999wm},
the theory has to be amplified with the so called color-octet
contribution, as it is commonly assumed in the literature
\cite{Kniehl:2003pc}.  Unlike the predictions of the color-singlet model,
the size of the color-octet matrix elements are not calculable within
the theory.  Therefore, the corresponding numerical results are always
shown with arbitrary normalizing factors (just chosen to fit the
experimental data when possible).

\begin{figure}
  \hskip -2cm\epsfig{figure=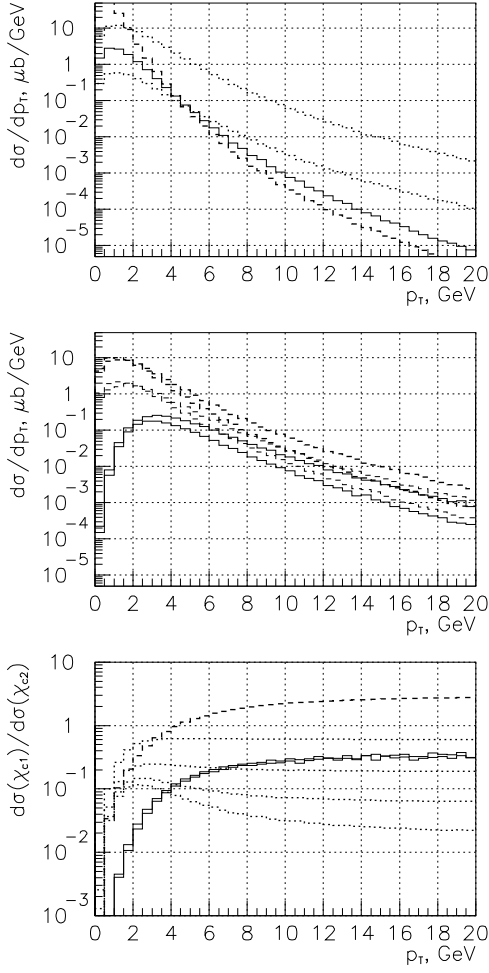,width=0.75\textwidth}
  \caption{\label{fig:chic} Theoretical predictions for the production
    of $\chi_c$ mesons at Tevatron conditions.
    {\bf Upper:}  Predictions of the collinear parton model.
    Solid histogram, $\chi_{1}$ production via color-singlet
    mechanism; dashed histogram, $\chi_{2}$ production via
    color-singlet mechanism; the lower and the upper dotted
    histograms, $\chi_{1}$ and $\chi_{2}$
    production via color-octet mechanism, respectively.
    {\bf Middle:}  Predictions of the \kf approach.  Solid
    histograms, $\chi_{1}$ production; thin and thick dashed
    histograms, $\chi_{0}$ and $\chi_{2}$ production, respectively.
    The upper and the lower histograms of each type correspond to the
    gluon densities of Refs.\ \protect\cite{Blumlein:1995eu} and
    \protect\cite{Gribov:1984tu}. Only the
    color singlet mechanism is assumed in all cases.
    {\bf Lower:}  Predictions on the ratio of the differential
    cross sections $d\sigma(\chi_{1})/d\sigma(\chi_{2})$.  Solid
    histograms, \kf approach with gluon densities of Refs.\ 
    \protect\cite{Blumlein:1995eu} and \protect\cite{Gribov:1984tu};
    dashed histogram, collinear parton model, color singlet
    contribution only; dotted histograms, collinear parton model with
    both singlet and octet production mechanisms taken into account.
    The different curves from top to bottom correspond to the
    color-octet $\chi_{1}/\chi_{2}$ suppression factor set to 1, 0.3,
    0.1 and 0.03, respectively.}

\end{figure}

The numerical predictions of the collinear parton model are summarized in
Fig. \ref{fig:chic} (upper panel).
At relatively low transverse momenta, the production of \xc states is
dominated by the color singlet mechanism. The differential cross section
$d\sigma/dp_T$ diverges when $p_T\to 0$ for $\chi_2$ states (dashed histogram),
while it remains finite for $\chi_1$ states (solid histogram). 
The production of $\chi_1$ states at zero $p_T$ is suppressed (in accord
with the \Landau theorem), because in the limit of very soft final state gluons
the $2\to 2$ gluon-gluon process degenerates into the $2\to 1$ process. The
shape of the $\chi_0$ spectrum is similar to that of $\chi_2$ (up to an
overall normalizing factor), and this spectrum is not shown in the figure.

The production of $\chi_c$ mesons at high $p_T$ is dominated by the
color-octet contribution, which mainly comes from the `gluon 
fragmentation' diagrams. Here, the perturbative production of $^3\!S_1$
color octet states,
    \begin{equation} g+g\to\;^3\!S_1^8 +g, \label{gg-8g} \end{equation}
is followed by a nonperturbative emission of soft gluons, which results
in the formation of physical color singlet $\chi_c$ mesons:
    \begin{equation} ^3\!S_1^8 \to\;^3\!P_J^1 +ng.\label{8-xg}
\end{equation}

As the co-produced gluons in eq.~(\ref{8-xg}) are assumed to be soft, the 
momentum distribution of $\chi_c$ mesons is taken identical to that of the 
color-octet $^3S_1$ state in eq.~(\ref{gg-8g}). The nonperturbative matrix 
elements responsible for the process  eq. (\ref{8-xg}) are related to the 
fictitious color-octet wave functions, which are used in calculations 
based on eq. (\ref{gg-8g}) in place of the ordinary color-singlet wave 
function: $<0|\O_8|0> =(9/2\pi)\;|\R_8(0)|^2$.

It should be noted that the fragmentation of an almost on-shell transversely 
polarized gluon into a $\chi_1$ state via the emission of a single additional 
gluon, $g\to\;^3\!S_1^8\to\chi_1 +g$, is suppressed in accord with \Landau 
theorem. In terms of the nonrelativistic approximation, it is equivalent 
to say that the formally leading color-electric dipole transitions are 
forbidden, and one must go to nonleading higher multipoles. As the degree 
of this suppression is not calculable within the color-octet model on its 
own, we rather arbitrarily set the suppression factor to 1/20, which 
corresponds to potential model expectations for the average value of $v^2$.

We now proceed with showing the results obtained in the \kf approach.
In this case the production of charmonium \xc states can be
successfully described within the color-singlet model alone
\cite{Baranov:2002cf}, or with only a minor admixture of color-octet
contributions \cite{Hagler:2000dd}. The consideration is based on the
$2\to 1$ partonic subprocess
   \begin{equation} g+g\to\chi, \label{gg-x} \end{equation}
which represents the true leading order in perturbation theory. The
nonzero transverse momentum of the final state meson comes from the
momenta of the initial gluons. The computational technique, which 
we are using here, is identical to the one described in detail in
Ref. \cite{Baranov:2002cf}%
\footnote{We use the FORTRAN code developed in \cite{Baranov:2002cf}.  
 This  code is public and is available from the author on request.}.  

\begin{figure}
  \hskip -2cm\epsfig{figure=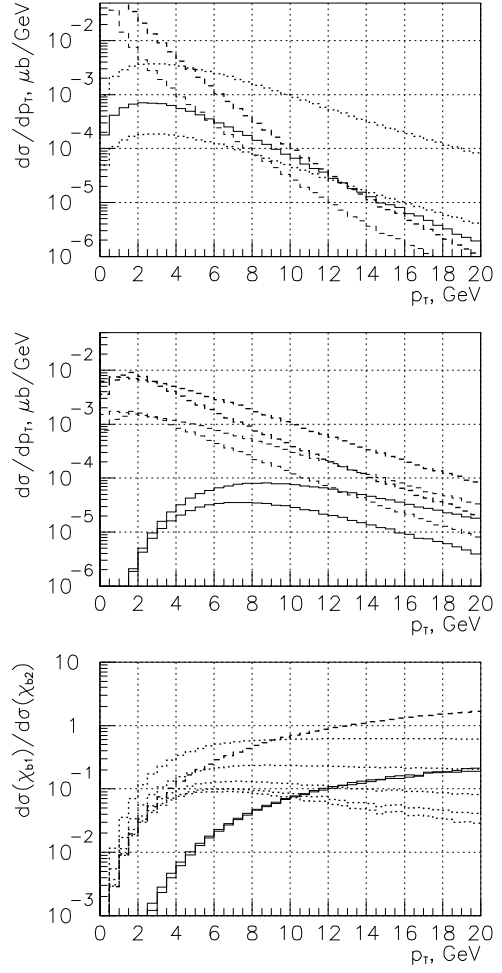,width=0.75\textwidth}
  \caption{Theoretical predictions on the production of $\chi_b$.
    The notations are the same as in Fig. \ref{fig:chic}.}
\label{fig:chib}
\end{figure}

In order to estimate the degree of theoretical uncertainty connected
with the choice of unintegrated gluon density, we also use the
prescription proposed in \cite{Gribov:1984tu}. In this approach, the
unintegrated gluon density is derived from the ordinary density
$G(x,q^2)$ by differentiating it with respect to $q^2$ and setting
$q^2=\kt^2$. Among the different parameterizations available on the
present-day theoretical market, this approach shows the largest
difference with Bl\"umlein's density \cite{Blumlein:1995eu}. Thus,
these two gluon densities
can represent a theoretical uncertainty band.

The numerical results are exhibited in Fig. \ref{fig:chic} (middle panel). 
In contrast with the collinear parton model, the differential cross sections 
are no longer divergent, even at very low $p_T$ values. This property emerges 
from the fact that the relevant $2\to 1$ matrix elements are always finite. 
One can see that the production of the $\chi_1$ state (solid histogram) at low 
$p_T$ is strongly suppressed (in comparison with the $\chi_{0}$ and $\chi_{2}$
states, short and long dashed histograms) because the initial gluons are 
almost on-shell. The suppression goes away at higher $p_T$, as the 
off-shellness of the initial gluons becomes larger.

In Fig. \ref{fig:chic} (lower panel) we compare the predictions of the
collinear and \kf approaches by showing the ratio of the differential
cross sections $d\sigma(\chi_{c1})/dp_T$ and $d\sigma(\chi_{c2})/dp_T$
plotted as a function of $p_T$. As long as the ratio of the nonperturbative 
color-octet matrix elements, $\O(^3S_1^8\to\chi_1)/\O(^3S_1^8\to\chi_2)$, 
is unknown, the predictions of the collinear parton model are very uncertain. 
The different dotted curves in Fig. \ref{fig:chic} from top to bottom 
correspond to the color-octet $\chi_{1}/\chi_{2}$ suppression factor set to 
1, 0.3, 0.1, and 0.03, respectively. The band between the two lowest histograms
may be considered as the most realistic case.
The predictions of the collinear and \kf approaches clearly differ
from each other in their absolute values, and show just the opposite trend
in the experimentally accessible region ($p_T>5$ GeV). 

We conclude our discussion with showing the predictions for the
bottomonium states. The calculations are performed with the parameter
setting given above, and with the value of the \xb wave function set
equal to $|\R'_{\chi_b}(0)|^2 = 1.4$\ GeV$^5$ \cite{Hagiwara:1986tu}.
The integration over the final state phase space is now restricted to
the pseudorapidity interval $-0.4<\eta(\chi_b)<0.4$, in accord with
the CDF experimental cuts
\cite{Abe:1992ww,Abe:1993vw,Abe:1995dv,Abe:1997yz,Affolder:2001ij,Affolder:1999wm}.

Our numerical results are displayed in Fig. \ref{fig:chib}. The
qualitative features of the differential cross sections are similar to
the ones, which we have seen in the case of charmonium. It is worth
recalling that the production of \Y~ mesons has been already measured
by the CDF collaboration
\cite{Abe:1992ww,Abe:1993vw,Abe:1995dv,Abe:1997yz,Affolder:2001ij,Affolder:1999wm}
at $p_T$ values close to zero. Although the $p_T$ dependence of the
direct ($\bar{p}p\to\Upsilon X$) and indirect ($\bar{p}p\to\chi_b
X\to\Upsilon\gamma X$) contributions have not been studied separately,
the net result seems to be at odds with collinear calculations. In
fact, the predicted magnitude of the indirect contribution coming from the
decays of $\chi_{b2}$ states at $p_T<2$ GeV exceeds the total measured
\Y production rate in this region. In contrast the measured
differential cross section $d\sigma(\Upsilon)/dp_T$ decreases with
decreasing $p_T$, in perfect agreement with the \kf predictions
\cite{Baranov:2002cf}.

In summary,
one major difference between the collinear and the \kf approaches 
is connected with the behavior of the differential
cross section $d\sigma(\chi_2)/dp_T$ at low transverse momenta. This
quantity remains finite in the \kf approach, while it diverges in the
collinear parton model when $p_T$ goes to zero. The latter prediction
seems to be not supported by the available experimental data on the
bottomonium production at the Tevatron.

Another well pronounced difference refers to the ratio between the production
rates $d\sigma(\chi_{1})/d\sigma(\chi_{2})$. The underlying physics is
connected with the off-shellness of the gluons. In the collinear parton
model the relative suppression of $\chi_1$ states becomes stronger with
increasing $p_T$ because of the increasing role of the color-octet
contribution. In this approach the leading-order fragmentation of an
on-shell transversely polarized gluon into a vector meson is forbidden.
In contrast with that, in the \kf approach the increase in the final
state $p_T$ is only due to the increasing transverse momenta (and corresponding
virtualities) of the initial gluons, and consequently the suppression
motivated by the \Landau theorem becomes weaker at large $p_T$.

In conclusion we see that quarkonium production can be regarded as a
direct probe of the gluon virtuality, and provides a direct test of the
need for a noncollinear parton evolution.
Our results seem especially promising in view of the fact that the
difference between the two theoretical approaches is clearly pronounced
at conditions accessible for direct experimental measurements.

\section{BFKL dynamics in jet-physics}
\label{sec:jetphysics}

\textit{Main author G.~Marchesini}\\

It has been generally taught that QCD dynamics in high-energy
scattering and in jet-physics are quite different.  However it has
been recently shown \cite{Marchesini:2003nh} that classes of jet
observables satisfy equations {\it formally} similar to the ones for
the high-energy $S$-matrix. The jet-physics observable here discussed
are the heavy quark-antiquark multiplicity (in certain phase-space
region) and the distribution in the energy emitted away from jets.
They satisfy equations formally similar to BFKL and Kovchegov
equations respectively. One may expect that by exploiting such a
formal similarity will bring new insights in both fields.

The common key feature shared by the observables in these two cases is
that enhanced logarithms come only from infrared singularities (no
collinear singularities).  The differences between the two cases is in
the relevant phase space for multi soft-gluon ensemble.  For the
$S$-matrix all transverse momenta of intermediate soft gluons are of
comparable order (no collinear singularities in transverse momenta).
For the considered jet-observables all angles of emitted soft gluons
are of comparable order (no collinear singularities in emission
angles).

We discuss first the $\QQ$ (heavy quark-antiquark) multiplicity in the
phase-space region where collinear singularities cancel and then the
distribution in the energy emitted away from jets.

\subsection{$\QQ$-multiplicity and BFKL equation \label{sec:jetI}}
The standard multiplicity in hard events has both collinear and
infrared enhanced logarithms which are resummed by the well known
expression \cite{Dokshitzer:1991wu,Ellis:1991qj}.
\begin{equation}
  \label{eq:mult}
 \ln N(Q)\sim \int_{Q_0}^{Q}\frac{dk_t}{k_t}\sqrt{2\,\bas(Q)}\,, \qquad
\bas=\frac{N_c\as}{\pi}\,.
\end{equation}
The $\QQ$-multiplicity introduced and studied in
\cite{Marchesini:2003nh} is, due to the peculiar phase space region
chosen, without collinear singularities.  In $\ee$ with center of mass
energy $Q$ one
considers the emission of a $\QQ$ system of mass $\cM$ and momentum
$\vec{k}$.  In the calculation one takes: small velocity
$v=|\vec{k}|/E_k$ so that there are no collinear singularities;
$Q\!\gg\!\cM$ so that perturbative coefficients are enhanced by powers
of $\ln Q/\cM$; and studies the process near threshold.
In this region, the leading logarithmic contributions ($\as^{n}\,\ln^n
Q/\cM$) are obtained by considering soft secondary gluons $q_1,\cdots
q_n$ emitted off $p\bar p$, the primary quark-antiquark.  The $\QQ$
system originates from the decay of one of these soft gluons, actually
the softest one, we denote by $k$,
\begin{equation}
  \ee \to p\bar p+q_1\ldots q_n\,k\,,\qquad { k}\to \QQ\,.
\end{equation}
As shown in \cite{Marchesini:2003nh}, to leading logarithmic order,
the $\QQ$-multiplicity distribution factorizes into the inclusive
distribution $I$ for the emission of the soft off-shell gluon of mass
$\cM$ and momentum $|\vec{k}|$ and the distribution for its successive
decay into the $\QQ$ system
\begin{equation}
  \label{eq:origine}
\frac{E_k\,dN}{d{\cM}^2\,d{|\vec{k}|}}
=\frac{\as^2\,C_F}{3\pi^2\cM^2}
\sqrt{\frac{\cM^2-4M^2}{\cM^2}}\frac{\cM^2+2M^2}{\cM^2}\cdot I\,,
\end{equation}
where $M$ is the heavy quark mass. The Born distribution is 
\begin{align}
  I^{(0)}&=v^2\int\frac{d\Om_k}{4\pi}\,w_{ab}(k)\nonumber\\
  w_{ab}(k)&=\frac{(ab)}{(ak)(kb)}\nonumber\\
  &=\frac{(1-\cos\theta_{ab})}
  {(1-v\cos\theta_{ak})(1-v\cos\theta_{kb})},
  \label{eq:Born}
\end{align}
with $w_{ab}(k)$ the (angular part of the) distribution for the off
soft gluon emitted off the $ab$-dipole (for $\ee$ in center of mass
$\theta_{ab}=\pi$).  For $v\!<\!1$ the Born contribution is finite.

For $Q\gg\cM$, secondary radiation contributes. Since the Born
contribution is regular, only soft logarithms ($\as^n\ln^nQ/\cM$) are
generated which need to be resummed by recurrence relation. To
understand the structure of the resulting equation and appreciate the
similarity with the BFKL equation we consider the first non trivial
contribution in which, besides the off-shell soft gluon $k$, there is
an additional massless soft gluon either emitted or virtual.

The real emission contribution is given by
\begin{equation}
\begin{split}
  w^{\rm R}_{ab}({ k};q)
  =&\frac{(ab)}{(aq)(qk)(kb)}+\frac{(ab)}{(ak)(kq)(qb)}\\
  =&\Theta(q\!-\!k)\>w_{ab}(q)\cdot[w_{aq}(k)+w_{qb}(k)]\\
  +&\Theta(k\!-\!q)\>w_{ab}(k)\cdot[w_{ak}(q)+w_{kb}(q)]\,,
\end{split}
\end{equation}
where, for massless $q$,
\begin{equation}
w_{ab}(q)=\frac{(ab)}{(aq)(qb)}=\frac{1\!-\!\cos\theta_{ab}}
{(1\!-\!\cos\theta_{aq})(1\!-\!\cos\theta_{qb})}\,.
\end{equation}
The corresponding virtual correction is obtained by integrating over the
massless momentum $q$ in the expression (softest gluon emitted off
external legs)
\begin{equation}
  \begin{split}    
    w^{\rm V}_{ab}({ k};q)=
    &-\Theta(q\!-\!k)\>w_{ab}(q)\cdot w_{ab}(k)\\
    &-\Theta(k\!-\!q)\>w_{ab}(k)\cdot[w_{ak}(q)+w_{kb}(q)]\,.
  \end{split}
\end{equation}
By summing the two contributions one finds
\begin{equation}
  \begin{split}
    w^{\rm R+V}_{ab}({ k};q)=&  \Theta(q\!-\!k)\>w_{ab}
    (q)\\
    &\times[w_{aq}(k)\!+\!w_{qb}(k)\!-\!w_{ab}(k)]\,,
  \end{split}
\end{equation}
which shows that $k$ is the softest gluon. From this we derive the
first iterative structure giving $I^{(1)}$ in terms of the Born
contribution \eqref{eq:Born}
\begin{equation}
  \begin{split}
    I^{(1)}(\rho_{ab},\tau)=&\int^Q_\cM\!\frac{d{ q_t}}{
      q_t}\bas(q_t)\!\int\!  \frac{d\Om_{ q}}{4\pi}\,w_{ab}(q)\\
    &\Big[I^{(0)}(\rho_{aq})\!+\!I^{(0)}(\rho_{qb})\!-
    \!I^{(0)}(\rho_{ab})\Big]\,,\\
    &\rho_{ij}\!=\!\frac{1\!-\!\cos\theta_{ij}}{2}\,,
  \end{split}
\label{eq:I1}
\end{equation}
with
\begin{equation}
  \label{eq:tau}
  \tau\>=\>\int_{\cM}^{Q} \frac{dq_t}{q_t}\bas(q_t)
  =\frac{2N_c^2}{11N_c-2n_f}
  \ln\left(\frac{\ln Q/\Lambda}{\ln \cM/\Lambda}\right)\,.
\end{equation}
Here the running coupling in $q_t$ is restored so $\tau$ is given by
an expansion in $\as(Q)\ln Q/\cM$.  The measure in \eqref{eq:I1} is
the branching distribution for a massless soft gluon $q$ emitted off
the $ab$-dipole.  One generalizes this branching structure as
successive dipole emission of softer and softer gluons and one deduces
\cite{Marchesini:2003nh}
\begin{equation}
  \begin{split}
    &\partial_\tau I(\rho_{ab},\tau)=\\
    &\int\!\frac{d\Om_q}{4\pi}\,w_{ab}(q)\,
    \Big[I(\rho_{aq},\tau)\!+\!I(\rho_{qb},\tau)\!-\!I(\rho_{ab},\tau)\Big]\,.
\end{split}
\label{eq:MM1}
\end{equation}
This recurrence structure is very similar to the one obtained in the
dipole formulation of the BFKL equation
\cite{Mueller:1993rr,Mueller:1994jq,Forshaw:1997dc}.  The fundamental
difference is that here the inclusive distribution $I$ depends on the
angular variable $\rho$ (with the limitation $\rho<1$), while in the
high energy scattering one deals with the $S$-matrix as a function of
the impact parameter $b$ (which is not bounded).

The similarity with the BFKL equation can be made even more evident if one
performs the azimuthal integration. One obtains
\cite{Marchesini:2003nh}
\begin{equation}
  \begin{split}
    \partial_{\tau}I(\rho,\tau)=&\int_0^1\frac{d\eta}{1-\eta}
    \left(\eta^{-1}I(\eta\rho,\tau)-I(\rho,\tau)\right)\\
    +&\int_{\rho}^1\frac{d\eta}{1-\eta}
    \left(I(\eta^{-1}\rho,\tau)-I(\rho,\tau)\right)\,.
\end{split}
\label{eq:eveqMM}
\end{equation}
The lower limit $\eta\!>\!\rho$ in the second integral ensures that
the argument of $I(\rho/\eta,\tau)$ remains within the physical region
$\rho/\eta<1$. The presence of this lower bound is the only formal
difference with respect to the BFKL equation for the high energy
elastic amplitude $T$ in the impact parameter representation
\begin{equation}
  \begin{split}
    \partial_{\tau}T(\rho,\tau)=&\int_0^1\frac{d\eta}{1-\eta}
    \left(\eta^{-1}T(\eta \rho,\tau)-T(\rho,\tau)\right)\\
    +&\int_{0}^1\frac{d\eta}{1-\eta}
    \left(T(\eta^{-1}\rho,\tau)-T(\rho,\tau)\right).
  \end{split}
\label{eq:eveqBFKL}
\end{equation}
Here $\rho=b^2$ is the square of the impact parameter and
$\tau=\bas\,Y$ with $Y$ the rapidity with the QCD coupling fixed.  We
discuss now the differences in the two solutions.

Recall first the solution for the high-energy scattering case.  Since
$b$ has no infrared bound we change the variable
\begin{equation}
\label{eq:x}
b^2=e^{-x}\,,\qquad -\infty <x<\infty\,.
\end{equation}
The BFKL equation \eqref{eq:eveqBFKL} satisfies translation invariance
and the area conservation law
\begin{equation*}
\partial_{\tau}\int_{-\infty}^{\infty}dx\,e^{\half x}\,T(e^{-x},\tau)\,
e^{-4\ln2\,\tau}=0\,.
\end{equation*}
This allows us to obtain the solution and then its asymptotic
behavior (using $D=28\zeta(3)=33.6576\ldots$)
\begin{equation}
  \begin{split}
    T(b,\tau)=&\int_{-\infty}^{\infty}\!\frac{dk}{2\pi}\,\tilde T(k)\,
    { e^{(ik-\half)\,x}}\,e^{\chi(k)\,\tau}\\
    \simeq& \tilde T(0)\,
    \frac{e^{4\ln2\,\tau}\,e^{-\half x}\,
      e^{-\frac{x^2}{2D\tau}}}{\sqrt{2\pi D\tau}}\,
  \end{split}
    \label{eq:T}
\end{equation}
with $\tilde T$ determined by the initial condition and
$\chi(k)\!=\!2\psi(1)\!-\!\psi(\half\!+\!k)\!-\!\psi(\half\!-\!k)$ the
BFKL characteristic function.

In the $Q\bar Q$-multiplicity case, the crucial difference is that the
angular variable $\rho$ is bounded.  Introducing the $x$-variable as
in \eqref{eq:x} one observes that translation invariance is lost and,
instead of area conservation, one has absorption
\begin{equation*}
  \begin{split}
    &\rho\!=\!\frac{1\!-\!\cos\theta}{2}=e^{-x}\,,\quad
    0<x<\infty\,,\\
    &\partial_{\tau}\int_{0}^{\infty}dx\,e^{\half
      x}\,I(e^{-x},\tau)\, e^{-4\ln2\,\tau} <0\,.
  \end{split}
\end{equation*}
The exact solution of \eqref{eq:eveqMM} was obtained in
\cite{Marchesini:2004ne}
\begin{equation}
  \begin{split}
    I(\rho,\tau)&=\int_{0}^{\infty}\!dk\,\tilde I(k)\,
    P_{-\half\!+\!ik}\left(\frac{2\!-\!\rho}{\rho}\right)
    \,e^{\chi(k)\,\tau}\\
    &\sim\,
    \frac{(x\!+\!x_0)e^{4\ln2\,\tau}\,e^{-\half x}\,
      e^{-\frac{x^2}{2D\tau}}}{\tau\sqrt{2\pi D\tau}}\,,
  \end{split}
\label{eq:solution}
\end{equation}
with $P_{\alpha}(z)$ the Legendre function (well known in Regge
theory) and $\tilde I$ given by initial condition.

\begin{figure}
  \includegraphics[width=0.45\textwidth]{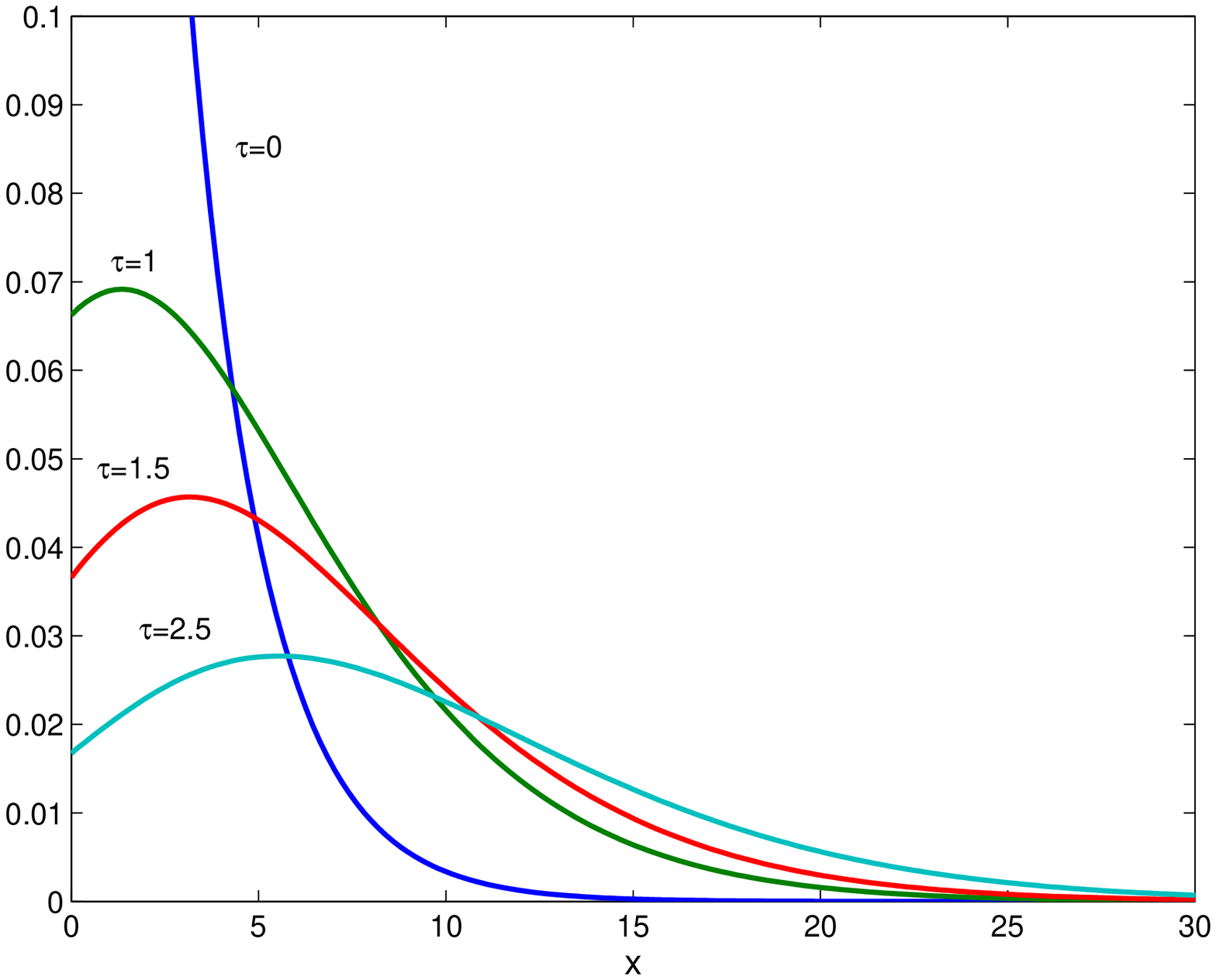}\\
  \includegraphics[width=0.45\textwidth]{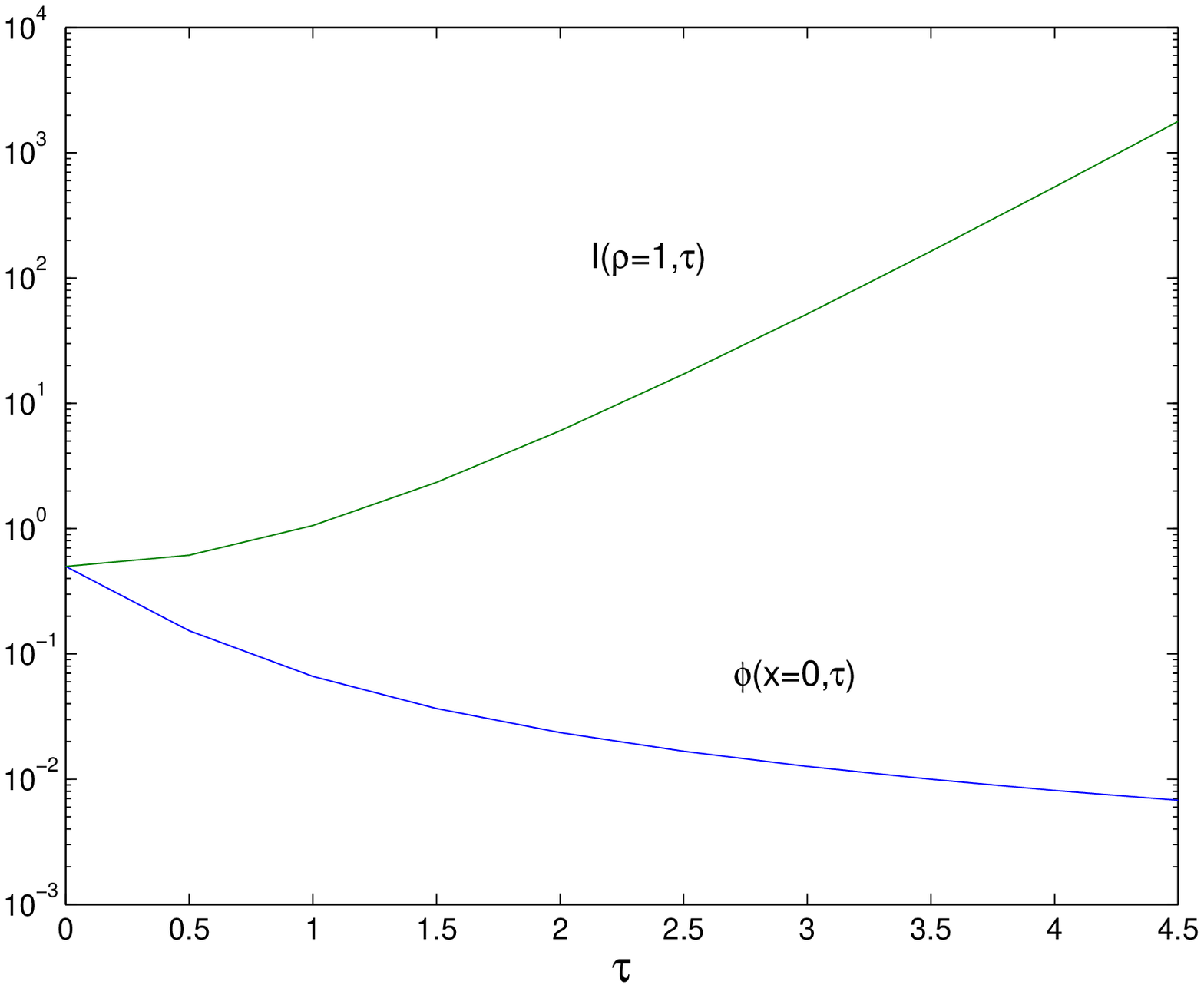}
  \caption{
    Plot of $\phi(x,\tau)=e^{-\half x}\,e^{-4\ln2\,\tau}
    I(e^{-x},\tau)$ solution of \eqref{eq:eveqMM} with initial
    condition $I(\rho,0)=\half\rho$.}
\label{pino-fig}
\end{figure}

From \eqref{eq:solution} and from the upper plot
of Fig.~\ref{pino-fig}, one has that the
inclusive distribution vanishes at the non-physical point $x=-x_0$
which is slowly varying with $\tau$. The asymptotic shape is developed
already at relatively small $\tau$. At $x=0$, corresponding to the
physical value $\rho=1$ for $\ee$ in center of mass, the function
$\phi(x,\tau)$ is decreasing, however, thanks to the $e^{4\ln2\tau}$
factor the inclusive distribution $I(\rho=1,\tau)$ is increasing as
shown in the lower  plot of Fig.~\ref{pino-fig}.

\subsection{Away-from-jet energy flow in $\ee$ \label{sec:jetII}}
Consider in $\ee$ annihilation the distribution in the energy emitted outside
a cone around the jets, $\Eout$: 

  \begin{center}
    \includegraphics[width=0.35\textwidth]{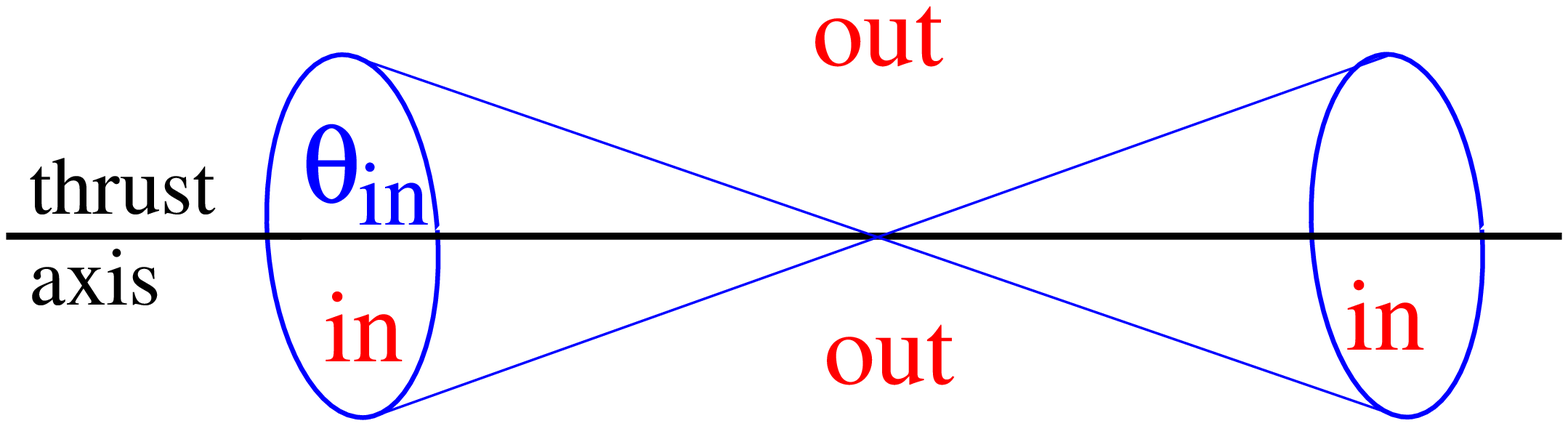}
  \end{center}
  \begin{equation*}
    \Sigma_{\ee}(\Eout)\!=\!\sum_n\!\int\frac{d\sigma_n}{\sigma_T}\,
    \Theta\!\left(\!\Eout\!-\!\sum_{\out}q_{ti}\right).
  \end{equation*}

This is the simplest (in principle) observable involving {\it
  non-global} single logarithms which were (re)discovered by Mrinal
Dasgupta and Gavin Salam
\cite{Dasgupta:2001sh,Dasgupta:2002dc,Dasgupta:2002zf,Dasgupta:2002bw}.
These enter all jet-shape observables which involve only a part of
phase space and therefore are present in a number of distributions
such as: Sterman-Weinberg distribution (energy in a cone); photon
isolation; away from jet radiation; rapidity cuts in hadron-hadron
(e.g. pedestal); DIS jet in current hemisphere.  As for the observable
previously discussed, these non-global logs originate from multiple soft
gluon emissions at large angles (i.e. not in collinear configuration).

$\Sigma_{\ee}(\Eout)$ contains only single logarithms
($\as^n\ln^nQ/\Eout$) coming from soft singularities so that
$\sigma_n/\sigma_T$ can be taken as the distribution in the number of
soft gluons emitted off the primary $p\bar p$ quark-antiquark pair which is known
\cite{Bassetto:1984ik} in the large $N_c$ limit.
$\Sigma_{\ee}(\Eout)$ was first studied
\cite{Dasgupta:2001sh,Dasgupta:2002dc,Dasgupta:2002zf,Dasgupta:2002bw}
numerically by a Monte Carlo method and then studied
\cite{Banfi:2002hw} analytically by deriving the following evolution
equation
\begin{equation}
  \label{eq:BMS}
  \begin{split}
    \partial_\tau \Sigma_{ab}&=-(\partial_\tau R_{ab})\, \Sigma_{ab}
    \!+\!\int_{ {\rm in}} \frac{d\Om_q}{4\pi} \,w_{ab}(q)
    \left[\Sigma_{aq}\cdot\Sigma_{qb}-\Sigma_{ab}\right],\\
    R_{ab}&=\tau\int_{ \out}\!\frac{d\Om_q}{4\pi}\,w_{ab}(q)\,,
\end{split}
\end{equation}
where $\tau$ is the single logarithmic variable previously introduced
\eqref{eq:tau}. As before, to set up a recurrence relation, one needs
to generalize the problem by introducing distribution
$\Sigma_{ab}=\Sigma_{ab}(\Eout)$ for the emission off $ab$-dipole
forming an angle $\theta_{ab}$. The physical distribution
$\Sigma_{\ee}(\Eout)$ for $\ee$ in the center of mass is obtained by
setting $\theta_{ab}=\pi$.

As shown in \eqref{eq:BMS}, the dipole directions $a$ and $b$ are
always inside the jet region ($q$ in the integral is bounded inside
the jet region). If $a,b$ are in opposite semicones, then either $a,q$
or $q,b$ are in the same semicone.  There are many properties of this
jet-physics equation (see
\cite{Dasgupta:2001sh,Dasgupta:2002dc,Dasgupta:2002zf,Dasgupta:2002bw}
and \cite{Banfi:2002hw}). What concerns us here as far as the
connection with high-energy physics is the case in which $a,b$ are in
the same semicone and we consider $a$ very close to $b$. In the small
angle limit we introduce the $2$-dimensional variable $\vec{\theta}$
for the $ab$-dipole ($\Sigma_{ab}\to\Sigma(\vec{\theta})$).  For small
$\theta$ we can neglect the linear term ($R_{ab}\sim \vec{\theta}^2$)
so that the evolution equation \eqref{eq:BMS} becomes
\begin{equation}
  \begin{split}
    \partial_{\tau}\Sigma(\tau,\vec{\theta})\!=\!
    \int&\frac{d^2\theta'}{2\pi}\frac{\theta^2}
    {{\theta'}^2(\vec{\theta}\!-\!\vec{\theta'})^2}\\
    &\Big[\Sigma(\tau,\vec{\theta'})\Sigma(\tau,\vec{\theta}\!-\!\vec{\theta'})-
    \Sigma(\tau,\vec{\theta})\Big]\,,
  \end{split}
\label{eq:BMS-small}
\end{equation}
with $\vec{\theta}'$ ranging in the full plane. The initial condition
is $\Sigma(0,\vec{\theta})=1$.  This equation is formally the same as
the Kovchegov equation \cite{Kovchegov:1999yj} for the S-matrix
\begin{equation}
  \begin{split}
    \partial_{\tau}S(\tau,\vec{b})=
    \int&\frac{d^2b'}{2\pi}\frac{b^2}{{b'}^2(\vec{b}\!-\!\vec{b'})^2}\\
    &\Big[S(\tau,\vec{b'})S(\tau,\vec{b}\!-\!\vec{b'})-S(\tau,\vec{b})\Big],
\end{split}
\label{eq:Kov}
\end{equation}
where $\vec{b}$ is the impact parameter ranging in the full plane and
$\tau=\bas Y$ as before.  Here the initial condition is
$1\!-\!S(0,\vec{b})\sim\as^2$ corresponding to the two gluon exchange.

The asymptotic properties of the solutions are well known.  Both
solutions undergo well known saturation for the variable $\theta^2$
or $b^2$ larger than a critical value with asymptotic behavior
$e^{-c\tau}$ with $c\simeq 4.88\cdots$ determined from the BFKL
characteristic function. Beyond such a critical value the solution
decreases in $\tau$ as a Gaussian, $\Sigma\sim S\sim e^{-c\tau^2/2}$.

The difference in the initial condition makes a difference in the way
the saturation regime is asymptotically reached in the two cases.  In
the high-energy case ($1\!-\!S(0,\vec{b})\sim\as^2$) the saturation
regime is reached after a critical time $\tau_c\sim\ln\as^{-2}/4\ln2$.
In the jet-physics case ($\Sigma(0,\vec{\theta})=1$) there is not a
critical $\tau$ and the solution goes without impediment into the
saturation regime.

In addition to the different initial conditions, an important
difference is that the variables in \eqref{eq:BMS} are angular
variables ranging in compact regions. We have seen in the previous
analysis that even at small angle it is not fully correct to neglect
the compactness affecting the integration limits. This question will
be further studied \cite{Onofri:2005XX}.
 
\subsection{Physics differences}
\label{sec:jetIII}
The basis for the two classes of equations,
\eqref{eq:eveqMM},\eqref{eq:BMS} in jet-physics and
\eqref{eq:eveqBFKL},\eqref{eq:Kov} in high-energy scattering, is of
course the (same) multi-soft gluon-distribution. However the dominant
contributions for the two classes of observables ($I_{ab},\Sigma_{ab}$
and $T,S$) are obtained from very different kinematical configurations
as we discuss now.

{\bf Jet-physics case:} Here all angles $\theta_{i}$ of emitted gluons are
of same order. This is due to the fact that this observable does not
contain collinear singularities for $\theta_{ij}\to0$.  Moreover, in
the (leading) infrared limit soft gluon energies can be taken ordered
so that also the emitted transverse momenta $q_{ti}$ are ordered. The
ordered variables $q_{ti}$ enter the argument of the running coupling.
The distribution $I_{ab}$ or $\Sigma_{ab}$ are functions of the
angular variable $\theta_{ab}$ (which ranges in a compact region) and
$\tau$, the logarithmic integral of the running coupling in
\eqref{eq:tau}. We are then interested in the solution for finite
$\theta_{ab}$ (e.g. $\theta_{ab}=\pi$ in $\ee$ center of mass) and for
$\tau$ never too large.
  
{\bf High-energy scattering case:} Here all intermediate soft gluon
transverse momenta $q_{ti}$ are of same order (no singularities for
vanishing transverse momentum differences). On the other hand, energy
ordering implies in this case that intermediate gluon angles
$\theta_i$ are ordered. Contrary to the previous case, the running
coupling is a function of the variables $q_{ti}$ which all are of same
order.  Therefore, in first approximation, one can take $\as$ fixed.
The high-energy $S$-matrix is a function of the impact parameter
(which has no bound at large $b$) and $\tau=\bas\,Y$.  In this case we
are then interested in the solution for small $\rho$ (the short
distance region) and for $\tau$ large.

As discussed in section~\ref{sec:jetI}, the fact that the variable $\rho$
entering the jet-observable ranges in a compact region affects the
prefactor of the asymptotic behavior and the shape of the
distribution at finite angles. In the non linear case discussed in
section~\ref{sec:jetII}, even neglecting compactness at small angle,
the difference in the initial conditions affects the ranges in $\tau$
at which the asymptotic behavior (saturation) is developing.

Concluding, by exploiting similarities and differences in the dynamics
of high energy scattering and jet-physics (with non-global logs) one
hopes that new insights in both fields could be developed.


\section{Saturation}
\label{sec:saturation}

\textit{Main authors M.~Lublinsky and K.~Kutak}\\

\def\ugdf{{f}}
\def\Ker{\mbox{\textit{Ker}}}

A parton evolution equation which attempts to describe saturation
phenomena was originally proposed by Gribov, Levin and
Ryskin~\cite{Gribov:1984tu} (GLR equation) in momentum space and
proven in the double log approximation of perturbative QCD by Mueller
and Qiu~\cite{Mueller:1985wy}. In the leading $\ln 1/x$ approximation
it was derived by Balitsky in the Wilson Loop Operator
Expansion~\cite{Balitsky:1995ub}. In the form presented later it was
obtained by Kovchegov \cite{Kovchegov:1999yj} (now called the
Balitsky-Kovchegov, or BK equation) in the color dipole approach
\cite{Mueller:1993rr} to high energy scattering in QCD.  This equation
was also obtained by summation of the BFKL pomeron fan diagrams by
Braun~\cite{Braun:2000wr} and most recently Bartels, Lipatov, and
Vacca~\cite{Bartels:2004ef}.  In the framework of Color Glass
Condensate it was obtained by Iancu, Leonidov and
McLerran~\cite{Iancu:2000hn}.

\subsection{Basic facts about the BK equation} 

Because the transverse coordinates are unchanged in a high energy
collision, unitarity constraints are generally more easy to take into
account in a formalism based on the transverse coordinate space
representation, and several suggestions for how to include saturation
effects in such a formalism have been proposed. Golec--Biernat and
W\"usthoff \cite{Golec-Biernat:1998js} formulated a dipole model, in
which a virtual photon is treated as a $q\bar{q}$ or $q\bar{q}g$
system impinging on a proton, and this approach has been further
developed by several authors (see e.g.\ \cite{Forshaw:1999uf} and
\cite{Bartels:2002cj}).  Mueller
\cite{Mueller:1993rr,Mueller:1994jq,Mueller:1994gb} has formulated a
dipole cascade model in transverse coordinate space, which reproduces
the BFKL equation, and in which it is also possible to account for
multiple sub-collisions.  Within this formalism Balitsky and Kovchegov
\cite{Balitsky:1995ub,Kovchegov:1999yj} have derived a non-linear
evolution equation (BK equation), which also takes into account these saturation
effects from multi-pomeron exchange 
 and which is the best presently available tool
to study saturation phenomena at high energies. Contrary to many
models the BK equation has solid grounds in perturbative QCD. The equation
reads
\begin{equation}   
  \begin{split}
    & \frac{d\,N({\mathbf{x}_{01}},y;b)}{d\,y}\,=\,
    \,\frac{N_c\,\as}{2\pi} \,\, \int_{\rho} \, d^2 {\mathbf{x}_{2}}
    \frac{{\mathbf{x}^2_{01}}}{{\mathbf{x}^2_{02}}\,
      {\mathbf{x}^2_{12}}} \,\,\times \\
    &\hspace{0.5cm}\left(\,2\, N({\mathbf{x}_{02}},y;{ \mathbf{b}- \frac{1}{2}
          \mathbf{x}_{12}})
      - \,N({\mathbf{x}_{01}},y;\mathbf{b})\right. \\
    &\hspace{0.7cm}\left.-N({\mathbf{x}_{02}},y;
      { \mathbf{b} - \frac{1}{2} \mathbf{x}_{12}})
      N({\mathbf{x}_{12}},y;{ \mathbf{b}- \frac{1}{2} \mathbf{x}_{02}})\right)
  \end{split}
  \label{EQ}   
\end{equation}   
The function $N(r_{\perp},x; b)$ is the imaginary part of the
amplitude for a dipole of size $r_{\perp}$ elastically scattered at an
impact parameter $b$.
   
In the equation (\ref{EQ}), the rapidity $y\equiv-\ln x$.  The ultraviolet
cutoff $\rho$ is needed to regularize the integral, but it does not
appear in physical quantities. We also use the large $N_c$ limit (number of
colors) value of $C_F=N_c/2$.

Eq.~(\ref{EQ}) has a very simple meaning: The dipole of size
$\mathbf{x}_{01}$ decays in two dipoles of sizes $\mathbf{x}_{12}$ and
$\mathbf{x}_{02}$ with the decay probability given by the wave
function $| \Psi|^2
\,=\,\frac{\mathbf{x}^2_{01}}{\mathbf{x}^2_{02}\,\mathbf{x}^2_{12}}$.
These two dipoles then interact with the target. The non-linear term
takes into account a simultaneous interaction of two produced dipoles
with the target. The linear part of eq.~(\ref{EQ}) is the LO BFKL
equation \cite{Kuraev:1977fs,Balitsky:1978ic}, which describes the
evolution of the multiplicity of the fixed size color dipoles with
respect to the energy $y$. For the discussion below we 
introduce a short notation for eq.~(\ref{EQ}):
\begin{equation}\label{BKE} 
\frac{d\,N}{d\, y}\,\,=\,\alpha_s\,\,\Ker\, \,\otimes\,\,(\,N\,\,- 
\,\,N\,\,N)\,.
\end{equation} 
The BK equation has been studied both analytically
\cite{Kovchegov:1999ua,Levin:1999mw,Levin:2000mv,Kovner:2001bh,Iancu:2002tr,Munier:2003sj,Munier:2003vc}
and numerically
\cite{Braun:2000wr,Gotsman:2002yy,Armesto:2001fa,Golec-Biernat:2001if,Lublinsky:2001yi,Rummukainen:2003ns,Golec-Biernat:2003ym,Gotsman:2004ra}.
The theoretical success associated with the BK equation is based on the
following facts:
\begin{itemize} 
\item The BK equation is based on the correct high energy dynamics which is
  taken into account via the LO BFKL evolution kernel.
\item The BK equation restores the $s$-channel unitarity of partial waves
  (fixed impact parameter) which is badly violated by the linear BFKL
  evolution.
\item The BK equation describes gluon saturation, a phenomenon expected at high
  energies.
\item The BK equation resolves the infrared diffusion problem associated with
  the linear BFKL evolution. This means that the equation is much more
  stable with respect to possible corrections coming from the
  non-perturbative domain.
\item The BK equation has met with phenomenological successes when confronted
  against DIS data from HERA
  \cite{Gotsman:2002yy,Lublinsky:2001yi,Gotsman:2003br,Bartels:2003aa,Bartels:2002uf,Levin:2001et,Levin:2001yv,Lublinsky:2001bc,Iancu:2003ge}.
\end{itemize}

The BK equation is not exact and has been derived in several approximations.
 
\begin{itemize}
\item The LO BFKL kernel is obtained in the leading soft gluon
  emission approximation and at fixed $\alpha_s$.
\item The large $N_c$ limit is used in order to express the nonlinear
  term as a product of two functions $N$. This limit is in the
  foundation of the color dipole picture. To a large extent the large
  $N_c$ limit is equivalent to a mean field theory without dipole
  correlations.
\item The BK equation assumes no target correlations. Contrary to the large
  $N_c$ limit, which is a controllable approximation within
  perturbative QCD, the absence of target correlations is of pure
  non-perturbative nature. This assumption is motivated for
  asymptotically heavy nuclei, but it is likely not to be valid for
  proton or realistic nucleus targets.
\end{itemize}

There are several quite serious theoretical problems which need to be
resolved in the future.
\begin{itemize} 
\item The BK equation is not symmetric with respect to target and
  projectile.  While the latter is assumed to be small and
  perturbative, the former is treated as a large non-perturbative
  object. The fan structure of the diagrams summed by the BK equation
  violates the $t$-channel unitarity. The $t$-channel unitarity is a
  completeness relation in the $t$-crossing channel.  It basically
  reflects a projectile-target symmetry of the Feynman diagrams.  The
  down-type fan graphs summed by the BK equation, obviously violate the
  symmetry. A first step towards restoration of the $t$-channel
  unitarity would be an inclusion of Pomeron loops.
\item Though the BK equation respects the $s$-channel unitarity
  \footnote{There was a recent claim of Mueller and Shoshi
    \cite{Mueller:2004se} that the $s$-channel unitarity is in fact
    violated during the evolution.} the exchange of massless gluons
  implies that it violates the Froissart bound for the energy
  dependence of the total cross section.  In order to respect the
  Froissart bound, gluon saturation and confinement are needed.  On
  one hand, the BK equation provides gluon saturation at fixed and
  large impact parameters. On the other hand, being purely
  perturbative, it cannot generate the mass gap needed to ensure a
  fast convergence of the integration over the impact parameter $b$.
  Because of this problem, up to now all the phenomenological
  applications of the BK equation were based on model assumptions
  regarding the $b$-dependence.  It is always assumed that the
  $b$-dependence factorizes and in practice the BK equation is usually
  solved without any trace of $b$. At the end, the $b$-dependence is
  restored via an ansatz with a typically exponential or Gaussian
  profile.  An attempt to go beyond this approximation has been
  reported in Ref.  \cite{Golec-Biernat:2003ym,Gotsman:2004ra}.
\item It is very desirable to go beyond the BK equation and relax all
  underlying assumptions outlined above. The higher order corrections
  are most needed. In particular it is important to learn how to
  include the running of $\alpha_s$, though in the phenomenological
  applications the running of $\alpha_s$ is usually implemented.
\item The LO BFKL kernel does not have the correct short distance
  limit responsible for the Bjorken scaling violation. As a result the
  BK equation does not naturally match with the DGLAP equation. Though several
  approaches for unification of the BK equation and DGLAP equations were
  proposed
  \cite{Kwiecinski:1997ee,Lublinsky:2001yi,Gotsman:2002yy,Kimber:2001nm,Kutak:2003bd},
  the methods are not fully developed. All approaches deal only with low
  $x$ and only with the gluon sector. We would like to have a unified evolution
  scheme for both small and large $x$ and with quarks included.
 \end{itemize} 
 
\subsection{Phenomenology with the BK equation} 
 
The deep inelastic structure function $F_2$ is related to the dipole
amplitude $N$ via
\begin{equation} \label{F2T}   
  \begin{split}
    F_2(x,Q^2)&=\\
    \frac{Q^2}{4\pi^2 }&\int d^2 r_{\perp} \int d z
    P^{\gamma^*}(Q^2; r_{\perp}, z) \sigma_{\rm dipole}(r_{\perp}, x),
\end{split}
\end{equation}   
with the dipole cross section given by the integration over the impact
parameter:
\begin{equation}
\label{TOTCX}
\sigma_{\rm dipole}(r_{\perp},x) \,\,=\,\,2\,\,\int\,d^2 b\, N(r_{\perp},x;b).
\end{equation}   
 
The physical interpretation of eq.~(\ref{F2T}) is transparent.  It
describes the two stages of DIS \cite{Gribov:1968gs}. The first stage
is the decay of a virtual photon into a colorless dipole ($ q \bar q $
-pair). The probability of this decay is given by $P^{\gamma^*}$ known
from QED
\cite{Mueller:1993rr,Mueller:1989st,Nikolaev:1990ja,Levin:1996vf}.
The second stage is the interaction of the dipole with the target
($\sigma_{\rm dipole}$ in eq.~(\ref{F2T})). In the large $N_c$ limit a
color charge has a well-defined anti-charge partner in a color dipole.
Eq.~(\ref{F2T}) illustrates the fact that in this limit these color
dipoles are the relevant degrees of freedom in QCD at high energies
\cite{Mueller:1993rr}.
 
For the phenomenological applications one may use the function
$N(r_{\perp},x;b)$ or $\sigma_{\rm dipole}(r_{\perp},x)$ obtained
directly from the solutions of the BK equation (\ref{EQ}). With additional
DGLAP corrections this approach was adopted by Gotsman et al. in
Ref.~\cite{Gotsman:2002yy}.
 
Alternatively one can relate $N$ to an unintegrated gluon distribution
function ${\cal F}(x,k^2)=\ugdf(x,k^2)/k^2$. The dipole cross section can be expressed via
$\ugdf$ \cite{Barone:1993sy,Bialas:2000xs}:
\begin{equation}\label{sf} 
  \begin{split}
    \sigma_{\rm dipole}(r_{\perp},x)&=\\
    \frac{8\,\pi^2}{N_c}&\int \frac{d\,k^2}{k^4}\,
    \left[1\,-\,J_0(k\,r_\perp)\right ]\,\alpha_s(k^2)\,\ugdf(x,k^2)
\end{split}
\end{equation} 
The inversion of eq.~(\ref{sf}) is straightforward
\begin{equation}\label{f} 
  \ugdf(x,k^2)\,=\,\int d^2b\,h(k^2,x,b) ,
\end{equation} 
\begin{equation}\label{h} 
  \begin{split}
    h(k^2,x,b)\,&=\,\frac{N_c}{4\,\alpha_s\,\pi^2}\,
    k^4\,\Delta_k\,\tilde N(k^2,x,b)\\
    &=\,
    \frac{N_c}{\alpha_s\,\pi^2}\,\frac{k^2\,\partial
      ^2}{\partial\,(\ln k^2)^2}\, \tilde N(k^2,x,b)\,.
\end{split}
\end{equation}
Here $\Delta_k$ is the 2-dimensional Laplace operator.  
The function $\tilde N$ is
related to the Fourier transform of $N$
\begin{equation}\label{tN} 
  \tilde N(k^2,x,b)\,=\,\int \frac{d^2\,\mathbf{r}_\perp}{2\,\pi \,r_\perp^2}
  \,e^{\,i\,\mathbf{k}\,\mathbf{r}_\perp}\,N(r_\perp,x,b)\,. 
\end{equation} 
In fact, $\tilde N$ obeys a nonlinear version of the LO BFKL evolution
equation in momentum space. The function $\tilde N$ can be interpreted as
an unintegrated gluon distribution.
$\tilde N$ and $h$ coincide at large momenta but differ at small ones.
On one hand, within the dipole picture it is rather the function
$\tilde N$ which gives the probability to find a gluon with a given
transverse momentum and at a given impact parameter. On the other
hand, it is the function $\ugdf$ (or $h$) which enters the $\kt$ (high
energy) factorization formula. In what follows we will concentrate on
the unintegrated gluon distribution $\ugdf$ only.
 
\begin{figure}
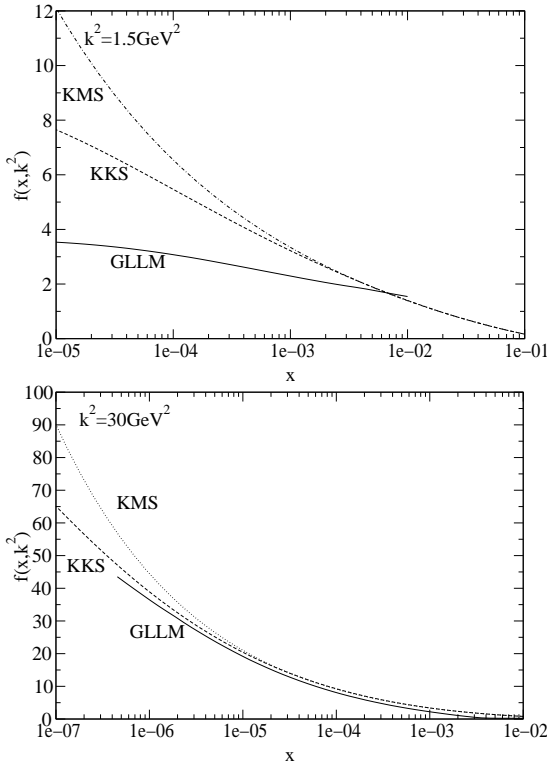
 
  \begin{center}
    \includegraphics*[width=0.40\textwidth]{fig/1_5.eps}
    \includegraphics*[width=0.40\textwidth]{fig/30.eps}
  \end{center}
\caption{The unintegrated gluon distribution $\ugdf(x,k^2)$ 
  as a function of $x$ for different values $k^2 = 1.5 \, GeV^2$ (top)
  and $k^2=30 \,GeV^2$ (bottom). Solid lines correspond to the
  solution of the nonlinear equation using GLLM
  \protect\cite{Gotsman:2002yy} parameterization whereas dashed
  lines (KKS) correspond to the approximate solution of
  (\ref{eq:fkovres})\protect\cite{Kutak:2003bd,Kutak:2004ym}. For
  reference we also present linear BFKL/DGLAP evolution
  (KMS)\cite{Kwiecinski:1997ee} .}
  \label{fig:misha1}   
\end{figure}

\begin{figure}
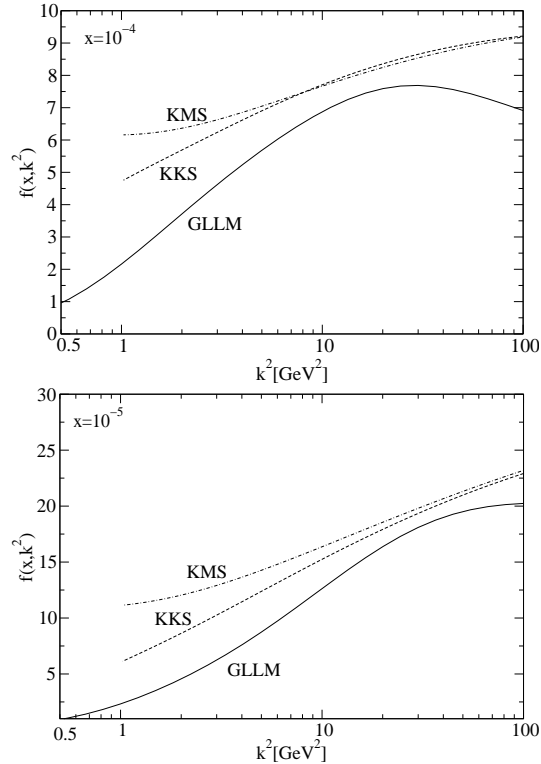
 
  \begin{center}
    \includegraphics*[width=0.39\textwidth]{fig/x10_4.eps}
    \includegraphics*[width=0.39\textwidth]{fig/x10_5.eps}
  \end{center}
  \caption{The unintegrated gluon distribution $\ugdf(x,k^2)$ 
    as a function of $k^2$ for different values $x = 10^{-4}$ (left)
    and $x=10^{-5}$ (right). The lines are the same as in
    fig.~\ref{fig:misha1} .}
  \label{fig:misha2}
\end{figure}

Instead of solving the BK equation (\ref{EQ}) and then inverting the relation
(\ref{sf}) one can adopt another strategy and reformulate the problem
directly in terms of the unintegrated gluon density $\ugdf$. This approach was
adopted in work by Kutak-Kwiecinski \cite{Kutak:2003bd}
and Kutak-Stasto \cite{Kutak:2004ym}. Using relations
(\ref{TOTCX}) and (\ref{sf}) one can transform (\ref{EQ}) into an
equation for the unintegrated gluon distribution
\begin{equation}   
  \begin{split}
    \ugdf(x,&k^2) = \tilde f^{(0)}(x,k^2) \\
    +&\,\frac{N_c\alpha_s(k^2)}{\pi} k^2 \int_x^1 \frac{dz}{z}
    \int_{k_0^2} \frac{dk'^2}{k'^2}\\
    &\bigg\{\frac{\ugdf(\frac{x}{z},k^{\prime 2}) \, -\,
      \ugdf(\frac{x}{z},k^2)}{|k'^2-k^2|}+
    \frac{\ugdf(\frac{x}{z},k^2)}{|4k^{\prime 4}+k^4|^{\frac{1}{2}}} \,
    \bigg\}\, \\
    -&\left(1-k^2\frac{d}{dk^2}\right)^2\frac{k^2}{R^2}
    \int_x^1\frac{dz}{z}\\
    &\left[\int_{k^2}^{\infty} {dk^{\prime 2}\,
        k^{\prime 4}}\alpha_s(k^{\prime 2})\ln\left( {k^{\prime 2}
          k^2}\right)\ugdf(z,k^{\prime 2})\right]^2.
  \end{split}
  \label{eq:fkov}    
\end{equation}  
Here it is written as an integral equation, corresponding to the
BFKL equation in momentum space supplemented by the negative nonlinear
term. The input $\tilde
f^{(0)}(x,k^2)$ is given at the scale $k_0^2=1GeV^2$.
This equation was derived under the following factorization
ansatz:
\begin{equation}
\tilde N(k^2,x,b)=\tilde n(k^2,l,x)\,S(b)
\label{eq:ANZ}   
\end{equation}
with normalization conditions on the profile function $S(b)$
\begin{equation}
\int d^2 {\bf b} \, S(b) =  1 ; \qquad
\int d^2 {\bf b} \, S^2(b)  =   \frac{1}{\pi R^2} .  
\end{equation}
The assumption (\ref{eq:ANZ}) is crude and corresponds to a situation
where the projectile size (color dipole) is neglected compared to the
target size (proton). A simple way to improve (\ref{eq:fkov}) is to
implement NLO corrections in the linear term of the equation.  It can
be done within the unified BFKL-DGLAP framework which is presented
below. The final equation (eq.~(\ref{eq:fkovres}) below) can be
used for phenomenological applications.
Figs. \ref{fig:misha1}, \ref{fig:misha2} 
display the unintegrated
gluon distributions $\ugdf$ obtained in Refs.
\cite{Gotsman:2002yy,Kutak:2004ym}.


\subsection{The saturation scale}
\begin{figure}     
  \center\includegraphics*[width=0.4\textwidth]{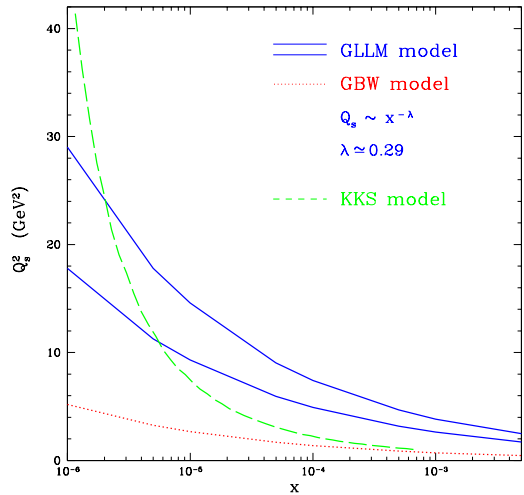} 
  \caption{Saturation scale from various models. The solid lines defines
    a band of possible saturation scales coming from the GLLM model
    \protect\cite{Gotsman:2002yy}.  The dashed line (KKS) is from
    Ref.\ \protect\cite{Kutak:2004ym}.  The dotted line (GBW) is the
    Golec-Biernat W\"usthoff model
    \protect\cite{Golec-Biernat:1998js}.}
  \label{fig:satscale}
\end{figure}

In order to quantify the strength of effects that slow down the gluon
evolution one introduces the saturation scale $Q_s(x)$. It divides the
$(x,k^2)$-space into regions of dilute and dense partonic systems.
In the case when $k^2\, <\, Q_s^2(x)$ the solution of the BK equation 
exhibits geometrical scaling, which means that it depends on one
variable only, $N(r,x)=N(r \,Q_s(x))$ or in momentum space $\tilde
N(k^2,x) = \tilde N(k/Q_s(x))$.  In Fig.  \ref{fig:satscale} 
we
present saturation scales obtained from (\ref{eq:fkov})  in \cite{Kutak:2004ym}
and the corresponding result obtained from Ref.~\cite{Gotsman:2002yy}.
Note, however, that the saturation scale is defined differently in these two
approaches.
In ref.~\cite{Kutak:2004ym} the saturation scale is defined quantitatively
as a relative difference between the solutions to the linear and
nonlinear equations, while in ref.~\cite{Gotsman:2002yy} it is defined by the 
requirement that $N(2/Q_s,x)$ is constant equal to $1/2$ or $1/e$.
  Note that both the KKS~\cite{Kutak:2004ym} and the GLLM models predict a
saturation scale much bigger than the one from the GBW model.

\subsection{Beyond the BK equation}  

Saturation effects are most easily studied in the coordinate space
representation in which it has been difficult to include non-leading
effects, and the non-leading effects have mainly been studied in
momentum space, where it is hard to include saturation.  We now
present a short (not complete) review of recent theoretical activities
which attempt to go beyond the leading order BK equation.  An
important issue relating to the NLO corrections is energy--momentum
conservation, which was already addressed in section~\ref{sec:updf}
and will be further discussed in more detail in section
\ref{section:energy-momentum-conservation}.

\subsubsection{Beyond leading order}
 The BFKL kernel is known at next-to-leading order.
Nevertheless, a nonlinear equation at NLO has not been derived yet.
I.~Balitsky and A.~Belitsky \cite{Balitsky:2001mr} have been able to compute
a single NLO contribution which has maximal nonlinearity, namely the
$N^3$ term:
\begin{equation}\label{BB} 
  \frac{d\,N}{d\, y}=\alpha_s \Ker\otimes(N-N\cdot N)- 
  \alpha_s^2\overline{\Ker}\otimes N\cdot N\cdot N. 
\end{equation} 
The new kernel $\overline{\Ker}$
can be found in Ref~\cite{Balitsky:2001mr}.  Triantafyllopoulos
\cite{Triantafyllopoulos:2002nz} has considered NLO BFKL in the
presence of a saturation boundary. The results show a decrease in the
saturation scale growth as a function of rapidity towards the value
$\lambda\,\simeq \,0.3$ observed experimentally (GBW
\cite{Golec-Biernat:1998js} and GLLM \cite{Gotsman:2002yy} models).

Another approach \cite{Kimber:2001nm,Kutak:2003bd} to partially
include the NLO corrections into the BK equation is to implement in the linear
term of eq.~(\ref{eq:fkov}) the unified BFKL-DGLAP framework developed
in \cite{Kwiecinski:1997ee}.  In this scheme the BFKL kernel also gets
modified by the consistency constraint
\cite{Andersson:1995jt,Kwiecinski:1996td,Ciafaloni:1988ur}
\begin{equation}   
  k'^2 < k^2 / z.
\label{eq:kincon}   
\end{equation}   
The origin of this constraint is the requirement that the virtuality
of the exchanged gluon is dominated by its transverse momentum
$|k^{\prime 2}|\simeq k_T^{\prime 2}$ (see also section~\ref{sec:updf}). 
The constraint
(\ref{eq:kincon}) resums a large part of the subleading corrections in
$\ln 1/x$, and it is also connected to the conservation of the
\emph{negative} lightcone component $p_-=E - p_L$ (\textit{cf.}
section \ref{section:energy-momentum-conservation}). Additionally, the
non-singular part of the leading order DGLAP splitting function, which
influences the normalization of the unintegrated gluon distribution, is
included into the evolution and $\alpha_s$ is assumed to run with
the scale $k^2$ . The final improved nonlinear equation for the
unintegrated gluon density becomes

\begin{equation}   
  \begin{split}
    &\ugdf(x,k^2) \; = \; \tilde f^{(0)}(x,k^2) + \\
    &+\,\frac{N_c\alpha_s(k^2)}{\pi} k^2 \int_x^1 \frac{dz}{z}
    \int_{k_0^2}
    \frac{dk'^2}{k'^2} \,  \\
    &\bigg\{ \, \frac{\ugdf(\frac{x}{z},k^{\prime 2}) \,
      \Theta(\frac{k^2}{z}-k'^2)\, -\, \ugdf(\frac{x}{z},k^2)}{|k'^2-k^2|}
    +\,
    \frac{\ugdf(\frac{x}{z},k^2)}{|4k^{\prime   4}+k^4|^{\frac{1}{2}}} \, \bigg\}\\
    &+ \, \frac{\alpha_s(k^2)}{2\pi} \int_x^1 dz \, \bar{P}_{gg}(z)
    \int^{k^2}_{k_0^2} \frac{d k^{\prime 2}}{k^{\prime 2}} \,
    \ugdf(\frac{x}{z},k'^2) \\
    &-\left(1-k^2{d\over dk^2}\right)^2{k^2\over R^2}
    \int_x^1\frac{dz}{z}\\
    &\hspace{1cm}\left[\int_{k^2}^{\infty} {dk^{\prime 2}\over k^{\prime
          4}}\alpha_s(k^{\prime 2})\ln\left( {k^{\prime 2}\over
          k^2}\right)\ugdf(z,k^{\prime 2})\right]^2 \; ,
\end{split}
\label{eq:fkovres}    
\end{equation}    
with the input distribution $\tilde f^{(0)}(x,k^2)$.

\subsubsection{JIMWLK} 
The $N_c$ corrections can be accounted for through the JIMWLK
functional equation
\cite{Jalilian-Marian:1996xn,Jalilian-Marian:1997dw,Jalilian-Marian:1998cb,Kovner:2000pt,Weigert:2000gi},
which is equivalent to Balitsky's original infinite chain of equations
\cite{Balitsky:1995ub}.  Introducing $N$ as a target expectation value
of a certain operator (product of two Wilson lines),
$N\,\,\equiv\,\,\langle\,W\,\rangle_{target}$, the first couple of
equations of the Balitsky chain are
\begin{equation} 
 \hspace{1cm} 
\frac{d\,\langle\,W\,\rangle}{d\, y}\,\,=\,\alpha_s\,\Ker \,\otimes\,(\, 
\langle\,W\,\rangle\,\,- 
\,\,\langle\,W\,W\,\rangle),
\end{equation}  
\begin{equation}  
\frac{d\,\langle\,W\,W\,\rangle}{d\, y}\,=\alpha_s 
\,\Ker\, \otimes\,( 
\langle\,W\,W\,\rangle\,- 
\,\langle\,W\,W\,W\,\rangle).
\end{equation} 
The large $N_c$ limit and the absence of the target correlations used
by Kovchegov \cite{Kovchegov:1999yj} is equivalent to a mean field
approximation which allows to express a correlator of a product as a
product of correlators:
$$
\langle\,W\,W\,\rangle\,\,
=\,\,\langle\,W\,\rangle\,\langle\,W\,\rangle\,\,=\,\,N\,N\,;\,\,
N_c\,\rightarrow\,\infty.
$$
Thus the first equation of the Balitsky chain closes to the BK
equation.
 
Rummukainen and Weigert \cite{Rummukainen:2003ns} have produced a
first numerical solution of the JIMWLK equation. They do not find any
qualitative deviation from solutions of the BK equation. The $N_c$ corrections
were found to be at a level of a few percents.
 
Bartels, Lipatov, and Vacca \cite{Bartels:2004ef} have considered
$N_c$ corrections to the triple Pomeron vertex:
$$
\hspace{1cm} \frac{d\,N}{d\, y}\,\,=\,\alpha_s\,\,\Ker\,
\,\otimes\,\,(\,N\,\,- \,\,N\,\,N\,\,-\,\,{ \frac{1}{N_c^2}\,n})\,
$$ 
where the function $n$ has to satisfy a separate equation.

\subsubsection{Target correlations} 
For proton and realistic (not very dense) nucleus targets a
systematic approach towards inclusion of target correlations has been
developed by Levin and Lublinsky \cite{Levin:2003nc}.  Target
correlations can be introduced via a certain linear functional
differential equation.  In general, this linear functional equation
cannot be reformulated as a non-linear equation.  However, in a
particular case when all $n$-dipole correlations can be accounted for
by a single correlation parameter, the equation can be brought to a
modified version of the BK equation:
\begin{equation}\label{LLcor}  
\frac{d\,N}{d\, y}\,=\,\alpha_s\,\Ker\, \otimes\,(\,N\,- 
\,\kappa \,N\,N); 
\end{equation} 
In Eq.~(\ref{LLcor}) $\kappa \ge 1$ is the correlation parameter to be
found from a model for the target.

\subsubsection{Pomeron loops} 
Pomeron loops are the first steps towards restoration of the
$t$-channel unitarity. Iancu and Mueller \cite{Iancu:2003zr} have
considered rare fluctuations which were interpreted by Kozlov
and Levin \cite{Kozlov:2004sh} as pomeron loop contributions.
Unfortunately, it looks as if contributions from the pomeron loops are
difficult to incorporate in a framework of a single equation. They are
known to modify the asymptotic behavior of the amplitude $N$ in the
deep saturation limit, where they give the following asymptotic
behavior:
$$ N(Y)\,\,=\,\,1\,\,-\,\,e^{\,-\,c\,(Y\,-\,Y_0)^2};\quad
Y\,\rightarrow\,\infty \quad c\,=\,2\,\bar\alpha_s\quad \mbox{BKE} 
$$  
$$ 
N(Y)\,\,=\,\,1\,\,-\,\,e^{\,-\,c\,(Y\,-\,Y_0)^2/2};  
\quad
Y\,\rightarrow\,\infty \quad  Pom\,\, Loops 
$$
Recently there has been a lot of activity in attempting to
consistently include Pomeron loops into high energy evolutions
\cite{Iancu:2004iy,Mueller:2005ut,Levin:2005au,Kovner:2005nq,Kovner:2005jc,Kovner:2005en}.

\subsubsection{Local multi-pomeron exchange} 
It is claimed that the BK equation sums all possible contributions
which are not suppressed either by $\alpha_s$ or $N_c$. For example,
the cubic term in Eq.~(\ref{BB}) appears at next-to-leading
$\alpha_s$ order only.  In particular it is implied that all
multi-pomeron exchanges and multi-pomeron vertices are either absorbed
by the triple pomeron vertex of the BK equation or suppressed.  Levin and
Lublinsky \cite{Levin:2003nc} have argued that this might be not true.
They argue that in addition to a possibility for a pomeron to split
into two, there exists a process of multi-pomeron
exchange, which is local in rapidity. 
After these contributions were resummed in the eikonal
approximation, a new modification of the BK equation was proposed:
\begin{equation}\label{LL} 
\frac{d\,N}{d\, y}\,\,=\,\,(1\,\,-\,\,N)\,\alpha_s\,\,\Ker\, \, 
\otimes\,\,(\,N\,\,-\,\,N\,\,N).
\end{equation}


\subsection{Energy conservation aspects}
\label{section:energy-momentum-conservation}

\textit{Main author G.~Gustafson}\\

\subsubsection{Rapidity veto}

It is well known \cite{Salam:1999cn} that a major fraction of the
higher order corrections to (not only) BFKL is related to energy
conservation.  The large effect of energy-momentum conservation is
also clearly demonstrated by the numerical analyses by
Andersen-Stirling \cite{Andersen:2003gs} and Orr-Stirling
\cite{Orr:1997im}.  Conservation of energy and momentum implies the
conservation of both the positive and the negative lightcone
components, $p_\pm = E \pm p_L$. Although most analyses have
concentrated on the conservation of $p_+$, as being more important, we
will see below that also conservation of $p_-$ has a very significant
effect. In LLA the steps in $\ln(1/x)$ are assumed to be large, and
the necessary recoils due to energy conservation are neglected. The
main effect of conservation of the positive lightcone component $p_+ =
E + p_L$, is that small steps in $\ln(1/x)$ with corresponding large
recoils are suppressed.  One way to take this into account is to
introduce a veto, not allowing steps in $\ln(1/x)$ smaller than a cut
$\eta$. (This is called a rapidity veto also if the evolution variable
is defined as $y=\ln(1/x)$ and not the true rapidity.) The effect of
such a veto is studied in
refs.~\cite{Andersson:1996ju,Schmidt:1999mz,Forshaw:1999xm}, and at
high energies it has a similar effect as the higher order corrections,
reducing the growth at small $x$.

A recent study of the BK equation in the presence of a rapidity veto
is presented by Chachamis, Lublinsky and Sabio Vera
\cite{Chachamis:2004ab}. The application of this method to the BK equation
makes it non-local in rapidity:
$$ 
\frac{d\,N(y)}{d\, y}\,\,=\,\alpha_s\,Ker\, \,\otimes\,\,(\,N(y\,{ -\,\eta}) 
\,\,-\,\,N(y\,{-\,\eta})\,\,N(y\,{ -\,\eta})) \,.
$$
The veto somewhat delays saturation in accordance with the
expectations associated with the next-to-leading order corrections. If
the veto is put on top of the BK equation with running $\alpha_s$ then
the effect of NLO corrections is significantly reduced. This
observation gives support to the phenomenological studies of
Refs.~\cite{Gotsman:2002yy,Kutak:2003bd}.

An similar approach to this problem is presented by Gotsman, Levin,
Maor, and Naftali \cite{Gotsman:2005vc}. The effects of the cut in
$\ln(1/x)$ is taken into account in a modified BK equation:

\begin{equation}
  \begin{split}
    &\frac{\partial N(r,Y;b)}{\partial Y} = \frac{C_F \alpha_s}{\pi^2}
    \int \frac{d^2 r' r^2}{(\mathbf{r} - \mathbf{r'})^2 r'^2} \left(1
      - \frac{\partial}{\partial Y} \right)\\
    & \times\left[2 N \left(
        r',Y;\mathbf{b} - \frac{1}{2} (\mathbf{r} -
        \mathbf{r'})\right) -N\left(r,Y;\mathbf{b}\right)\right.\\
    &\left.\hspace{5mm}-N\left(r',Y;\mathbf{b} - \frac{1}{2} (\mathbf{r} -
        \mathbf{r'})\right) N\left(\mathbf{r} -
        \mathbf{r'},Y;\mathbf{b} - \frac{1}{2} \mathbf{r'}\right)
    \right].
\end{split}
\end{equation}
\label{GLMNeq}

The derivative under the integral is related to a cut in $\ln(1/x) \propto
\ln p_+$. The modification of the pole at
$\gamma = 1$, which is related to conservation of the \emph{negative}
lightcone component $p_- = E-p_L$ (or the inverse $k_\perp$ ordering) 
and
the consistency constraint, is not included. The motivation for this is that
this effect is not important once the dipole density has reached saturation,
that is for $x$ so small that $Q_s^2(x) > Q^2$.

We note, however, that the non-leading effects can significantly
reduce the value of $Q_s^2(x)$, and thus delay the onset of
saturation, as discussed in e.g.
refs.~\cite{Triantafyllopoulos:2002nz}.
An essential result of the analysis discussed in the next subsection
is that also the conservation of $p_-$ has an important effect, and
contributes significantly to pushing the $x$-values, where saturation
becomes essential, to smaller values. We note also that an estimate of
the relative importance of saturation and non-leading effects for the
reduced growth rate is very important for reliable extrapolations to
higher energies at LHC and high energy cosmic rays.

\subsubsection{Full energy-momentum conservation}

A different approach to energy-momentum conservation is presented in
ref.~\cite{Avsar:2005iz}. As discussed above non-leading effects are
most easily studied in momentum space, while unitarity or saturation
effects are easier analyzed when formulated in transverse coordinate
space. In ref.~\cite{Avsar:2005iz} similarities between the Linked
Dipole Chain model (LDC)~\cite{Andersson:1996ju,Andersson:1998bx} in
momentum space and the Mueller dipoles in transverse coordinate space
\cite{Mueller:1993rr,Mueller:1994jq,Mueller:1994gb} are used to derive
a scheme for implementing energy momentum conservation in Mueller's
dipole formalism. It is conjectured that only those gluon emissions,
which satisfy energy-momentum conservation, can correspond to real
final state gluons, and that keeping only these (with a corresponding
modification of the Sudakov form factor) will not only give a better
description of the final states, but also account for essential parts
of the NLO corrections to the BFKL equation. The approach is based on
the observation that the emission of a dipole with a very small
transverse size, $r$, corresponds to having two very well localized
gluons, and such gluons must have large transverse momenta of the
order $p_\perp \sim 1/r$. By in this way assigning a transverse
momentum to each emitted gluon, and also taking into account the
recoils of the emitting gluons, it is possible to make sure that each
dipole splitting is kinematically allowed.

\begin{figure}
\begin{center}
\scalebox{0.6}{\mbox{
\begin{picture}(400,100)(0,0)
\Vertex(50,100){2}
\Vertex(50,0){2}
\Vertex(150,100){2}
\Vertex(150,0){2}
\Vertex(180,30){2}
\Vertex(285,100){2}
\Vertex(285,0){2}
\Vertex(315,30){2}
\Vertex(315,90){2}
\Line(50,0)(50,100)
\DashLine(150,100)(150,0){2}
\Line(150,100)(180,30)
\Line(150,0)(180,30)
\DashLine(285,100)(285,0){2}
\DashLine(285,100)(315,30){2}
\Line(285,100)(315,90)
\Line(315,90)(315,30)
\Line(285,0)(315,30)
\Text(40,100)[]{$Q$}
\Text(40,0)[]{$\bar{Q}$}
\Text(60,100)[]{$1$}
\Text(60,0)[]{$0$}
\Text(140,100)[]{$1$}
\Text(140,0)[]{$0$}
\Text(140,50)[]{$\mathbf{r}_{01}$}
\Text(190,30)[]{$2$}
\Text(175,78)[]{$\mathbf{r}_{12}$}
\Text(175,2)[]{$\mathbf{r}_{02}$}
\Text(275,100)[]{$1$}
\Text(275,0)[]{$0$}
\Text(325,30)[]{$2$}
\Text(325,90)[]{$3$}
\LongArrow(75,50)(110,50)
\LongArrow(210,50)(245,50)
\LongArrow(360,25)(360,50)
\LongArrow(360,25)(385,25)
\Text(360,60)[]{$y$}
\Text(395,25)[]{$x$}
\end{picture}
}}
\end{center}
\caption{A quark-antiquark dipole in transverse coordinate space is split 
into successively more dipoles via gluon emission.}
\label{figdipolesplit}
\end{figure}
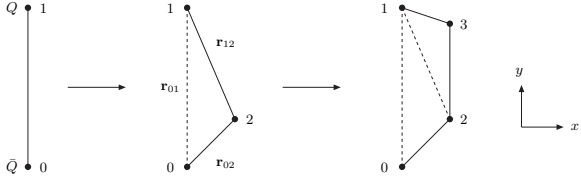
\paragraph{Formalism}
In the process $\gamma^* \rightarrow Q\bar{Q} \rightarrow Q g \bar{Q}
\rightarrow Q g g \bar{Q} \rightarrow \ldots$, a virtual photon 
is
split into a $Q\bar{Q}$ color dipole, which is first split into two
dipoles by the emission of a gluon, then into three dipoles by a
second gluon, etc. The process is illustrated in transverse coordinate
space in fig.  \ref{figdipolesplit}. The probability for such a dipole
splitting is given by the expression
\cite{Mueller:1993rr,Mueller:1994jq,Mueller:1994gb} (for notation see
fig.~\ref{figdipolesplit})
\begin{equation}
  \begin{split}
    \frac{dP}{dy} &= \frac{\bar{\alpha}}{2\pi} d^2 \textbf{r}_2
    \frac{r_{01}^2}{r_{02}^2 \,r_{12}^2} \cdot S;\\
    S &= \exp\left[-\frac{\bar{\alpha}}{2\pi} \int dy\int
      d^2\textbf{r}_2 \frac{r_{01}^2}{r_{02}^2 \,r_{12}^2}\right].
\label{splitprob}
\end{split}
\end{equation}
Here $S$ denotes a Sudakov form factor. We note that the integral 
over $d^2\textbf{r}_2$ in the exponent diverges for small values of $r_{02}$ 
and $r_{12}$. Therefore Mueller introduced a cutoff $\rho$, such that
the integration region satisfies $r_{02}>\rho$ and $r_{12}>\rho$. 
A small cutoff value $\rho$ will here imply that we get very many dipoles 
with small $r$-values.

\begin{figure}
\begin{center}
\scalebox{0.7}{\mbox{
\begin{picture}(400,83)(0,13)
\Vertex(20,50){1}
\Vertex(250,70){1}
\Vertex(240,30){1}
\Vertex(260,35){1}
\Vertex(260,25){1}
\Vertex(290,25){1}
\Vertex(335,70){1}
\Line(20,50)(250,70)
\Line(20,50)(240,30)
\Line(250,70)(240,30)
\Line(250,70)(260,35)
\Line(240,30)(260,35)
\Line(260,35)(260,25)
\Line(240,30)(260,25)
\Line(260,25)(290,25)
\Line(260,35)(290,25)
\Line(290,25)(335,70)
\Line(260,35)(335,70)
\Text(135,70)[]{$a$}
\Text(135,30)[]{$\approx a$}
\Text(10,50)[]{$0$}
\Text(240,50)[]{$b$}
\Text(263,56)[]{$\approx b$}
\Text(250,78)[]{$1$}
\Text(236,24)[]{$2$}
\Text(265,40)[]{$3$}
\Text(260,18)[]{$4$}
\Text(293,19)[]{$5$}
\Text(335,78)[]{$6$}
\Text(250,36)[]{$c$}
\Text(250,22)[]{$c$}
\Text(265,30)[]{$d$}
\Text(279,35)[]{$e$}
\Text(316,44)[]{$f$}
\end{picture}
}}
\end{center}
\caption{A dipole cascade, where a chain of smaller and smaller dipoles is
followed by a set of dipoles with increasing sizes. This is interpreted as
one $k_\perp$-ordered cascade from the left and one from the right, up to a 
central hard subcollision, which is represented by the dipole with minimum 
size and therefore maximum $k_\perp$.
}
\label{figdglap}
\end{figure}
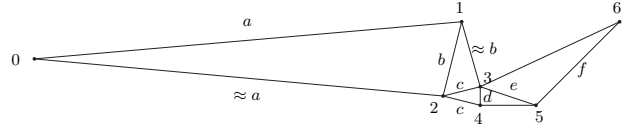

If a dipole size, $\textbf{r}$, is small, it means that the gluons are well 
localized, which must imply that transverse momenta are correspondingly large.
This implies that not only the new gluon gets a large $k_\perp \sim 1/r$, 
also the original gluon, which is close in coordinate space, gets a 
corresponding recoil. Let us study the example in fig. \ref{figdglap}.
For the emissions of the gluons marked 2, 3, and 4 the dipole sizes 
become smaller and smaller, $a \gg b \gg c \gg d$, in each step of the evolution.
The corresponding $k_\perp$ therefore
become larger and larger in each step. After the minimum dipole, with size $d$,
the subsequent emissions, 5, and 6, give
again larger dipoles with correspondingly lower $k_\perp$ values.
The probability for this chain is proportional to 
\begin{equation}
\frac{d^2 \textbf{r}_2\, a^2}{a^2 \,b^2} \cdot 
\frac{d^2 \textbf{r}_3\, b^2}{b^2\, c^2} 
\cdot \frac{d^2 \textbf{r}_4\, c^2}{c^2 \,d^2} \cdot 
\frac{d^2 \textbf{r}_5\, d^2}{e^2 \,e^2} \cdot
\frac{d^2 \textbf{r}_6\, e^2}{f^2 \,f^2}
\label{dglap}
\end{equation}
For the first emissions, 2 and 3, in this expression we recognize the product 
of factors 
$d^2\textbf{r}_i/r_i^2\, \sim \prod d^2\textbf{k}_i/k_i^2$, 
just as is expected from a ``DGLAP evolution'' of a chain with monotonically
increasing $k_\perp$. Emission number 4 corresponds to the minimum dipole size,
$d$, and we here note that the factors of $d$ cancel in eq.~(\ref{dglap}). 
We therefore get the weight $d^2 \textbf{r}_4 
\sim d^2\textbf{k}_{\mathrm{max}}/k_{\mathrm{max}}^4$, which corresponds to a 
hard gluon-gluon collision. When the dipole sizes get larger again, this gives
factors corresponding to a ``DGLAP chain'' from the other end of the chain,
up to the central hard subcollision. 

It is also easy to see that for a chain with increasing dipole sizes
up to a maximum value, $r_{\mathrm{max}}$, which thus corresponds to a
minimum transverse momentum, $k_{\perp \mathrm{min}}$, we get the
weight $d^2 \textbf{r}_{\mathrm{max}}/r_{\mathrm{max}}^4 \sim d^2
\textbf{k}_{\mathrm{min}}$. Therefore there is no singularity for the
minimum $k_\perp$-value.  This result agrees exactly with the result
in the Linked Dipole Chain model, LDC
\cite{Andersson:1996ju,Andersson:1998bx}, which is a reformulation of
the CCFM model \cite{Catani:1990yc,Ciafaloni:1988ur}, interpolating
between DGLAP and BFKL for non-$k_\perp$-ordered chains.
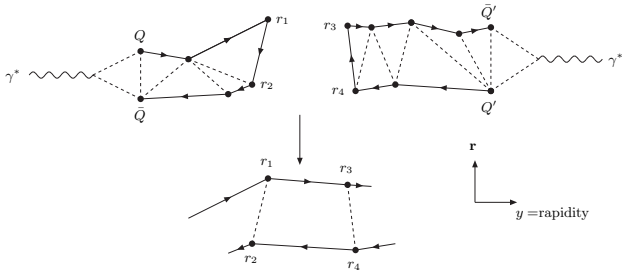
\begin{figure}
\begin{center}
\scalebox{0.6}{\mbox{
\begin{picture}(400,175)(0,5)
\Photon(20,130)(60,130){2}{4}
\Photon(380,140)(340,140){2}{4}
\DashLine(60,130)(90,145){2}
\DashLine(60,130)(90,115){2}
\DashLine(340,140)(310,160){2}
\DashLine(340,140)(310,120){2}
\DashLine(90,145)(90,115){2}
\DashLine(310,120)(310,160){2}
\DashLine(90,115)(120,140){2}
\DashLine(120,140)(145,118){2}
\DashLine(160,124)(120,140){2}
\DashLine(310,120)(290,156){2}
\DashLine(310,120)(260,163){2}
\DashLine(260,163)(250,124){2}
\DashLine(250,124)(235,160){2}
\DashLine(235,160)(225,120){2}
\ArrowLine(90,145)(120,140)
\ArrowLine(120,140)(170,165)
\ArrowLine(145,118)(90,115)
\ArrowLine(170,165)(160,124)
\ArrowLine(160,124)(145,118)
\ArrowLine(120,140)(170,165)
\ArrowLine(290,156)(310,160)
\ArrowLine(260,163)(290,156)
\ArrowLine(310,120)(250,124)
\ArrowLine(225,120)(220,161)
\ArrowLine(250,124)(225,120)
\ArrowLine(220,161)(235,160)
\ArrowLine(235,160)(260,163)
\Vertex(310,160){2}
\Vertex(310,120){2}
\Vertex(90,145){2}
\Vertex(90,115){2}
\Vertex(120,140){2}
\Vertex(170,165){2}
\Vertex(145,118){2}
\Vertex(160,124){2}
\Vertex(290,156){2}
\Vertex(260,163){2}
\Vertex(310,120){2}
\Vertex(250,124){2}
\Vertex(235,160){2}
\Vertex(260,163){2}
\Vertex(220,161){2}
\Vertex(225,120){2}
\Text(10,130)[]{$\gamma^*$}
\Text(390,140)[]{$\gamma^*$}
\Text(90,155)[]{$Q$}
\Text(90,105)[]{$\bar{Q}$}
\Text(310,170)[]{$\bar{Q}'$}
\Text(310,110)[]{$Q'$}
\Text(210,161)[]{$r_3$}
\Text(215,120)[]{$r_4$}
\Text(180,165)[]{$r_1$}
\Text(170,124)[]{$r_2$}
\LongArrow(190,105)(190,75)
\DashLine(170,65)(160,24){2}
\ArrowLine(160,24)(145,18)
\ArrowLine(120,40)(170,65)
\DashLine(225,20)(220,61){2}
\ArrowLine(250,24)(225,20)
\ArrowLine(220,61)(235,60)
\Vertex(220,61){2}
\Vertex(225,20){2}
\Vertex(170,65){2}
\Vertex(160,24){2}
\ArrowLine(170,65)(220,61)
\ArrowLine(225,20)(160,24)
\Text(170,75)[]{$r_1$}
\Text(160,14)[]{$r_2$}
\Text(220,71)[]{$r_3$}
\Text(225,10)[]{$r_4$}
\LongArrow(300,50)(300,75)
\LongArrow(300,50)(325,50)
\Text(300,85)[]{$\mathbf{r}$}
\Text(350,43)[]{$y=$rapidity}
\end{picture}
}}
\end{center}
\caption{A symbolic picture of a $\gamma^* \gamma^*$ collision in 
rapidity-$\mathbf{r}_\perp$-space. The two dipole chains interact and 
recouple with probability $f$ given by eq.~(\ref{dipoltvarsnitt}).}
\label{figgammagamma}
\end{figure}

To study $\gamma^* \gamma^*$ scattering we imagine that the two virtual photons
split up into quark-antiquark pairs, which develop into dipole cascades 
as schematically illustrated in fig.~\ref{figgammagamma}. When the two central 
dipoles collide and interact via gluon exchange, it implies a recoupling 
of the color charges,
as indicated by the arrow,
and the probability for this is given by 
the expression \cite{Salam:1995uy}
\begin{equation}
f = \frac{\alpha_s^2}{2} \left\{\ln\left[
\frac{|\textbf{r}_1 -\textbf{r}_3|\cdot|\textbf{r}_2 -\textbf{r}_4|}
{|\textbf{r}_1 -\textbf{r}_4|\cdot|\textbf{r}_2 -\textbf{r}_3|}\right]\right\}^2.
\label{dipoltvarsnitt}
\end{equation}

As the dipole cascades from the two virtual photons branch out, it is also
possible to have {\em multiple collisions}, when more than one pair of
dipoles from the left and the right moving cascades are interacting.
The total cross section is then given by
\begin{equation}
\sigma \sim \int d^2b (1-e^{-\sum f_{ij}}),
\label{multint}
\end{equation}
where $b$ denotes the impact parameter.

With a small cutoff $\rho$ ($r>\rho$) we get, as mentioned above, very many 
small dipoles. If these are interpreted as real emissions, it would imply a 
violation of energy-momentum conservation. The emission of these small dipoles 
must be compensated by virtual emissions. Thus the result in eq.~(\ref{multint})
will describe the inclusive cross section, but the many dipoles produced in
all the branching chains
will not correspond to the production of exclusive final states.

The main feature
of the LDC model is the observation that both the total cross section
and the final state structures are determined by chains consisting of
a subset of the gluons appearing in the final state. These gluons were
called ``primary gluons'' in ref.~\cite{Andersson:1996ju} and later
called ``backbone gluons'' in ref.~\cite{Salam:1999ft}. Remaining real
final state gluons can be treated as final state radiation from the
primary gluons. Such final state emissions do not modify the total
cross sections, and give only small recoils to the parent emitters.
The primary gluons have to satisfy energy-momentum conservation, and
are ordered in both positive and negative light-cone momentum
components, $p_+$ and $p_-$. We saw above that in
Mueller's cascade the emission probabilities for gluons, which satisfy
the conditions for primary gluons in LDC, have exactly the same
weight, when the transverse momenta are identified with the inverse
dipole size, $2/r$. This inspires the conjecture that with this
identification an appropriate subset of the emissions in Mueller's
cascade can correspond to the primary gluons in the momentum space
cascade, meaning that they determine the cross sections while the
other emissions can be regarded as either virtual fluctuations or
final state radiation.

A necessary condition for this subset of gluons is that energy and
momentum is conserved. Therefore we expect that keeping 
only emissions which satisfy energy-momentum
conservation can correspond to real emissions, and keeping only these
emissions (with a corresponding modification of the Sudakov form
factor) will not only account for important NLO effects, but also give 
a closer correspondence between the generated dipole
chains and the observable final states.

A very important consequence of energy-momentum conservation is also
that it implies a {\em dynamical cutoff}, $\rho(\Delta y)$, which is
large for small steps in rapidity, $\Delta y$, but gets smaller
for larger $\Delta y$. (Alternatively it could be described as a
cutoff for $\Delta y$ which depends on $r$. Note that in this formalism
  $y$ is the true rapidity and not $\log(1/x)$.)
Conserving also the negative light-cone momentum, $p_-$, implies that
in a similar way we may also get a maximum value for $r$ in each emission.

The net result of conservation of both $p_+$ and $p_-$ is that the
number of dipoles grows much more slowly with energy. Besides its physical 
effects, this also simplifies the implementation in a MC program. It is here
straight forward to calculate cross sections and to study saturation effects,
by comparing the unitarized expression $\int d^2b (1-e^{-\sum f_{ij}})$ 
in eq.~(\ref{multint}) with $\int d^2b \sum f_{ij}$ representing single
$I\!\!P$ exchange.
(The large numerical complications in MCs without energy conservation, 
discussed in ref.~\cite{Salam:1995uy}, are not present.)

\paragraph{Results}
Below we show some results obtained with a fixed coupling $\bar{\alpha} = 0.2$.

\begin{figure}[t] 
  \includegraphics[angle=270, scale=0.45]{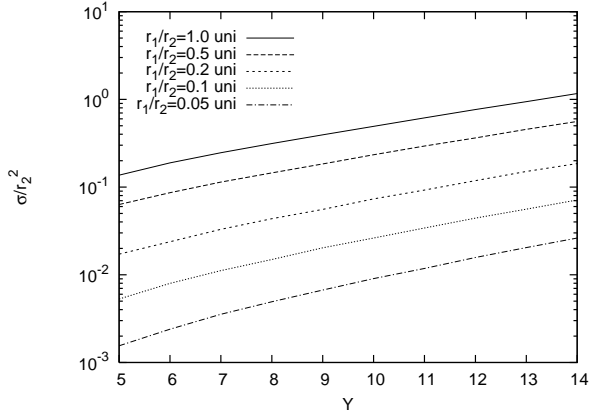}
  \caption{\label{ECuni}The scaled unitarized
    dipole--dipole cross section, $\sigma/r_2^2$, as a function of $Y$
    for different initial conditions.}
\end{figure}

{\em Dipole-dipole scattering}.
The cross section for scattering of two dipoles with sizes $r_1$ and 
$r_2$ is shown in fig.~\ref{ECuni}. With a fixed coupling the 
scaled cross section, $\sigma/r_2^2$,
depends only on the ratio $r_1/r_2$. We can imagine a target with size
$r_2 \sim 1/M$, and a varying projectile size $r_1 \sim 1/\sqrt{Q^2}$. 
The results show that the cross section 
grows faster with the total rapidity range, $Y \sim \ln s$, for smaller $r_1$ 
(larger $Q^2$), in a way qualitatively similar to the 
behavior of the proton structure function. 

\begin{figure}[t] 
  \includegraphics[angle=270, scale=0.45]{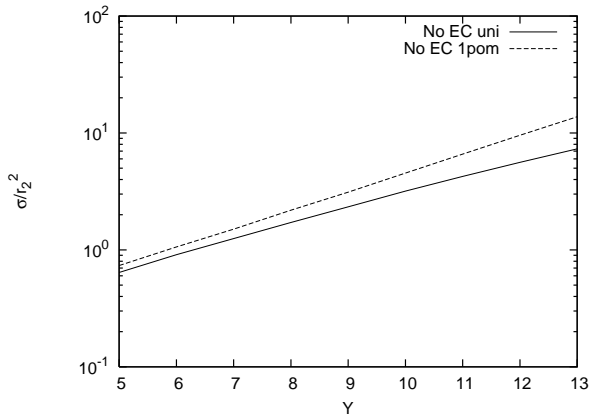}
  \caption{\label{noEC}The scaled unitarized (full line) and
    one-pomeron (dashed line) dipole--dipole cross
    sections calculated \emph{without} energy conservation.}
\end{figure}

The effect of energy conservation is demonstrated in 
fig.~\ref{noEC} by the results obtained for the case $r_1\!=\!r_2$,
with a constant cutoff, 
$\rho=0.02 \,r_i$. Comparing with fig.~\ref{ECuni} we see that energy 
conservation
has a very strong effect, reducing $\sigma$ by almost an order of magnitude 
for $Y\sim13$.

In fig.~\ref{noEC} we also see that without energy-momentum conservation the 
effect of multiple $I\!\!P$ 
exchange (saturation) is about a factor 2 for $r_1 = r_2$ and $Y=13$.
The much smaller cross section obtained with energy-momentum conservation
implies that the saturation effect is much less important, being only
$\approx 20\%$ for the same parameter values.

{\em Dipole-nucleus and dipole-proton collisions}.
Dipole-nucleus collisions have been studied using a toy model nucleus, 
with a Gaussian distribution in dipole size $r$ 
and impact parameter $b$. The dipole density is given by
\begin{equation}
dN=B\cdot  d^2\textbf{r}\, e^{-r^2/r_0^2}\cdot d^2\textbf{b}\, e^{-b^2/b_0^2}
\label{toynucleus}
\end{equation}
The widths of the distributions are taken to be $r_0=1 \,\mathrm{fm}$ and 
$b_0=A^{1/3}\cdot1\,\mathrm{fm}$ (where $A$ is the mass number of the nucleus),
and the normalization constant $B$ is adjusted so that the transverse energy 
is given by $A\!\cdot\!1\, \mathrm{GeV}$.

\begin{figure}[t]
  \includegraphics[angle=270,scale=0.45]{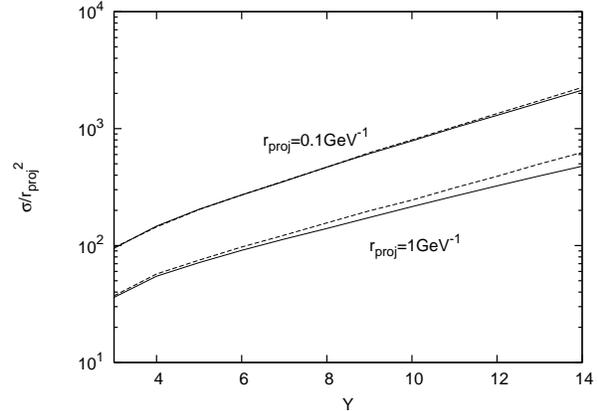}
  \caption{\label{sprobe}The dipole--nucleus cross section for 
    $r_{\mathrm{proj}}=0.1$ and $1$~GeV$^{-1}$ and $A = 200$. The
    unitarized result is shown by the solid lines, and the one-pomeron
    contribution by the dashed lines.}
\end{figure}
The results for $A = 200$ and projectile sizes $r_{\mathrm{proj}}=0.1$ 
and
$1$~GeV$^{-1}$ are shown in figure \ref{sprobe}. Results are presented
both for single pomeron exchange and including unitarization. The
effect of unitarization grows with nuclear size and with the size of
the projectile.  For a small projectile of size $0.1$~GeV$^{-1}$ we
can see the effect of color transparency, as the cross sections for
the unitarized and the one pomeron calculations are almost identical.
For a larger projectile we do see a clear effect from unitarization,
but even for $r_{\mathrm{proj}}=1$~GeV$^{-1}$ and a nucleus with $A=200$ this 
effect is only about $20\%$ in the rapidity interval
$10-14$. For smaller nuclei the effect will be correspondingly
smaller.

When the same toy model is applied to  deep inelastic ep scattering
(with $A=1$ and simply identifying $Q^2$ with $4/r_\mathrm{proj}^2$), we want 
to emphasize that we here only want to study the qualitative behavior.
A quantitative comparison with HERA data has to wait for an improvement
of the crude toy model for the proton target (dipole
correlations may be important), and one should then also 
take into account the detailed effects of the photon wavefunction. 

\begin{figure}[t]
  \includegraphics[angle=270, scale=0.45]{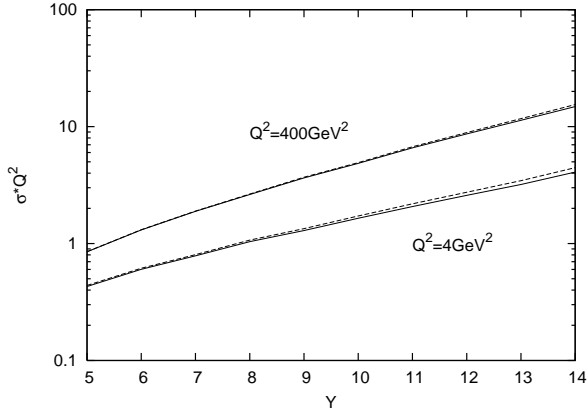}
  \caption{\label{labcm}The scaled dipole--p cross section as a
    function of log$1/x$, for $Q^2=4$~GeV$^2$ and $Q^2=400$~GeV$^2$.
    The unitarized results are shown by the solid lines while the
    dashed lines show the one-pomeron results.}
\end{figure}

The resulting dipole--nucleon cross section is shown in
figure \ref{labcm} for two different projectile sizes, corresponding to
$Q^2=4\, \mathrm{GeV}^2$ and $Q^2=400 \,\mathrm{GeV}^2$. The result for 
single pomeron exchange, i.e.\ 
without unitarization corrections, is shown by the dashed lines,
and we see that the effect from unitarization is quite small. 

\begin{figure}[t]
  \includegraphics[angle=270, scale=0.45]{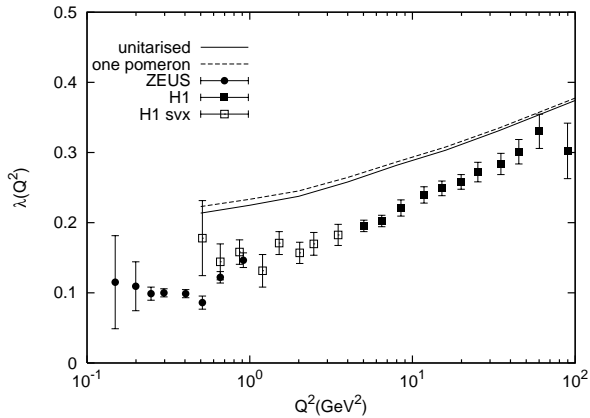}
  \caption{\label{lambdaQ22}The effective slope measured at different
    $Q^2$ compared to data from HERA. The full line is our model
    including unitarization, while the dashed line is without. Filled
    circles are data from ZEUS\protect\cite{Breitweg:2000yn},
    filled\protect\cite{Adloff:2000qk} and open\protect\cite{DIS04Petrukhin} squares
    are data from H1.}
\end{figure}

In figure \ref{labcm} we also see that the logarithmic slope
$\lambda_{\mathrm{eff}}=d (\log \sigma)/d (\log 1/x)$ is increasing
with increasing $Q^2$.  The effective slope, $\lambda_{\mathrm{eff}}$,
is not a constant for fixed $Q^2$, but depends on both $Q^2$ and $x$,
when unitarization and/or energy conservation is taken into account.
For the comparison with experimental data figure \ref{lambdaQ22} shows
$\lambda_{\mathrm{eff}}$ determined in the $x$-interval used in the
analysis by H1 \cite{Adloff:2001rw}, which varies from $x\approx
2\times 10^{-5}$ for $Q^2=1.5$~GeV$^2$ to $x\approx 3\times 10^{-2}$
for $Q^2=90$~GeV$^2$. We note that the result of our crude model
is not far from the experimental data, although the dependence on
$Q^2$ is somewhat weaker in the model calculations.  As in figure
\ref{labcm} we see that the effect of unitarization is small, and, as
expected, it gets further reduced for larger $Q^2$-values. 

Thus we find that the result of the simple model is
surprisingly close to experimental data from HERA. 
The effect of energy conservation is a suppression for
small $x$-values and small $Q^2$, which is qualitatively similar to
the effect expected from unitarization. This suppression is so strong
that the effect from adding unitarization is only a very small
correction, visible for small $Q^2$-values.

If we compare these results with those of ref.~\cite{Gotsman:2005vc}, 
we find a significantly larger
effect from energy-momentum conservation. One reason appears to be the
inclusion of $p_-$-conservation.  This is related
to the consistency constraint in eq.~(\ref{eq:kincon}), which orders the 
emissions in the negative lightcone momentum. In the formalism discussed here 
this 
is found to
have a noticeable effect. Thus we find that including only
conservation of $p_+$, and not of $p_-$, increases the cross section
by a factor 2 (3) for dipole--proton collisions at $Q^2=4$ (400)GeV$^2$.
Consequently we conclude that full energy-momentum conservation is 
very essential for the result and for the relative importance of saturation
and NLO effects.

\subsection{Outlook}

It is essential for the future phenomenological studies to eliminate
the model dependent treatments of the impact parameter.  Though the BK equation
has been solved numerically with the full $b$-dependence traced
\cite{Golec-Biernat:2003ym,Gotsman:2004ra}, these results are not yet
suitable for phenomenological applications.
 
A further study of the relation between the dipole picture \emph{vs}.  
traditional diagrammatics based on the $s$-channel unitarity is needed. 
In particular, it is not clear if the dipole picture survives at NLO.  
In general there is a quest for a {\bf simple} effective Reggeon field  
theory in QCD. 

The large effect of full energy-momentum conservation make further studies
of the relative importance of NLO effects and saturation important.

NLO effects and saturation both contribute to a reduction of the parton
distributions for small $x$. An improved understanding of these effects,
including the relation between them, is very important for extrapolations to 
higher energies at LHC or high energy cosmic ray events.

The discussions presented above concentrate on total or inclusive cross 
sections. More work is also needed to calculate the properties of the 
resulting final states.


\section{Multiple interactions, saturation and rapidity gaps}
\label{sec:agk}


\newcommand{\mbf}[1]{\mbox{\boldmath $#1$}}

\newcommand{\bk}{\mbf{k}}
\newcommand{\bq}{\mbf{q}}
\renewcommand{\bb}{\mbf{b}}

\subsection{AGK cutting rules}
\textit{Main author J.~Bartels}\\
\subsubsection{Introduction}
\begin{figure}
  \begin{center}
    \epsfig{file=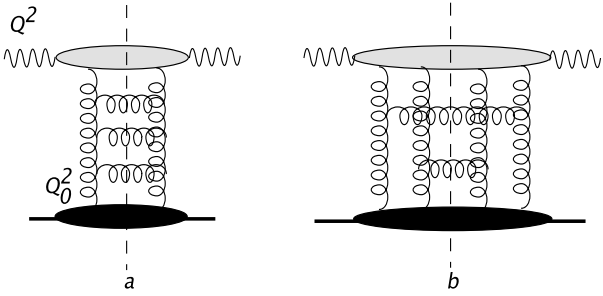,width=8cm,height=4cm}
  \end{center}
  \caption{\label{fig:jochen1}Contributions to the total cross section
    $\sigma_{tot}^{\gamma^*p}$: (a) the single chain representing the
    linear QCD evolution equations; (b) gluon production from two
    different gluon chains.}
\end{figure}

Multiple parton interactions play an important role both in electron
proton scattering at HERA and in high energy proton proton collisions
at the LHC. At HERA, the linear QCD evolution equations provide, for
not too small $Q^2$, a good description of the $F_2$ data (and of the
total $\gamma^* p$ cross section, $\sigma_{tot}^{\gamma^*p}$). This
description corresponds to the emission of partons from a single chain
(Fig.~\ref{fig:jochen1}a).  However, at low $Q^2$ where the transition
to nonperturbative strong interaction physics starts, this simple
picture has to be supplemented with corrections. First, there exists a
class of models
~\cite{Golec-Biernat:1998js,Golec-Biernat:1999qd,Bartels:2002cj} which
successfully describe this transition region; these models are based
upon the idea of parton saturation: they assume the existence of
multiple parton chains (Fig.~\ref{fig:jochen1}b) which interact with
each other,
and they naturally explain the observed scaling behavior, $F_2(Q^2,x)
\approx F_2(Q^2/Q_s^2(x))$ with $Q_s^2(x)= Q_0^2 (1/x)^{\lambda}$.
Next, in the photoproduction region, $Q^2 \approx 0$, direct evidence
for the presence of multiple interactions also comes from the analysis
of final states \cite{Gwenlan:2004iu}. A further strong hint at the
presence of multi-chain configurations comes from the observation of a
large fraction of diffractive final states in deep inelastic
scattering at HERA. In the final states analysis of the linear QCD
evolution equations, it is expected that the produced partons are not
likely to come with large rapidity intervals between them.  In the
momentum-ordered single chain picture (Fig.~\ref{fig:jochen1}a),
therefore, diffractive final states should be part of the initial
conditions (inside the lower blob in Fig.~\ref{fig:jochen1}a), i.e.
they should lie below the scale $Q_0^2$ which separates the parton
description from the nonperturbative strong interactions. This
assignment of diffractive final states, however, cannot be
\begin{figure}
  \begin{center}
    \epsfig{file=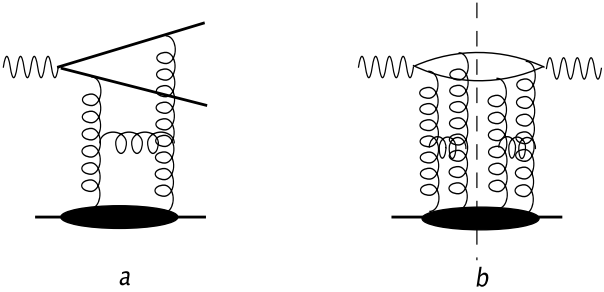,width=8cm,height=4cm}
  \end{center}
  \caption{\label{fig:jochen2}Hard diffractive final states.(a) dijet
    production; (b) the diffractive cross section as $s$-channel
    discontinuity of a two-ladder diagram.}
\end{figure}
complete. First, data have shown that the Pomeron which generates the
rapidity gap in DIS diffraction is harder than in hadron - hadron
scattering; furthermore, there are specific diffractive final states
with momentum scales larger than $Q_0^2$, e.g. vector mesons built
from heavy quarks and diffractive dijets (illustrated in
Fig.~\ref{fig:jochen2}): the presence of such final states naturally
requires corrections to the single chain picture
(Fig.~\ref{fig:jochen2}b).  From a $t$-channel point of view, both
Fig.~\ref{fig:jochen1}b and Fig.~\ref{fig:jochen2}b belong to the same
class of corrections, characterized by four gluon states in the
$t$-channel.
  
\begin{figure}
  \begin{center}
    \epsfig{file=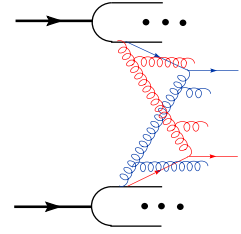,width=3cm,height=3cm}
  \end{center}
  \caption{\label{fig:jochen3}Jet production in $pp$ collisions from two
    different parton chains}
\end{figure}

In proton-proton collisions corrections due to multiple interactions
should be important in those kinematic regions where parton densities
for small momentum fractions and for not too large momentum scales are
being probed, e.g. jet production near the forward direction. Another
place could be the production of multijet final states
(Fig.~\ref{fig:jochen3}): multiple jets may come from different parton
chains, and these contributions
may very well affect the background to new physics beyond the standard
model.  Moreover, the modeling of multijet configurations will be
necessary for understanding the underlying event structure in $pp$
collisions (see \cite{Sjostrand:2004ef} and references therein).

From the point of view of collinear factorization, multiple
interactions with momentum ordered parton chains are higher-twist
effects, i.e they are suppressed by powers of the hard momentum scale.
At small $x$, however, this suppression is compensated by powers of
the large logarithms, $\ln 1/x$: multiple interactions, therefore, are
mainly part of small-$x$ physics. In this kinematic region the
Abramovsky-Gribov-Kanchelli (AGK) ~\cite{Abramovsky:1973fm} rules can
be applied to the analysis of multi-gluon chains, and it is the aim of
this article to present a brief overview about the current status of
the AGK rules in pQCD.
      
As we will discuss below, in the analysis of multiple parton chains
the couplings of $n$ gluons to the proton play an essential role.
Regge factorization suggests that these couplings should be universal,
i.e. the couplings in $\gamma^* p$ collisions at HERA are the same as
those in $pp$ scattering at the LHC. Therefore, a thorough analysis of
the role of multiple interactions in deep inelastic electron-proton
scattering at HERA should be useful for a solid understanding of the
structure of events at the LHC.

\subsubsection{Basics of the AGK cutting rules}

\begin{figure}
  \begin{center}
    \epsfig{file=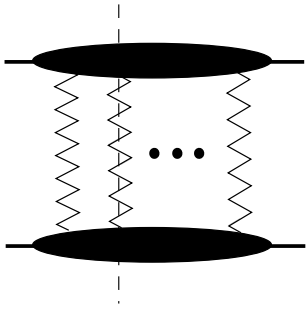,width=4cm,height=4cm}
  \end{center}
  \caption{\label{fig:jochen4}$s$-cut through a multi-Pomeron exchange:
    the zig-zag lines stand for nonperturbative Pomerons.}
\end{figure}

The original AGK paper ~\cite{Abramovsky:1973fm}, which was written
before the advent of QCD, addresses the question how, in the optical
theorem,
\begin{equation}
\sigma_{tot}^{pp} = \frac{1}{s} Im\, T_{2 \to 2}\,\, =\,\, 
\sum_f \int d \Omega_f | T_{i \to f}|^2,  
\label{optical}
\end{equation}
the presence of multi-Pomeron exchanges (Fig.~\ref{fig:jochen4})
in the total hadron-hadron cross section leads to observable effects
in the final states (rhs of eq.(\ref{optical})).  Based upon a few
model-independent assumptions on the couplings of multi-Pomeron
exchanges to the proton, the authors derived simple `cutting rules':
different contributions to the imaginary part belong to different cuts
across the multi-Pomeron diagrams, and each cut has its own, quite
distinct, final state characteristics.  As a result, the authors found
counting rules for final states with different particle
multiplicities, and they proved cancellations among rescattering
corrections to single-particle and double-particle inclusive cross
sections.

In the QCD description of hard (or semihard) final states a close
analogy appears between (color singlet) gluon ladders and the
nonperturbative Pomeron: multiple parton chains (for example, the two
chains in Fig.~\ref{fig:jochen1}b) can be viewed as cuts through two
perturbative BFKL Pomerons. In the same way as in the original AGK
paper, the question arises how different cuts through a QCD
multi-ladder diagram can be related to each other.  In the following
we briefly describe how AGK cutting rules can be derived in pQCD
~\cite{Bartels:1996hw,Bartels:2005wa}. Subsequently we will present a
few new results which come out from pQCD calculations, going beyond
the original AGK rules, followed by some numerical estimates of the
effects which can be expected.

\begin{figure}
  \begin{center}
    \epsfig{file=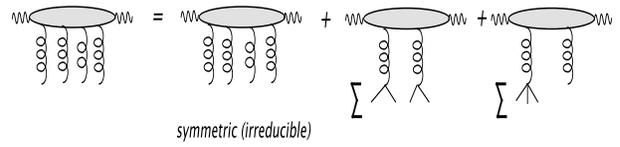,width=8cm,height=2cm}
  \end{center}
  \caption{\label{fig:jochen5}Decomposition of the coupling of four gluons to
    a virtual photon. In the last two terms on the rhs it is
    understood that we have to sum over different pairings of gluons
    at the lower end.}
\end{figure}

One of the few assumptions made in the original AGK paper states that
the coupling of the Pomerons to the external particle are (i)
symmetric under the exchange of the Pomerons (Bose symmetry), and (ii)
that they remain unchanged if some of the Pomerons are being cut.
These properties also hold in pQCD, but they have to be reformulated:
(i') the coupling of (reggeized) gluons to external particles is
symmetric under the exchange of reggeized gluons, and (ii') it remains
unchanged if we introduce cutting lines between the gluons.  In QCD,
however, the color degree of freedom also allows for another
possibility: inside the n-gluon state (with total color zero), a
subsystem of two gluons can form an antisymmetric color octet state:
in this case the two gluons form a bound state of a reggeized gluon
(bootstrap property). For the case of $\gamma^* \gamma^*$ scattering,
explicit calculations ~\cite{Bartels:1994jj} have shown that the
coupling of $n$ gluons to virtual photons can be written as a sum of
several pieces: the fully symmetric (`irreducible') one which
satisfies (i') and (ii'), and other pieces which, by using the
bootstrap property, can be reduced to symmetric couplings of a smaller
number of gluons (`cut reggeons'). This decomposition is illustrated
in Fig.~\ref{fig:jochen5}.
Since the bootstrap property is related to the reggeization of the gluon 
and, therefore, is expected to be valid to all orders of perturbation 
theory, also these properties of the couplings of multi-gluon states to 
external particles should be of general validity. 
In this short review we will mainly concentrate on the symmetric couplings.   
            
As an illustrative example, we consider the coupling of four gluons to 
a proton. The simplest model of a symmetric coupling is a
sum of three pieces, each of which contains only the simplest color 
structure:   
\begin{figure}
  \begin{center}
    \epsfig{file=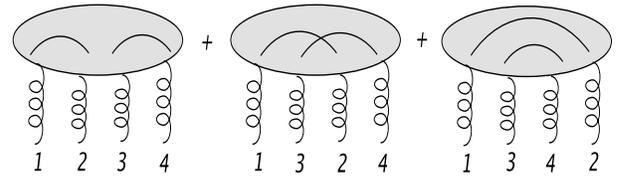,width=8cm,height=2.5cm}
  \end{center}
  \caption{\label{fig:jochen6}The symmetric coupling of four gluons to an
    external particle.  The lines inside the blob denote the color
    connection, e.g. the first term has the color structure
    $\delta_{a_1 a_2} \delta_{a_3 a_4}$.}
\end{figure} 
The best-known cutting rule for the four gluon exchange which follows
~\cite{Bartels:1996hw,Bartels:2005wa} from this symmetry requirement
is the ratio between the three different pairings of lines given in
Fig.~\ref{fig:jochen7}.
\begin{figure}
  \begin{center}
    \epsfig{file=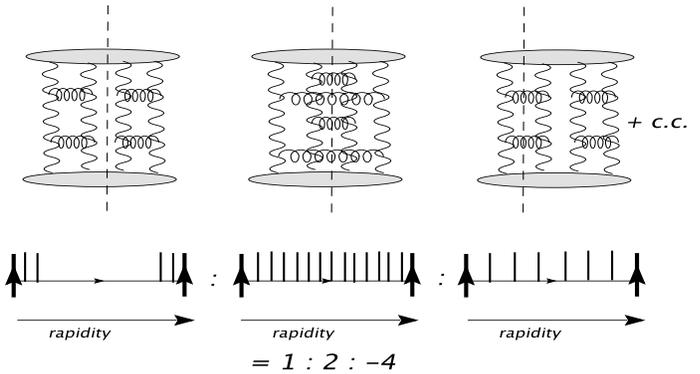,width=9cm,height=5cm}
  \end{center}
  \caption{\label{fig:jochen7}Different cutting lines in the four-gluon
    exchange.}
\end{figure}
Each term, on the partonic level, corresponds to a certain multiplicity 
structure of the final state: a rapidity gap (`zero multiplicity'), 
double multiplicity, and single multiplicity. Simple combinatorics
then leads to the ratio\cite{Abramovsky:1973fm}
\begin{equation}
1:2:-4. 
\label{famous}
\end{equation}
for the two-ladder contribution to the cross section.  In order to be
able to generalize and to sum over an arbitrary number of gluon
chains, it is convenient to use an eikonal ansatz:
\begin{equation}
\begin{split}
  N_{2n}^A(\bk_1,a_1;\ldots;&\bk_{2n},a_{2n};\omega)=\\
  \frac{1}{\sqrt{(N_c^2-1)^n}}
  \Biggl(& \sum_{Pairings} \phi^A(\bk_1,\bk_2;\omega_{12})
  \delta_{a_1a_2}
  \cdot\ldots\\
  & \cdot\phi^A(\bk_{2n-1},\bk_{2n};\omega_{{2n-1},{2n}})
  \delta_{a_{2n-1}a_{2n}}  
  \Biggr)\;.
\end{split}
\label{eq:jochen3}
\end{equation}
Inserting this ansatz into the hadron--hadron scattering amplitude, 
using the large-$N_c$ approximation, and switching to the impact parameter 
representation, one obtains, for the contribution of $k$ cut gluon ladders, 
the well-known formula:
\begin{equation}
Im A_k = 4 s \int d^2 \bb e^{i\bq \bb} P(s,\bb)
\label{eq:jochen4}
\end{equation}
where
\begin{equation}
P(s,\bb)\;=\;\frac{[\Omega(s,\bb)]^k}{k!} e^{-\Omega(s,\bb)},
\end{equation}
and $\Omega$ stands for the (cut) two-gluon ladder.

\begin{figure}
  \begin{center}
    \epsfig{file=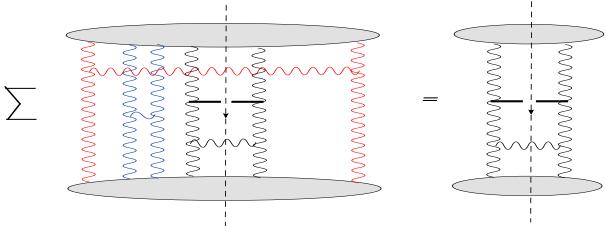,width=8cm,height=3cm}
  \end{center}
  \caption{\label{fig:jochen8}AGK cancellations in the one-jet inclusive
    cross section.}
\end{figure}

Another result ~\cite{Bartels:2005wa} which follows from the symmetry
properties of the $n$ gluon-particle coupling is the cancellation of
rescattering effects in single and double inclusive cross sections. In
analogy with the AGK results on the rescattering of soft Pomerons, it
can be shown that the sum over multi-chain contributions and
rescattering corrections cancels (Fig.~\ref{fig:jochen8}),
leaving only the single-chain contribution (in agreement with the
factorization obtained in the collinear analysis). This statement,
however, holds only for rescattering between the two projectiles: it
does not affect the multiple exchanges between the tagged jet and the
projectile (Fig.~\ref{fig:jochen9}) which require a separate
discussion (see below).
\begin{figure}
  \begin{center}
    \epsfig{file=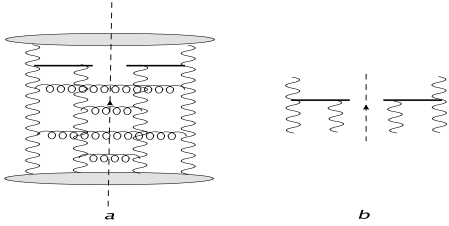,width=6cm,height=3cm}
  \end{center}
  \caption{(a) Nonvanishing rescattering corrections in
    the one-jet inclusive cross section; (b) a new vertex: $g\, +\, 2g
    \to jet$.}
  \label{fig:jochen9}
\end{figure}
All these results can be generalized to include also the soft Pomeron:
all one needs to assume is that the couplings of soft Pomerons and 
reggeized gluons are symmetric under interchanges, and they are not altered 
if cutting lines are introduced.

\subsubsection{New results}    

\begin{figure}
  \begin{center}
    \epsfig{file=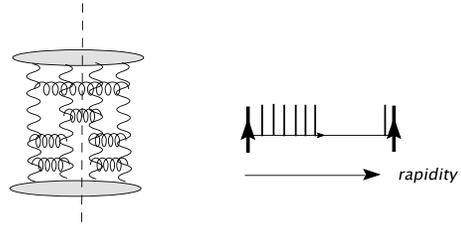,width=6cm,height=3cm}
  \end{center}
  \caption{Decomposition into two rapidity intervals:
    the upper (left) interval has double multiplicity, the lower
    (right) one corresponds to a rapidity gap.}
  \label{fig:jochen10}
\end{figure}

Explicit calculations in QCD lead to further results on multiple
interactions.  First, in the four gluon exchange there are other
configurations than those shown in Fig.~\ref{fig:jochen7}; one example
is depicted in Fig.~\ref{fig:jochen10}. Here the pairing of gluon
chains switches from $(14)(23)$ in the upper part (= left rapidity
interval) to $(12)(34)$ in the lower part (= right rapidity interval).
One can show that the ratio $1:2:-4$ holds for each rapidity interval.
In ~\cite{Bartels:2005wa} this has been generalized to an arbitrary
number of exchanged gluon lines.
   
Another remark applies to the applicability of the cutting rules to
rescattering corrections in the single jet inclusive cross section
(Fig.~\ref{fig:jochen9}). Below the jet vertex we, again, have an
exchange of four gluon lines, similar to the diagram in the middle of
Fig.~\ref{fig:jochen7}. As to the cutting rules, however, there is an
important difference between the two situations. In
Fig.~\ref{fig:jochen7}, the blob above the four gluons is totally
inclusive, i.e. it contains an unrestricted sum over $s$-channel
intermediate states, whereas in Fig.~\ref{fig:jochen9} the part above
the four gluon state is semi-inclusive, i.e. it contains the tagged
jet. This `semi-inclusive' nature destroys the symmetry above the four
gluon states, and the cutting rules have to be modified
~\cite{Kovchegov:2001sc,Braun:2005ma}.  In particular,
eqs.(\ref{eq:jochen3}) - (\ref{eq:jochen4}) are not applicable to the
rescattering corrections between the jet and projectile.  A further
investigation of these questions is in progress
~\cite{Bartels:2006prep}.
     
\begin{figure}[htbp]
  \begin{center}
    \epsfig{file=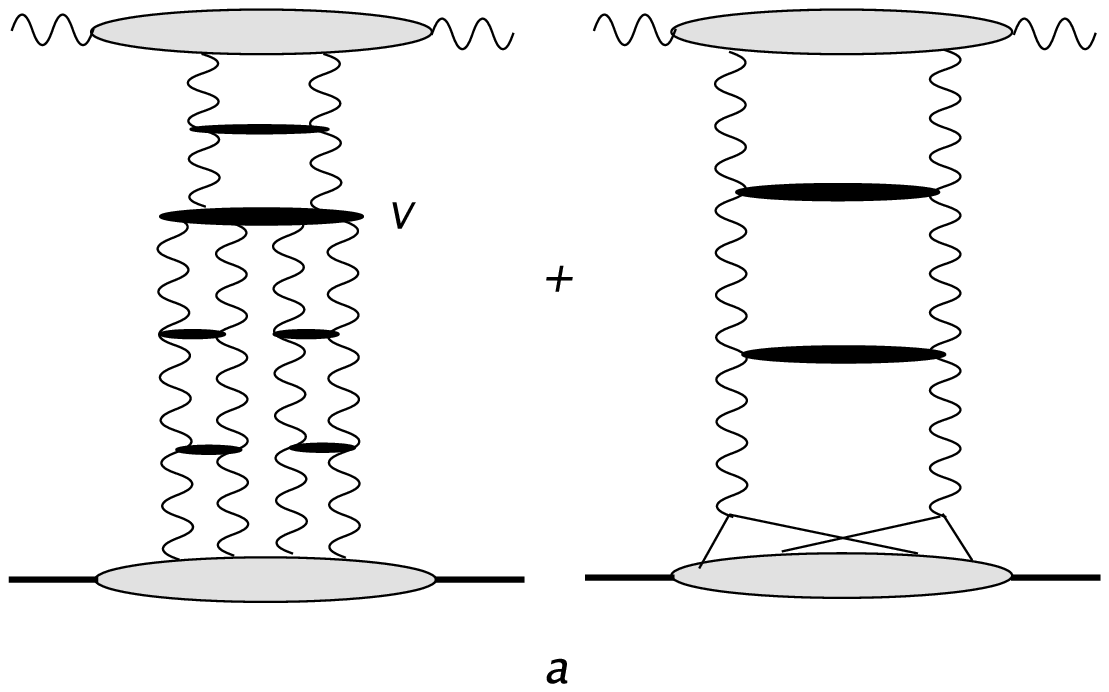,width=5cm,height=3cm}\hspace{1cm}
    \epsfig{file=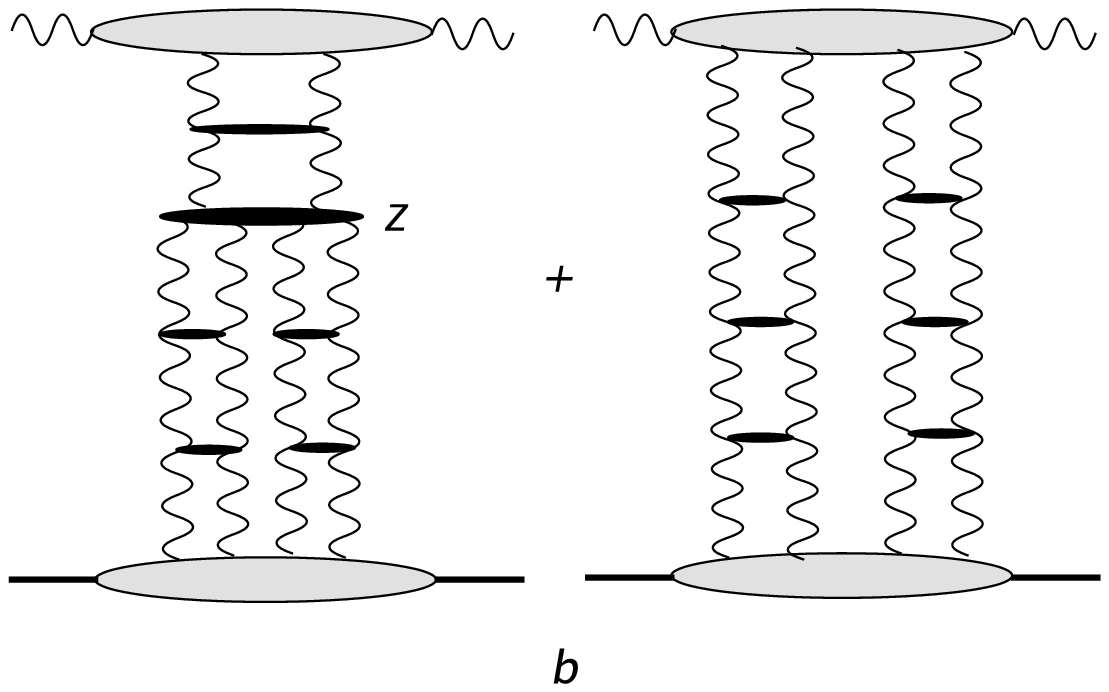,width=5cm,height=3cm}
  \end{center}
  \caption{Four-gluon contributions to $\gamma^*p$
    proton scattering: two equivalent ways of summing over all
    contributions.  (a) the decomposition of Fig.~\ref{fig:jochen5}
    with the pQCD triple Pomeron vertex.  (b) an alternative way of
    summation which explicitly shows the coupling of two Pomerons to
    the photon vertex and which leads to a new vertex $Z$.}
  \label{fig:jochen11}
\end{figure}

Finally a few comments on reggeization and cut reggeons. Clearly there
are more complicated configurations than those which we have discussed
so far; an example appears in $\gamma^* p$ scattering (deep inelastic
electron proton scattering). In contrast to $pp$ scattering, the
coupling of multi-gluon chains to the virtual photon can be computed
in pQCD, and the LO results, for the case of $n=4$ gluons, are
illustrated in Fig.~\ref{fig:jochen11}.
It turns out that we have two alternative possibilities: in the
completely inclusive case (total cross section), it is convenient to
chose Fig.~\ref{fig:jochen11}a, i.e. the sum of all contributions can
be decomposed into two sets of diagrams. In the first set, at the top
of the diagram two gluons couple to the quark-antiquark pair, and the
subsequent transition to the four-gluon state goes via the pQCD triple
Pomeron vertex. This vertex, as a function of the 4 gluons below, has
the symmetry properties described above. As a result, we can apply the
cutting rules to the four gluon state, as discussed before.  However,
there is also the second term in Fig.~\ref{fig:jochen11}a, which
consists of a two gluon state only: this is the reggeizing
contribution we have mentioned before. As indicated in the figure, the
splitting of the reggized gluons at the bottom amounts to a change in
the (nonperturbative) coupling. We want to stress that, because of the
inclusive nature of this set of diagrams, the triple Pomeron vertex
$V$ in Fig.~\ref{fig:jochen11}a, similar to the BFKL kernel, contains
both real and virtual contributions. For this reason, the
decomposition in Fig.~\ref{fig:jochen11}a is applicable to inclusive
cross sections, and it is not convenient for investigating specific
final states such as, for example, diffractive final states with a
fixed number of quarks and gluons in the final state.

There exists an alternative way of summing all contributions
(Fig.~\ref{fig:jochen11}b) which is completely equivalent to
Fig.~\ref{fig:jochen11}a but allows to keep track of diffractive
$q\bar{q}$, $q\bar{q}g$,\ldots final states: this form is illustrated
in Fig.~\ref{fig:jochen11}b. One recognizes the `elastic intermediate
state' which was not visible in Fig.~\ref{fig:jochen11}a, and the new
triple Pomeron vertex $Z$ which contains only real gluon production.
This vertex $Z$, as discussed in \cite{Bartels:2004hb} is no longer
symmetric under permutations of the gluons at the lower end;
consequently, we cannot apply the AGK cutting rules to the four gluon
states below. These findings for multiple scattering effects in DIS
imply, strictly speaking, that cross sections for diffractive
$q\bar{q}$ or $q\bar{q}g$ states cannot directly be inserted into the
counting rules (\ref{famous}).
 
\begin{figure}
  \begin{center}
    \epsfig{file=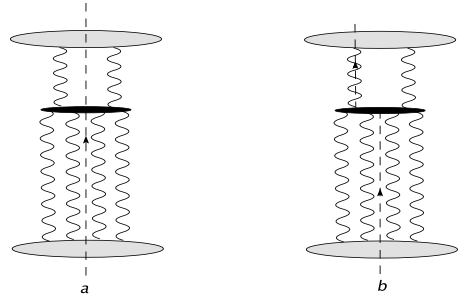,width=6cm,height=4cm}\\
  \end{center}
  \caption{A correction in which the number of lines changes. The
    black vertex denotes the $2 \to 4$ gluon vertex.}
  \label{fig:jochen12}
\end{figure}

Also $pp$ scattering will contain corrections due to multiple
interactions which are more complex. There are, for example, graphs
which contain the $2 \to 4$ gluon vertex $V$, leading to a change of
the number of gluon lines (Fig.~\ref{fig:jochen12}).
Since this $2 \to 4$ gluon vertex, as a function of the four gluons
below the vertex, satisfies the symmetry requirements listed above, we
can apply our previous analysis to the cutting lines below the vertex.
In addition, however, one can ask how the lines continue above the $2
\to 4$ gluon vertex: we show two examples, one of them containing a
cut (reggeized) gluon.  Concentrating on this two-gluon state (i.e. we
imagine that we have already summed over all possible cutting lines
below the vertex $V$), the counting rules are quite different: in
contrast to the even-signature Pomeron, the gluon is a odd-signature
reggeon. Consequently, the cut gluon is suppressed w.r.t. the uncut
gluon by one power in $\alpha_s$, and this suppression leads to the
following hierarchy of cutting lines: the cut between the gluons
belongs to leading order, the cut through one of the two reggeized
gluons is suppressed by one power in $\alpha_s$, the cut through both
reggeized gluons is double suppressed (order $\alpha_s^2$). A closer
analysis of this question is under investigation
\cite{Bartels:2006prep}.
\subsubsection{Conclusions}
Corrections due to multiple interactions seem to be important in DIS
at small $x$ and low $Q^2$; they are expected to play a significant
role also in multijet production in $pp$ scattering. The study of the
AGK rules to pQCD provides help in understanding the systematics of
multiple gluon chains. Results described in this review represent the
beginning of a systematic analysis. We have listed a few questions
which require further work.

As an immediate application, we believe that a quantitative analysis
of multiple scattering at HERA will provide a useful input to the
modeling of final states at the LHC.



\subsection{Experimental consequences}

\textit{Main author H.~Kowalski}\\

\begin{figure}
  \vspace*{-0.5cm} 
  \epsfig{file=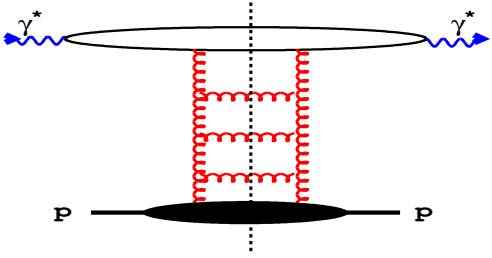,height=6cm,width=8cm}
  \vspace*{-1.5cm}
  \caption{The single gluon-ladder contribution to the 
    total $\gamma^*p$ cross section.  The blob at the lower end of the
    diagrams contains the physics below the scale $Q_0^2$ which
    separates hard from soft physics, whereas the blob at the upper end
    contains hard physics that can be described by pQCD. The dashed line
    denotes the cut.}
  \label{fig:Ael2}
\end{figure}

Experimentally it is easy to differentiate between diffractive and
{\it single} or {\it multiple} inclusive final states since
diffractive states exhibit large rapidity gaps. The {\it multiple}
inclusive final states should also be distinct from the {\it single}
inclusive ones since, at least naively, we would expect that in the
{\it multiple} case the particle multiplicity should be considerably
higher.  At low $x$, however, the relation between the number of
virtual states excited in the interaction (as measured by $F_2$) and
the final particle multiplicity cannot be straightforward since the
growth of $F_2$ with decreasing $x$ is faster than the multiplicity
increase\question{Shouldn't it be the other way around? For small $x$
  the total cross-section is saturating, but the multiplicity still
  increases due to multiple interactions}.  This may indicate that the
hadronization mechanism may be different from the string picture
commonly used in the hadronization procedure of single chain parton
showers.  The influence of multiple scattering on the particle
multiplicity of the final states should also be damped by the energy
conservation. The cut through several Pomerons leads clearly to more
gluons produced in the final state, but the available energy to
produce particles in the hadronization phase remains the same. A
detailed Monte Carlo program is therefore necessary to evaluate this
effect.

\begin{figure}
\vspace{-3.5cm}
\hspace{-0.7cm}
\epsfig{file=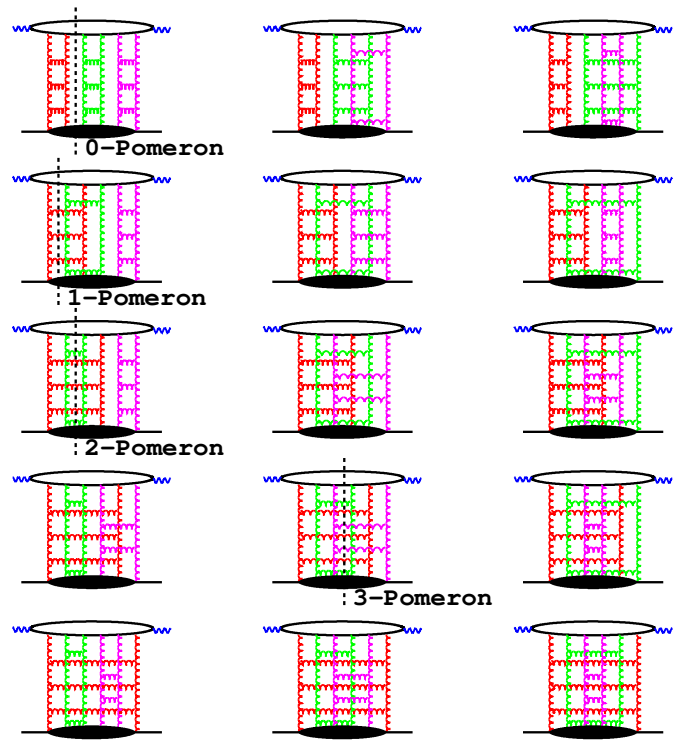,height=14cm,width=10cm}
\vspace{ 0.cm} 
\caption{3-Pomeron contributions to the 
  elastic $\gamma^*p$ amplitude. All 15 possible diagrams are shown
  with some examples of Pomeron cuts.  }
\label{fig:Ael3a}
\end{figure}
 
The number of diagrams contributing to 
the reaction amplitude increases very quickly with the number of 
Pomerons. For the 3-Pomeron amplitude 
the gluons can be paired in  15 possible ways, shown in 
Fig.~\ref{fig:Ael3a} with the examples of
0-Pomeron, 1-Pomeron, 2-Pomeron and 3-Pomeron cuts. For 
$m$-Pomerons the number of possible gluon pairs and also diagrams is:  
$$ (2m-1)(2m-3)(2m-5).... = (2m-1)!/(2^{m-1}(m-1)!).$$

Assuming that all the diagrams for a given multi-Pomeron exchange
amplitude contribute in the same way, the above analysis suggests that
the probability for different cuts to contribute should be given by
the combinatorial factors.  This is the content of the AGK rules which
were obtained from the analysis of field theoretical diagrams well
before QCD was established~\cite{Abramovsky:1973fm} and which relate
the cross-section, $\sigma_k$, for observing a final state with
$k$-cut Pomerons with the amplitudes for exchange of $m$ Pomerons,
$F^{(m)}$:
\begin{eqnarray}
\sigma_k = \sum_{m=k}^{\infty} (-1)^{m-k}\, 2^m \, \frac{m!}{k!(m-k)!} F^{(m)} .
\label{eq:agk}
\end{eqnarray}
The same result is also obtained from a detailed analysis of the
Feynman diagram contributions in QCD above with the oversimplified
assumption that only the symmetric part of the two-gluon couplings
contributes~\cite{Bartels:2005wa}.

\subsection{Multiple Interactions in the Dipole Model} 

\textit{Main author H.~Kowalski}\\

The properties of the multi-Pomeron amplitude and of the cut Pomeron
cross-sections can be quantitatively studied in a dipole model. Along
the lines which were discussed in section \ref{sec:saturation}
the $\gamma^* p$ interaction proceeds in three stages: first the
incoming virtual photon fluctuates into a quark-antiquark pair, then
the $q\bar{q}$ pair elastically scatters on the proton, and finally
the $q\bar{q}$ pair recombines to form a virtual photon.
The total cross-section for $\gamma^* p$ scattering, or equivalently
$F_2$, is obtained by averaging the dipole cross-sections with the
photon wave functions, $\psi(r,z)$, and integrating over the impact
parameter, $b$:
\begin{eqnarray}
F_2 = \frac{Q^2} {4\pi^2 \alpha_{em}}\int d^2r \int \frac{dz}{4\pi} \psi^*\psi \int d^2 b
\frac{d\sigma_{qq}}{d^2 b} .
\label{eq:siggp}
\end{eqnarray}
Here $\psi^* \psi$ denotes the probability for a virtual photon to 
fluctuate into a $q\bar{q}$ pair, summed over all flavors and helicity states.
The dipole cross-section is assumed to be 
 a function of the opacity $\Omega$:
\begin{eqnarray}
\frac{d\sigma_{qq}}{d^2b} = 2\, \left(1-\exp(-\frac{\Omega }{2})\right). 
\label{eq:xdip}
\end{eqnarray}
At small-$x$ the opacity $\Omega$ can be directly related 
to the gluon density, $xg(x,\mu^2)$, and the transverse profile 
of the proton, $T(b)$:  
\begin{eqnarray}
\Omega = \frac{\pi^2}{N_C}\, r^2\, \alpha_s(\mu^2)\, xg(x,\mu^2)\, T(b).
\label{eq:omega}
\end{eqnarray}
The parameters of the gluon density are determined from the fit to the
total inclusive DIS cross-section~\cite{Kowalski:2003hm}. 
The transverse profile
was determined from the exclusive diffractive $J/\Psi$ 
cross-sections~\cite{Kowalski:2003hm}. 
The opacity function $\Omega$ determined in
this way has predictive properties; it allows to describe other
measured reactions, e.g. charm structure function or elastic
diffractive $J/\Psi$ production.
\question{What is the difference
  between ``exclusive diffractive $J/\Psi$'' and ``elastic diffractive
  $J/\Psi$'' production?}

For a small value of $\Omega$ the dipole cross-section,
eg.~(\ref{eq:xdip}), is equal to $\Omega$ and therefore proportional to
the gluon density. This allows to identify the opacity with the single
Pomeron exchange amplitude of Fig.~\ref{fig:Ael2}.  The multi-Pomeron
amplitude is determined from the expansion:
\begin{eqnarray}
  \begin{split}
    \frac{d\sigma_{qq}}{d^2b} = &2\, \left(1-\exp(-\frac{\Omega
      }{2})\right)\\ = &2\, \sum_{m=1}^{\infty} (-1)^{m-1}\,
    \left(\frac{\Omega}{2}\right)^m \frac{1}{m!}
  \end{split}
  \label{eq:xdip1}
\end{eqnarray}
as
\begin{eqnarray}
F^{(m)} =  \left(\frac{\Omega}{2}\right)^m \frac{1}{m!},
\label{eq:fm}
\end{eqnarray}
since the dipole cross-section can be expressed as a sum of
multi-Pomeron amplitudes~\cite{Mueller:1996te} in the following way:
\begin{eqnarray}
\frac{d\sigma_{qq}}{d^2b} = 2\, \sum_{m=1}^{\infty} (-1)^{m-1}\, F^{(m)}.
\label{eq:dppom}
\end{eqnarray}
The cross-section for {\it k}~ cut Pomerons is then obtained from the
AGK rules, eq.~(\ref{eq:agk}), and from the multi-Pomeron amplitude,
eq.~(\ref{eq:fm}), as:
\begin{eqnarray}
  \begin{split}
    \frac{d\sigma_{k}}{d^2b} =& \sum_{m=k}^{\infty} (-1)^{m-k}\, 2^m \,
    \frac{m!}{k!(m-k)!}  \left(\frac{\Omega}{2}\right)^m \frac{1}{m!}\\
    = &\frac{\Omega^k} {k!} \sum_{m=k}^{\infty} (-1)^{m-k}\,
    \frac{\Omega^{m-k}} {(m-k)!}
\end{split}
\label{eq:sigcut1}
\end{eqnarray}
which leads to a simple expression:
\begin{eqnarray}
\frac{d\sigma_{k}}{d^2b} = \frac{\Omega^k} {k!} \exp(-\Omega).
\label{eq:sigcut}
\end{eqnarray}
The diffractive cross-section is given by the difference between the
total and the sum over all cut cross-sections:
\begin{eqnarray}
  \begin{split}
    \frac{d\sigma_{diff}}{d^2b} =& \frac{d\sigma_{tot}}{d^2b} -
    \sum_{k=1}^{\infty} \frac{d\sigma_{k}}{d^2b}\\
     =& 2\left(1-\exp\left(-\frac{\Omega}{2}\right)\right)-(1-\exp(-\Omega))\\
     =& \left(1-\exp\left(-\frac{\Omega}{2}\right)\right)^2
\end{split}
\label{eq:sigdiff}
\end{eqnarray}

\begin{figure}
\vspace{-1.0cm}
\hspace{-0.0cm}
\epsfig{file=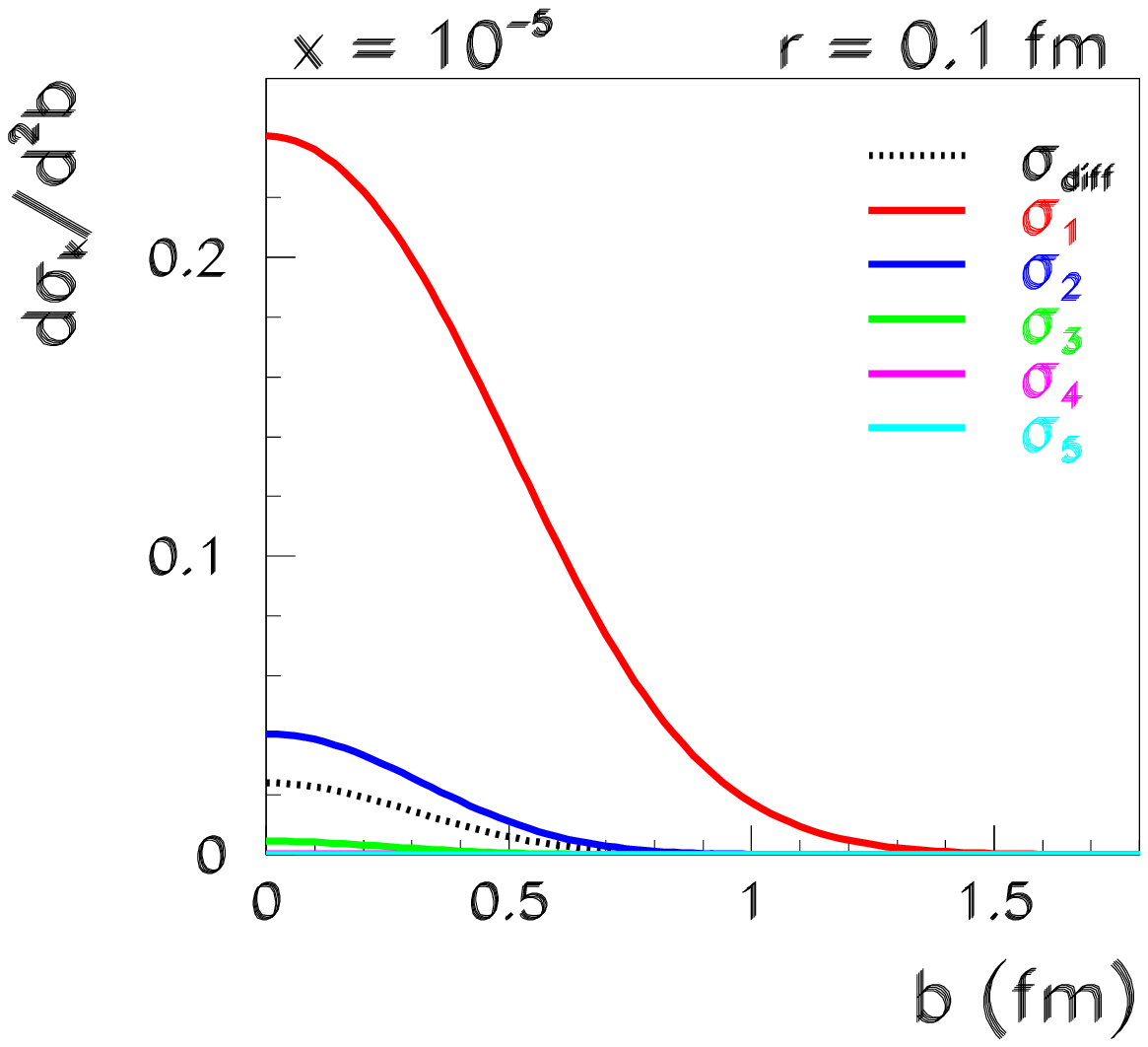,height=8cm,width=8cm}\\[-1cm]
\epsfig{file=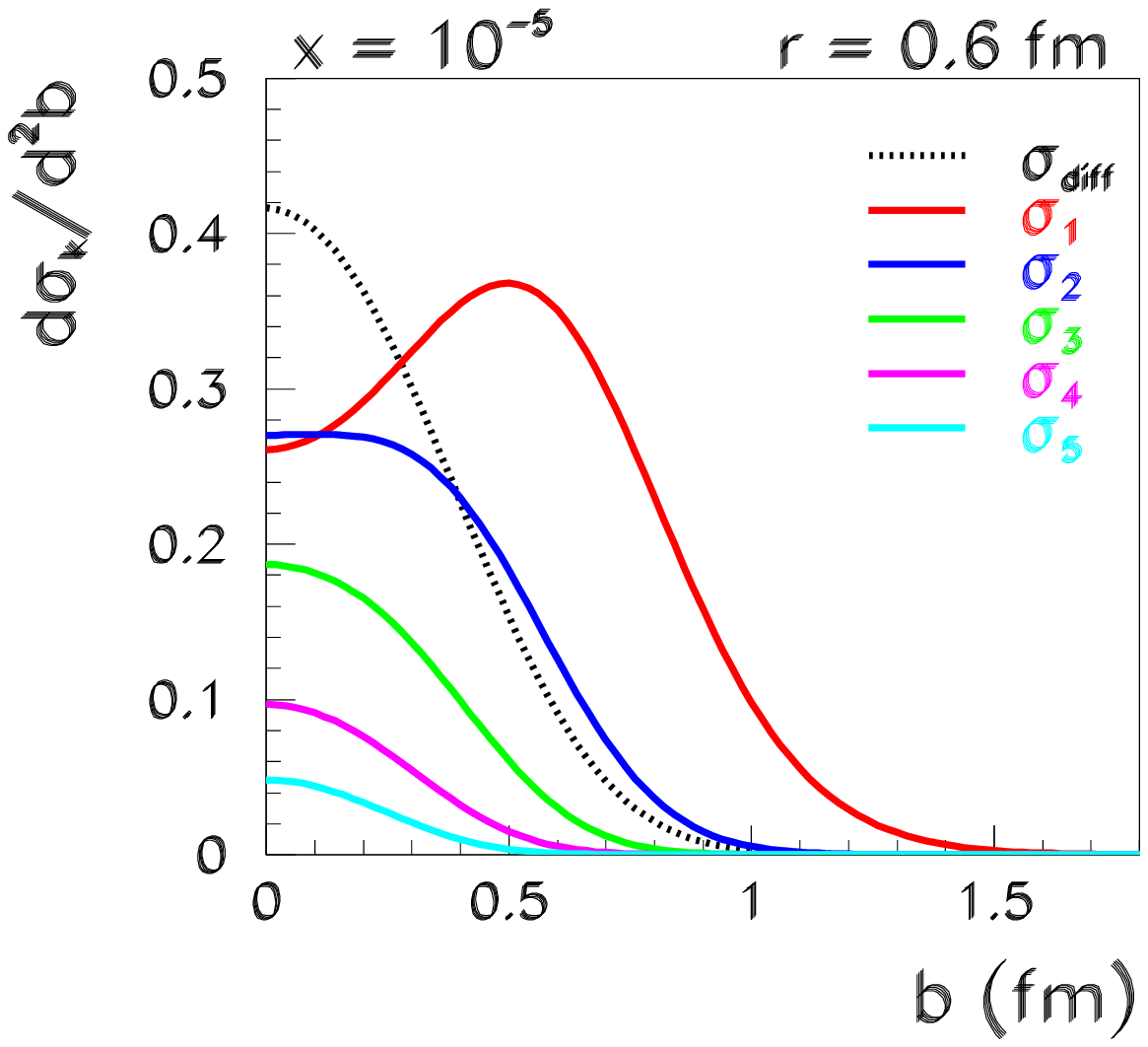,height=8cm,width=8cm}
\caption{Examples of $b$ dependence of various cut dipole 
  and diffractive cross-sections.  }
\label{fig:sighk}
\end{figure}
The cut cross-sections determined in the dipole model analysis of HERA
data have several interesting properties shown in
Fig.~\ref{fig:sighk}: for small dipoles ($r = 0.1$ fm) the opacity
$\Omega$ is also small, so the single cut cross-section, $\sigma_1$,
dominates.  This leads to particle production emerging only from the
one-cut pomeron, which should correspond, in the context of e.g. the
LUND model, to a fragmentation of only one string.  For larger dipoles
($r=0.6$ fm) the dipole cross-section starts to be damped in the
middle of the proton (at $b\approx 0$) by saturation effects.
Therefore, the single cut cross-section is suppressed in the middle
while the multiple cut cross-sections, $\sigma_2, \; \sigma_3$, etc,
become substantial and increasingly concentrated in the proton center.
These, fairly straightforward properties of dipoles indicate that in
the central scattering events the multiple scattering probability will
be enhanced, which may lead at the LHC to substantial effects in a
surrounding event multiplicity.

\begin{figure}
\vspace{-1.0cm}
\epsfig{file=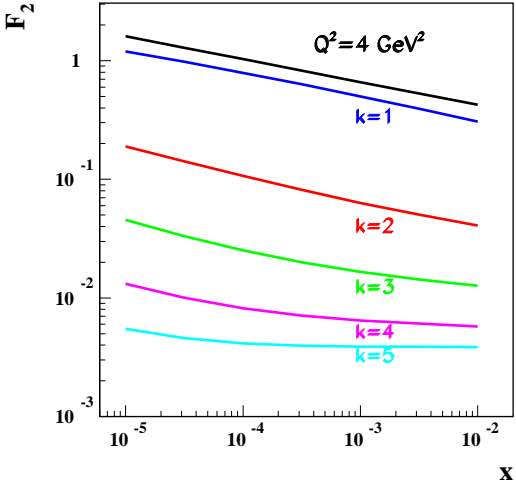,height=8cm,width=8cm}\\[-1cm]
\epsfig{file=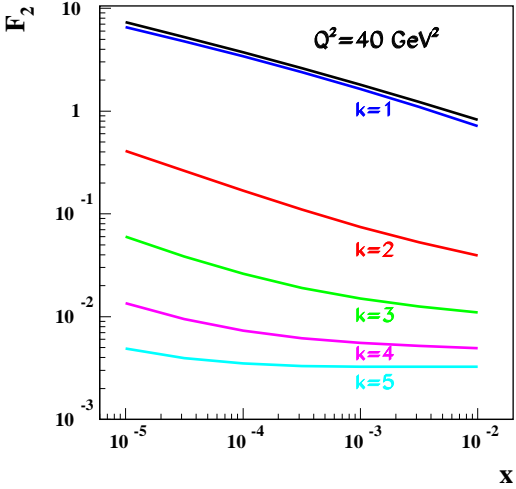,height=8cm,width=8cm}
\caption{$F_2$ and the contributions of k-cut Pomeron processes, $F_2^k$. }
\label{fig:f2k}
\end{figure}

The contribution to $F_2$ from the $k$-cut Pomeron exchanges are
computed in the analogous way to $F_2$:
\begin{eqnarray}
F_2^{k} = \frac{Q^2} {4\pi^2 \alpha_{em}}\int d^2r \int \frac{dz}{4\pi} \psi^*\psi \int d^2 b
\frac{d\sigma_k}{d^2 b}.
\label{eq:f2cut}
\end{eqnarray}
These contributions are shown, together with $F_2$, as a function of
$x$ for two representative $Q^2$ values in Fig.~\ref{fig:f2k}. One
finds that multiple interaction contributions, i.e. $k\ge 2$, in the
perturbative region, at $Q^2=4$ GeV$^2$, are substantial. In the
typical HERA range of $x \approx 10^{-3} - 10^{-4}$, the $k=2$
contribution is around 10\% of $F_2$ and the contributions of higher
cuts are also non-negligible.  For example, the contribution of the
5-cut Pomeron exchanges is still around 0.5\%, which means that at
HERA, many thousand events may come from this type of process.
\begin{figure}
\vspace{-1.0cm}
\epsfig{file=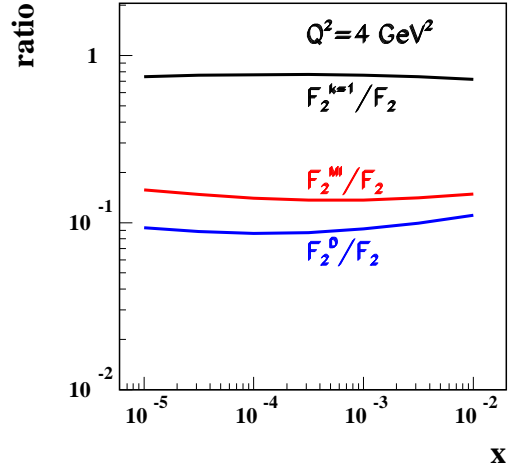,width=8cm}\\[-1cm]
\epsfig{file=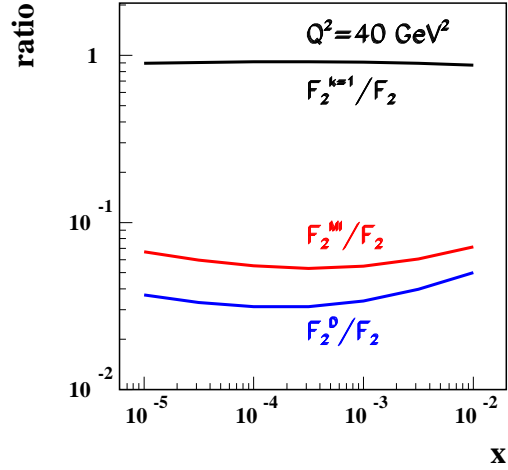,width=8cm}
\caption{Fractions of single (k=1), multiple interaction (MI) and
  diffraction (D) in DIS.}
\label{fig:rf2k}
\end{figure}
Figure~\ref{fig:rf2k} shows the fraction of the multiple interaction
processes, $F_2^{MI}=F_2^{k=2}+F_2^{k=3}+F_2^{k=4}+F_2^{k=5}$ in
$F_2$, at the same $Q^2$ values. At $Q^2=4$ GeV$^2$ the fraction of
multiple scattering events is around 14\% and at $Q^2=40$ GeV$^2$
around 6\%, in the HERA $x$ region, which indicates that the decrease
of multiple scattering with increasing $Q^2$ is only logarithmic. The
fraction of diffractive processes, shown for comparison, is of the
same order, and drops also logarithmically with $Q^2$. The logarithmic
drop of the diffractive contribution expected in the dipole model is
confirmed by the data~\cite{Chekanov:2005vv}.

\begin{figure}[htp]
\vspace{-1.0cm}
\epsfig{file=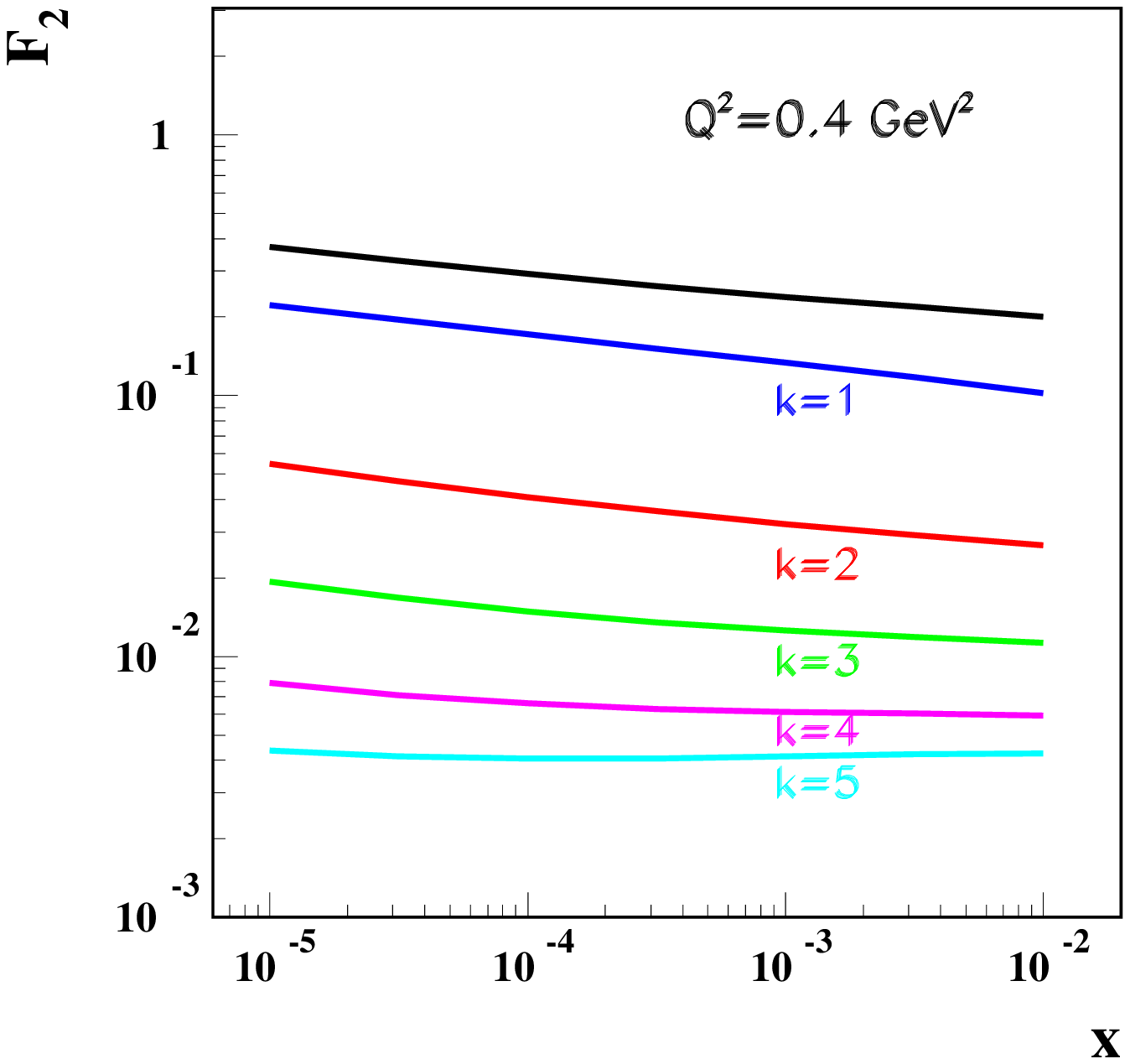,width=8cm}\\[-1cm]
\epsfig{file=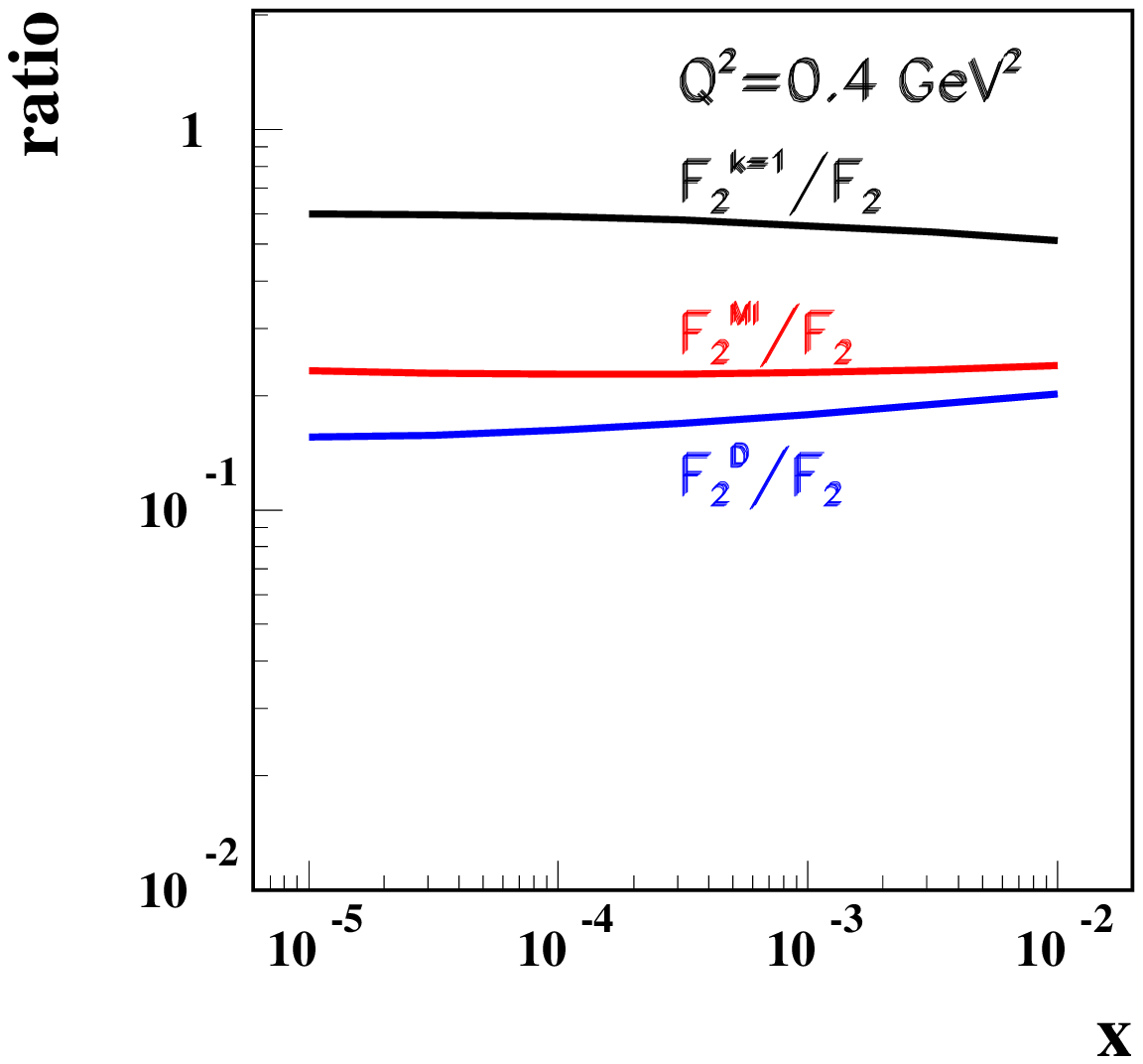,width=8cm}
\caption{Left: $F_2$ and the contributions of $k$-cut Pomeron processes.
  Right: Fractions of single (k=1), multiple interaction (MI) and
  diffraction (D) in DIS at $Q^2 = 0.4$ GeV$^2$.}
\label{fig:f2kl}
\end{figure}

The dipole model provides a straightforward extrapolation to the
region of low $Q^2$, which is partly perturbative and partly
non-perturbative.  Figure~\ref{fig:f2kl} shows the contribution to
$F_2$ of $k$-cut Pomeron processes and the fractions of multiple
interactions and diffractive processes at $Q^2 =0.4$ GeV$^2$.

Note also that, as a byproduct of this investigation, the ratio of
diffractive and inclusive cross-sections, $F^D_2/F_2$ is found to be
almost independent of $x$, in agreement with the data and also other
dipole model
predictions~\cite{Chekanov:2005vv,Golec-Biernat:1999qd,Bartels:2002cj}.
The absolute amount of diffractive effects is underestimated, since
the evaluation of diffraction through AGK rules is oversimplified. It
is well known~\cite{Bartels:2002cj}, that a proper evaluation of
diffraction should also take into account the $q\bar{q} g$
contribution which is missing in the simple AGK schema.

Hence, we find that the impact parameter dependent dipole saturation
model~\cite{Kowalski:2003hm} reproduces well the main properties of
the data and leads to the prediction that multiple interaction effects
at HERA should be of the order of diffractive effects, which are known
to be substantial. The multiple interaction effects should decrease
slowly (logarithmically) with increasing $Q^2$, similarly to the
diffractive contribution.



\section{Experimental comparisons}
\label{sec:exp}

With the luminosity collected at HERA during the past years very
precise measurements of the proton structure function, $F_2(x,Q^2)$,
have been performed over a large range in the fractional proton
energy, $x$, and in the photon virtuality, $Q^2$.  The measurements
are now limited by systematic errors rather than statistical.  Parton
density functions have been obtained mainly by fitting the DGLAP
equations, evolved from an input scale $Q_0^2$, to the
structure functions, measured at some scale $Q^2$.  Especially the
precision data at low $Q^2$ have provided an important input to
various QCD fit analyses.  It was recognized early that inclusive
measurements, like that of structure functions, are not very sensitive
to the new parton dynamics expected to appear in the low $x$ region.
Instead evidence from such dynamics has to be found from
investigations of hadronic final states in a phase space region where
the DGLAP governed evolution is suppressed.  Thus, a global fit, which
also includes data from more exclusive processes, would further
constrain the PDFs.  A problem is that measurements of the hadronic
final states suffer from much larger uncertainties than the inclusive
structure function measurements and therefore measurements of many
different complementary processes are desirable.

Forward jet production in DIS is expected to be sensitive to new
dynamics and early results indeed showed a deviation from the
predictions of the LO DGLAP model as well as of NLO calculations.
However, with the inclusion of resolved photon contributions, DGLAP
provided the same level of agreement as the colour diple model (CDM),
in which the parton emission follows the same scheme as in the new
dynamics proposed. Only recent studies of final states with a
'forward jet and two additional jets' give the first evidence for
parton dynamics in which there is additional breaking of the
$k_t$-ordering compared to that predicted by the resolved photon
model.

Dijet data may be used to gain better insight into the dynamics of the
parton evolution and for extracting updf's.  In the low $x$ region
boson-gluon fusion processes are dominating and in the LO DGLAP
description the gluon and the photon collide head on in the hadronic
center-of-mass system and thus will be produced back-to-back. 
 Deviations from this may arise from additional
radiation and if the parton propagator, entering the hard scattering
process, has significant transverse momentum, such that the two
partons produced in the hard intreraction are no longer balanced in
transverse momentum.
 Thus, the two jets produced will not be back-to-back in
azimuth. A measurement of the azimuthal correlation between the two
jets should be directly sensitive to the predictions of models based
on different evolution schemes.

The flavour composition of the final state can also provide important
information about the evolution and production mechanisms of partons.
This has motivated a measurement of final states with identified
strange particles.

Although $F_2$ data can be well described by the exchange of a
single gluon ladder, it is unlikely that a single chain generates
large rapidity gaps, which is the signature of diffractive processes.
The traditional picture of diffractive processes is scattering by the
virtual photon against a pomeron with a partonic structure.
 Over the past years significant progress in the understanding
of diffraction has been made at HERA, which has led to a modification
of this description. Data are much better described assuming
multi-gluon exchange, where a pair of gluons is the minimum to create
a colour singlet state. The multi-pomeron exchange model provides a
natural connection between inclusive scattering, diffractive
scattering and multiple scattering given by different cuts through the
ladder diagrams according to the so called AGK cutting rules, as
discussed in section \ref{sec:agk}. Rapidity gaps between
high-transverse energy jets have been observed at the Tevatron, at a
fraction that is in good agreement with BFKL predictions. Also
multiple scattering has been studied at the Tevatron, and found to
give significant contributions to the final state.  In $ep$-collisions
at HERA multiple interactions can occur in processes where the
exchanged photon interacts via its parton content.  Through the
possibility to control the fraction of the photon momentum,
$x_{\gamma}$, entering into the scattering process, more systematic
investigations of underlying events may be performed at HERA over a
wide energy range.

In general, measurements of final states provide information about the hard
scattering process,
parton evolution, initial and final state radiation and multiple interactions.
Thus, it is important to
measure, as accurately as possible, the final states in order to test the
theoretical models.

In the following the studies of multiple interactions, gaps between
jets, forward jets and strange particle production will be discussed
in more detail.


\subsection{Multiple interactions
  at the Tevatron and HERA}
\label{sec:expmi}

\textit{Main author J.~Turnau}\\

\noindent
Since hadrons are composite objects of quarks and gluons there is a
certain probability that collisions between hadrons involve more than
one parton interaction i.e. we have multiple interactions (MI). As a
consequence of the strong rise of the parton distribution at low $x$
the probability to have MI increases with the collision energy and the
effect at the Tevatron has turned out to be significant. At the LHC
the contribution from MI will be even larger.  In electron-proton
collisions at HERA MI may occur in processes where the exchanged
photon is resolved and interacts via its parton content. The final
state of collisions with MI will thus contain the products of the
primary hard collisions, those of additional soft or semihard parton
interactions, contributions from initial and final state radiation and
from the beam remnants.  All products not coming from the primary
interaction contribute to the so called underlying event (UE).
\par
Effects of MI will influence the total cross section, the
inclusive jet cross section, the jet multiplicity, the jet profile,
the jet pedestal (the level of transverse energy outside the jets),
the transverse energy flow and transverse energy correlations, the
hadron multiplicity, the multiplicity correlations and may cause large
multiplicity fluctuations.  Experimental data from HERA and the
Tevatron have been compared to various theoretical models containing a
description of MI.
 
\subsubsection{Monte Carlo models for description of multiple interactions}
\noindent
So far multiple interactions are theoretically not well understood.
The theoretical description is mainly based on QCD inspired models,
which assume a hard scattering process superimposed on soft or
semi-hard interactions. Various models differ in how initial and final
state radiation is taken into account as well as how the hadronization
process and the beam remnants are treated.
\par 
HERWIG \cite{Marchesini:1991ch,Corcella:2000bw} assumes that the UE is
a soft collision between the two beam ``clusters''. The parameters of
this model are tuned to describe experimental results on soft
hadron-hadron collisions.  Also the strength and frequency parameters
of the secondary interactions are subject to tuning. There is a
possibility to include multiparton interactions by employing an
interface to the JIMMY generator
\cite{Butterworth:1996zw,HERALHC:Jimmy}.  To some extent the formalism
that is used to describe MI in JIMMY is the same as in PYTHIA (see
below).
\par
PYTHIA \cite{Sjostrand:2000wi} assumes that each interacting beam
hadron (or resolved photon) leaves behind a beam remnant, which does
not radiate. In contrast to the original HERWIG and ISAJET generators
PYTHIA uses multiple parton interactions to enhance the activity of
the UE. In the simplest version of the PYTHIA multiple interaction
model, the transverse momentum cut-off of the hard interactions is
lowered to $p_t^{mia} < p_t^{min}$.  The mean number of (semi-) hard
interactions is given by $<n>=\sigma_{parton}(p_t^{mia})/\sigma_{nd}$,
where $\sigma_{nd}$ is the non-diffractive part of the total cross
section. The distribution of the number of interactions is not
uniquely determined. In the simplest approach the fluctuations are
calculated from a Poisson distribution.  In the more sophisticated
version the number of interactions are given by a Poisson distribution
for each given impact parameter, where the impact parameter dependence
is given by a double-Gaussian overlap function.  The number of
additional interactions is typically of order 1 -2 . The parton
process with the highest transverse momentum in the partonic final
state can be calculated by the quark/gluon $2 \to 2$ matrix element.
Additional parton interactions in the event are calculated from
perturbative gluon-gluon scattering processes.
\par 
Simulations of photon-hadron processes have frequently been performed
using the PHOJET generator\cite{Engel:1994vs}. PHOJET was designed to
simulate, in a consistent way, all components which contribute to the
total photoproduction cross section.  In contrast to PYTHIA, PHOJET
incorporates both multiple soft- and (semi-)hard parton interactions
on the basis of a dual unitarization scheme \cite{Capella:1986cm}.
\par
In their initial investigations of UE \cite{Affolder:2001xt} CDF used
the ISAJET Monte Carlo \cite{Paige:2003mg}, which does not include
multiple scattering a la PYTHIA or HERWIG. Instead the beam jets are
added assuming that they are identical to a minimum bias event at the
energy remaining after the hard scattering.  However, ISAJET did not
describe the UE data and has not been used in subsequent analyses.
\par
Generally speaking, the Monte Carlo models which include multiple
scattering have enough free parameters to describe the most important
features of data from HERA, the Tevatron and of other data found in
the JetWeb database \cite{Butterworth:2002ts}.  A program to tune the
model parameters is under way.

\subsubsection{Underlying events at the Tevatron}
\noindent
In the standard analysis of hard scattering events one measures jet
cross sections and jet properties, which in general are very well
described by QCD Monte Carlo models and NLO QCD calculations, provided
that jet pedestals are properly parameterized. The uncertainty in the
UE contribution to jet events is actually dominating the systematic
errors for inclusive jet measurements. In order to understand the
physics of UE, special studies which go far beyond a simple
parameterization of the energy flow outside the jets, are required.
\par
The CDF collaboration at the Tevatron has performed
\cite{Affolder:2001xt,Acosta:2004wq} detailed studies of the structure
and properties of the underlying event in two complementary analyses
of Run I data at $\sqrt(s) = 1800$ and $\sqrt(s)=630$ ~\gev .  The
overall event structure was investigated using global variables such
as charged particle multiplicities and the scalar sum of the
transverse momenta of charged particles as a function of the leading
jet momentum. The sensitivity to UE is expected to be the highest in
phase space regions perpendicular to the direction of the leading jet.
In the first analysis \cite{Affolder:2001xt} jets 
were defined
by applying the simple cone algorithm to charged particles only. Since
the lower limit of the jet transverse momenta (scalar $p_T$ sum) was
chosen as low as 0.5 \gev, UE could be studied in the transition
region from minimum bias events to events with high transverse
momentum jets. In a later analysis \cite{Acosta:2004wq} jets were
defined using the cone algorithm on calorimetric objects with $E_T >
15-20 \gev$.
\begin{figure}[htb]
\center
\epsfig{file=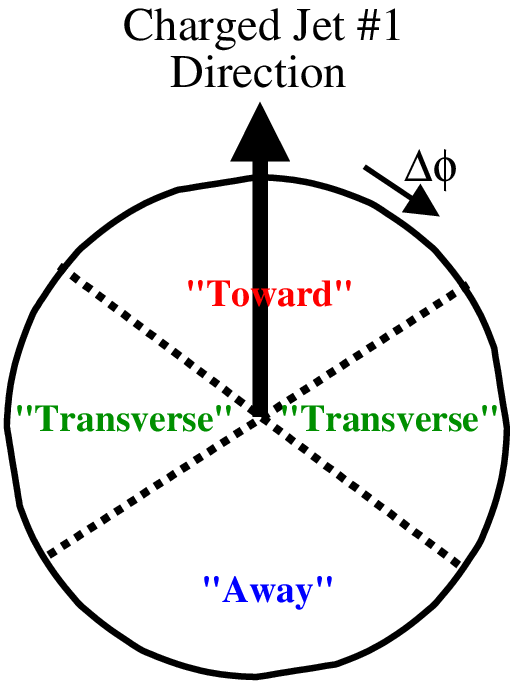,width=0.20\textwidth}%
\epsfig{file=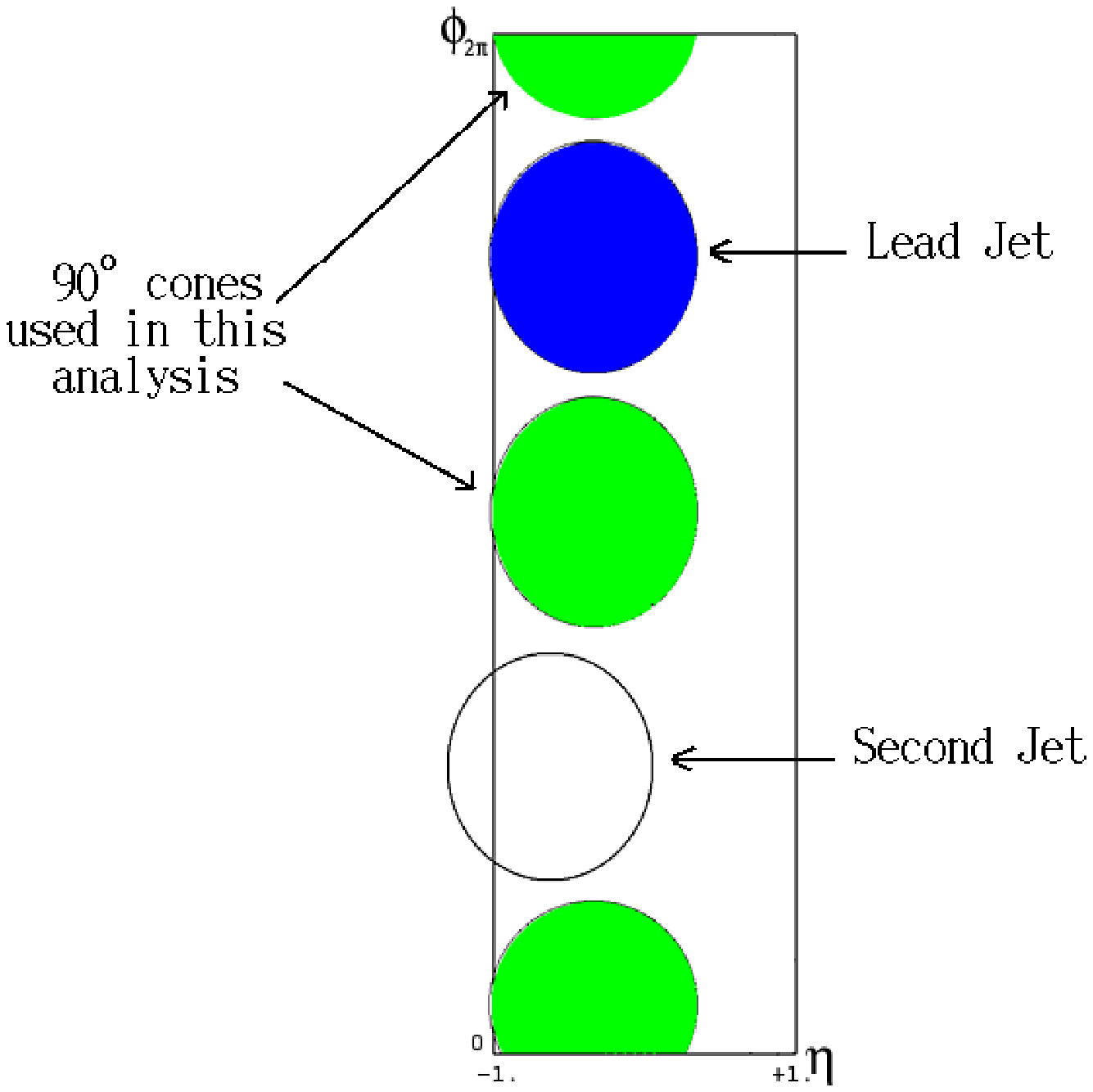,width=0.245\textwidth}
\caption[dummy]{{\bf LEFT}: Illustration of correlations in the azimuthal
  angle $\Delta\phi$ relative to the direction of leading charged jet
  in the event. The regions $\mid{\Delta\phi}\mid < 60 $,
  $\mid\Delta\phi\mid > 120$ and $60 <\mid{\Delta\phi}\mid < 120 $ are
  referred to as ``towards'', ``away'' and ``transverse''.  Each region
  covers the same range $\mid\Delta\eta\mid \times \mid\Delta\phi\mid
  =2 \times 120^{\circ}$.  On an event by event basis the regions
  ``transverse minimum/maximum'' are defined
  to be the ones containing the minimum/maximum transverse momentum. \\
  {\bf RIGHT}: The phase space regions, as defined in the analysis
  \cite{Affolder:2001xt}, shown in the $\eta-\phi$ plane, where the
  ``transverse'' regions are given by cones at $\pm 90^{\circ}$ to the
  leading jet direction.}
\label{fig:cheese} 
\end{figure}
As shown in Fig.~\ref{fig:cheese} (left) the direction of the leading
jet in each event is used to define different regions in $\eta - \phi$
space : ``toward'', ``away'' and ``transverse''.  The ``transverse''
region is particularly sensitive to the UE. In ref.~
\cite{Acosta:2004wq} the ``transverse'' region was defined as the area
in the $\eta - \phi$ plane covered by the two cones with radii
$R=\sqrt{(\Delta\eta)^2+(\Delta\phi)^2}=0.7 $ perpendicular to the
highest energy jet (Fig.~\ref{fig:cheese} right). On an event-by-event
basis the regions of ``minimal'' and ``maximal'' transverse momentum
were defined as the regions containing the smallest and largest scalar
$p_T$ sum of charged particles, respectively.  Such an investigation
of the UE helps separating the initial and final state radiation
component from the ``beam remnant'' components.  It can be argued that
transverse energy in the ``minimal transverse'' region
($P_T^{90,min}$) is due to multiple scattering while the difference in
transverse momentum between the ``minimal-'' and ``maximal
transverse'' regions $\Delta P_T^{90} = P_T^{90,max}$-$P_T^{90,min}$
is a measure of the hard initial/final state radiation connected to
the primary interaction.  The CDF analyses have established several
basic properties of UE, illustrated in Figs \ref{fig:chj} (from
\cite{Affolder:2001xt}) and \ref{fig:maxmin} (from
\cite{Acosta:2004wq}) and listed below.
\begin{itemize}
\item In the ``transverse'' regions most sensitive to UE, the average
  number of charged particles and the average charged scalar $p_T$ sum
  grow very rapidly with the momentum of the leading jet. At $p_T(jet)
  > 5~\gev$ an approximately constant plateau is observed (Fig.~
  \ref{fig:chj}). The height of this plateau is at least twice that
  observed in ordinary soft collisions at the corresponding energy.
  Although models including multiple scatterings (soft or semi-hard)
  predict a growth of both the average number of charged particles and
  the average charged scalar $p_T$ sum at low momenta of the leading
  jet, they are not able to describe the data in this region
  ($p_T(jet) < 5~\gev$).
\item For the leading jet above 50\gev, $P_T^{90,min}$ is almost
  independent on the momentum of the leading jet which is correctly
  described by HERWIG and PYTHIA.
\item The difference $\Delta P_T^{90}$ increases slowly.
\item Neither PYTHIA nor HERWIG are able to reproduce the $P_T$
  distribution of tracks in minimum bias events (not shown).
\end{itemize}
\begin{figure}
\center
\epsfig{file=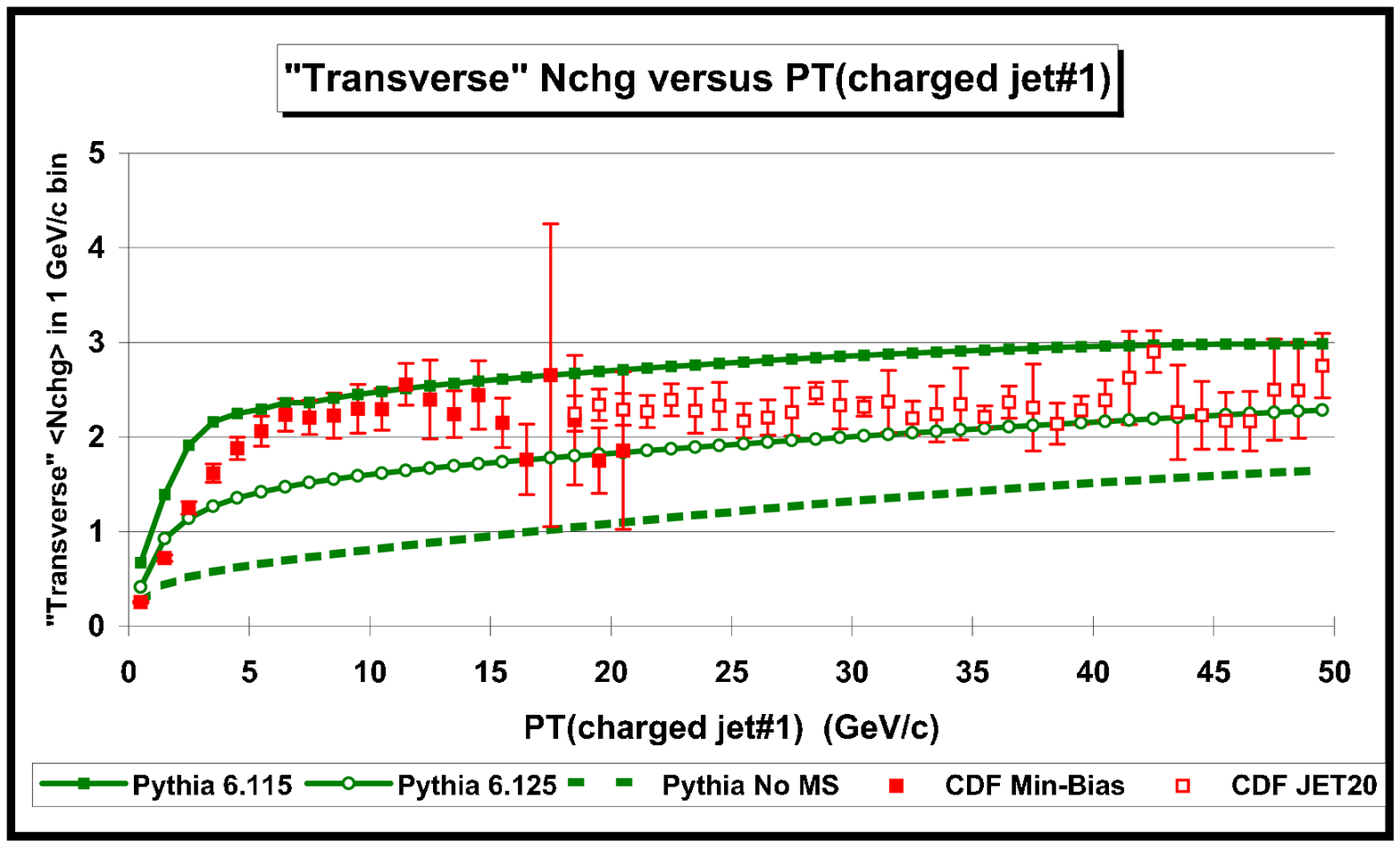,width=0.45\textwidth}\vspace*{-3cm}
\epsfig{file=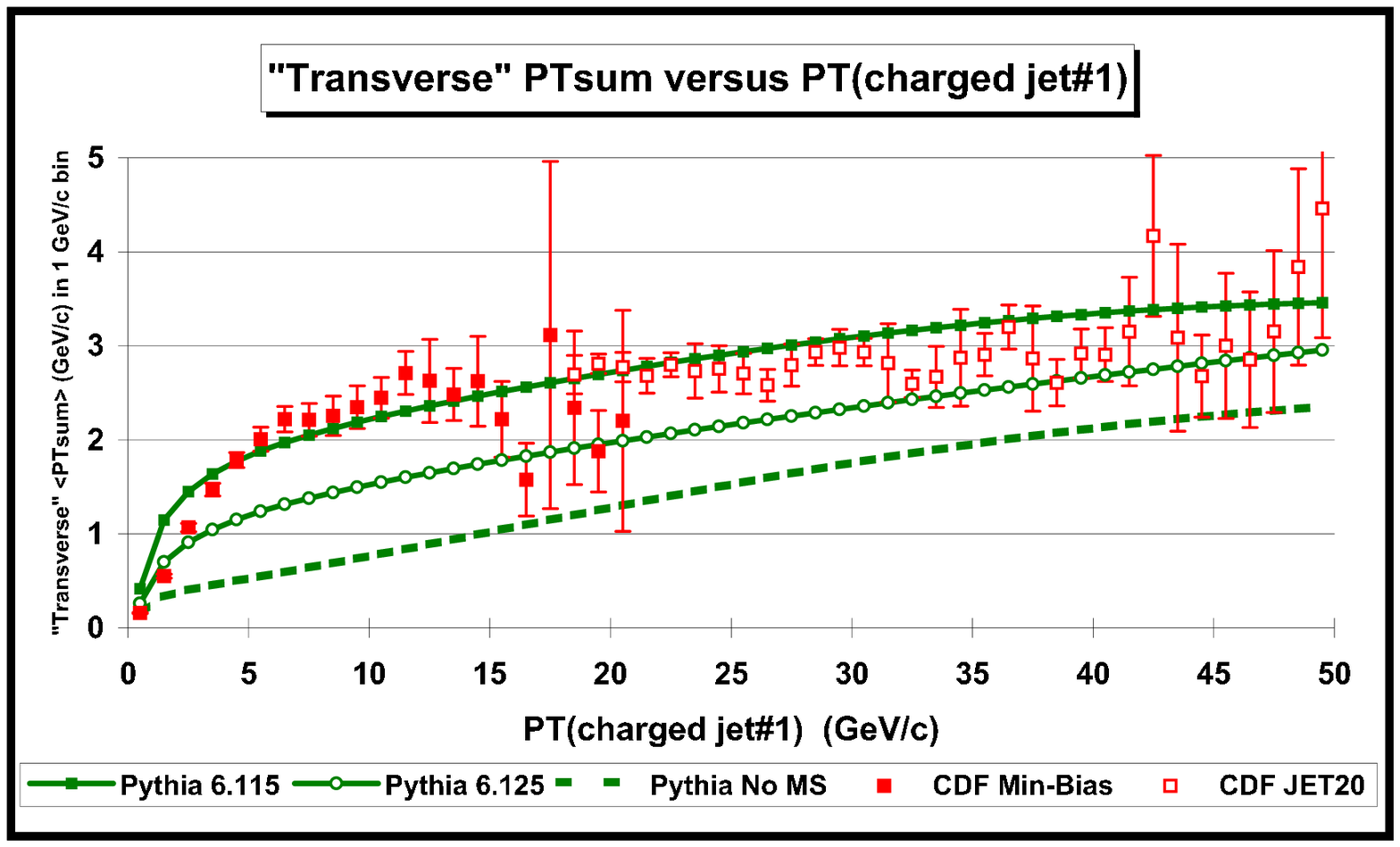,width=0.45\textwidth}\vspace*{-3cm}
\caption{Data (taken from \protect\cite{Affolder:2001xt})
  on the average number of charged particles
  ($p_T>0.5 \gev, \mid\eta\mid < 1$) (top) and the scalar $p_T$ sum of
  charged particles (bottom) in transverse region defined in
  Fig.~\ref{fig:cheese} as a function of transverse momentum of the
  leading charged jet compared with Monte Carlo Models.}
\label{fig:chj} 
\end{figure}
\begin{figure}
\center
\epsfig{file=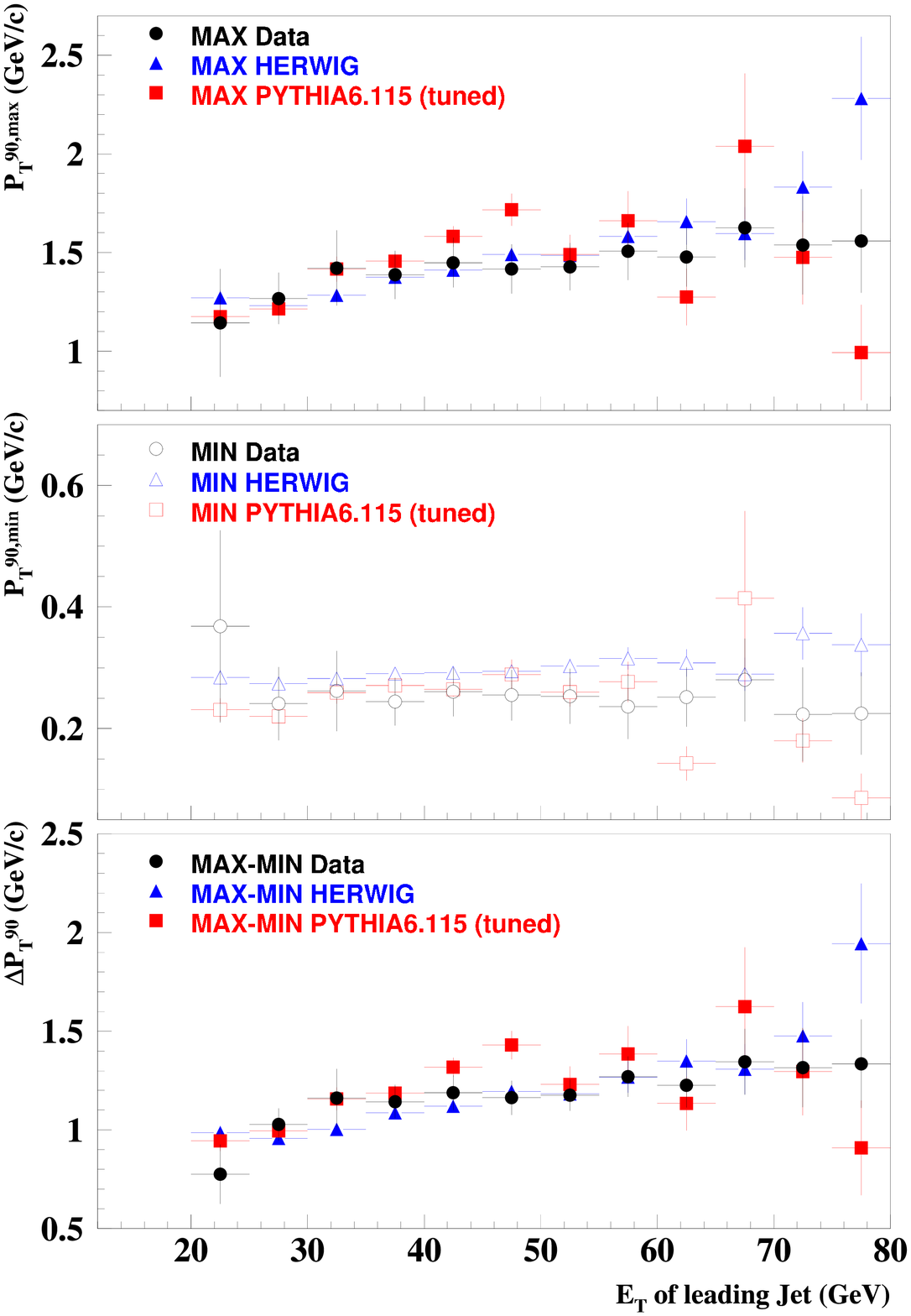, width=0.4\textwidth}
\epsfig{file=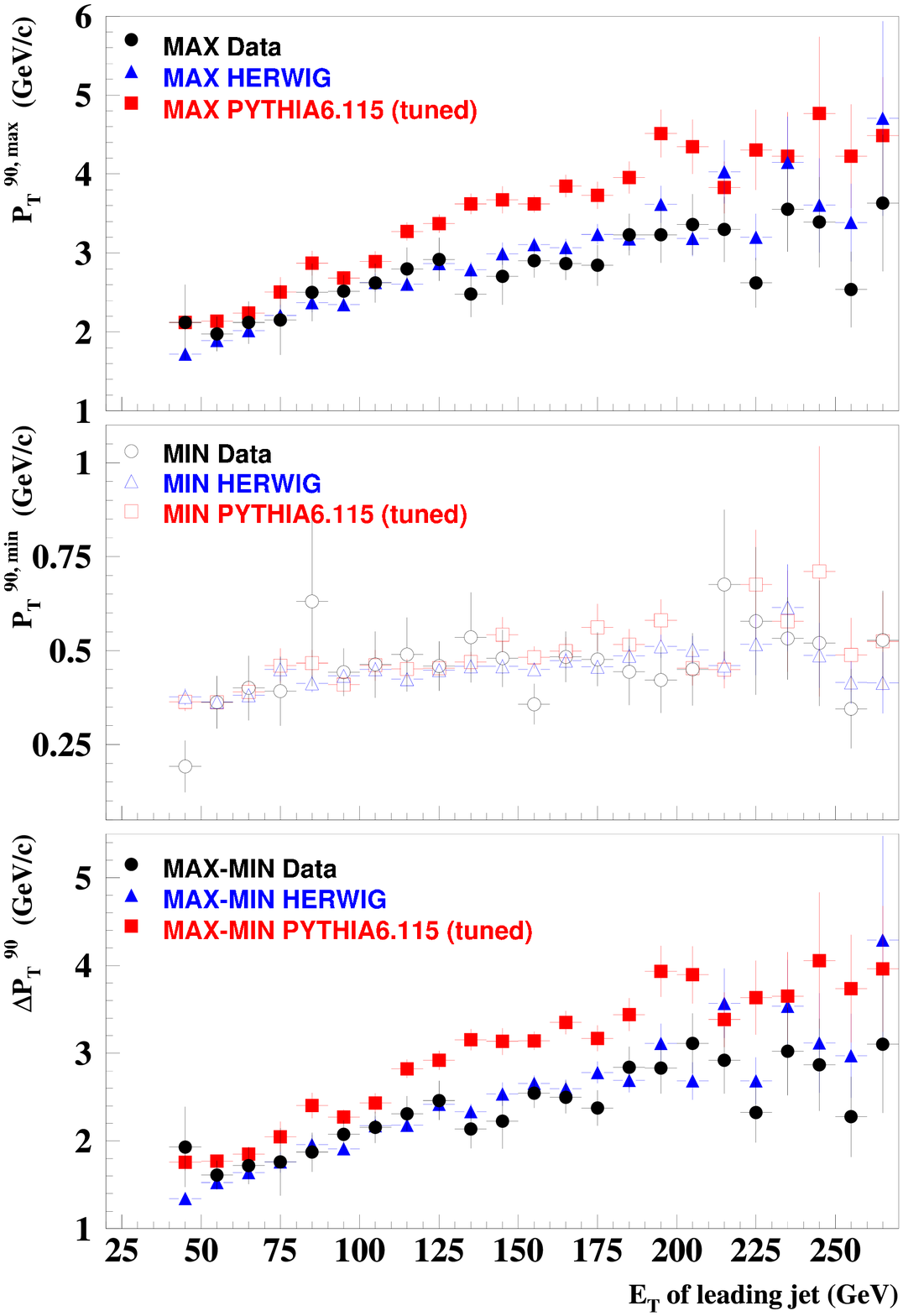, width=0.4\textwidth}
\caption{$P_T^{90,max}, P_T^{90,min} $ and $\Delta P_T^{90}$
  as a function of $E_T$ of the highest energy jet at $\sqrt{s}=1800
  ~\gev$ (bottom three plots) and $\sqrt{s}=630 ~\gev$ (top three
-  plots) taken from \protect\cite{Acosta:2004wq}. 
   PYTHIA has been tuned to describe the data.}
\label{fig:maxmin}  
\end{figure}
\par
In summary, the QCD models implemented in the PYTHIA and HERWIG Monte
Carlo programs are able to describe the most important features of the
UE from the Tevatron data. In both cases the agreement is reached only
after careful tuning of many parameters, in particular the
regularization scale of the transverse momentum.  Clearly the
experimental tests of the predictions from PYTHIA and HERWIG
concerning correlations and fluctuations in the UE will be an
important challenge over the coming years\cite{HERALHCButtar}.

\subsubsection{Underlying event energy at HERA}
\begin{figure}[htb]
\center
\epsfig{file=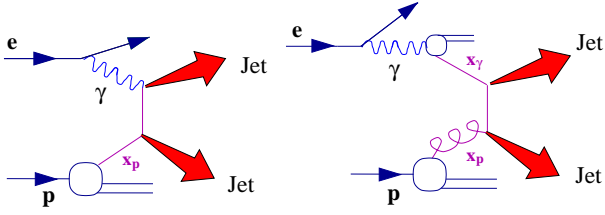, width=0.50\textwidth}
\caption{Examples of LO QCD diagrams for photoproduction of
  inclusive jets in direct (a) and resolved (b) photon interactions.}
\label{fig:dirres} 
\end{figure}

\noindent
At HERA, the interaction of electrons and protons via the exchange of
a quasi-real photon can result in the production of jets. The photon
may interact as a point-like particle in so called direct process
(Fig.~\ref{fig:dirres}a) or it may interact via its partonic structure
such that a parton carrying a fraction $x_{\gamma}$ of the photon
momentum interacts with a parton in the proton.
In resolved processes the photon remnant can interact with the proton
remnant very much like in hadron hadron collisions. The center of mass
energy in the $\gamma p$ system extends up to 300~\gev, much below the
reach of the Tevatron. Thus the effects of MI at HERA are certainly
weaker and more difficult to study. On the other hand studies of the
photon properties from measurements of UE at HERA are interesting and
complementary to the measurements at hadron-hadron colliders. The
experimental results presented in this section have been published by
the H1 collaboration \cite{Aid:1995ma}. They are based on a sample
where photoproduction events are tagged by detecting the scattered
electron and it contains 3 sub-samples : the minimum bias sample
(charged track with momentum$>0.3$ \gev + reconstructed vertex), the
high- $E_T$ sample (total transverse energy in the pseudorapidity
range $-0.8 < \eta <3.3$; $E_T>20$\gev ) and the jet sample ( at least
1 jet with $E_T>7$\gev).
\par
\begin{figure}
\center
\vspace*{-1cm}
\epsfig{file=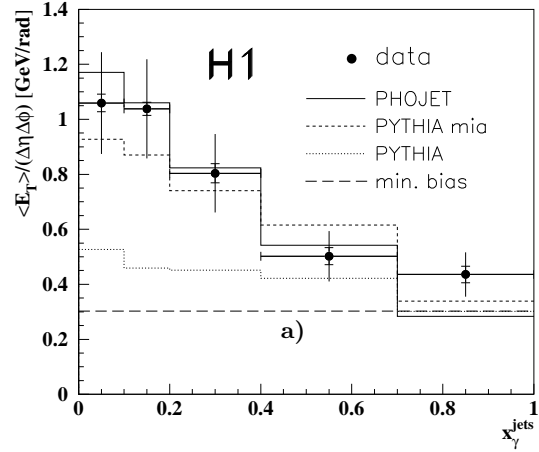, width=0.44\textwidth}\vspace*{-1cm}
\epsfig{file=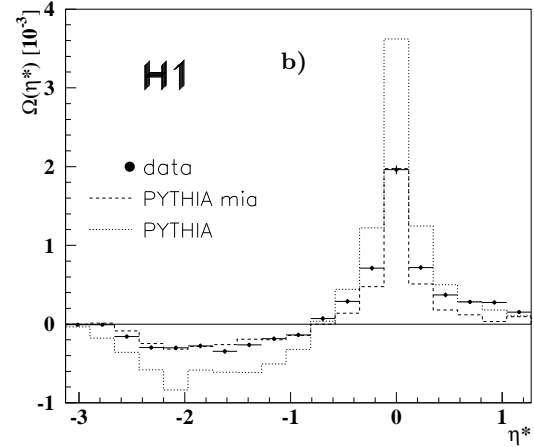, width=0.45\textwidth}
    \setlength{\unitlength}{\textwidth}
    \begin{picture}(0,0)
      \put(-0.25,0.49){\bfseries a)}
      \put(-0.25,0.3){\bfseries b)}
    \end{picture}
\caption[dummy]{{\bf a}: The transverse energy density outside
    jets in the central rapidity region, $\mid\eta_{\gamma p}\mid <1$,
    of $\gamma p$ collisions as a function of the momentum fraction
    $x_{\gamma}^{jet}$ of the parton entering the hard scattering
    process from the photon side. The data 
    (taken from \protect\cite{Aid:1995ma})
    are compared to models with
    multiple interactions (PYTHIA mia, PHOJET) and without (PYTHIA).
    The dashed horizontal line marks the energy density
    level of minimum bias events. \\
    {\bf b}: the observed rapidity correlations with respect to
    the central rapidity of $\gamma p$ collisions, $\eta^{\star}=0$.
    The dashed (dotted) histogram represents calculations of the QCD
    generator PYTHIA with (without) interactions of the beam remnants.
  }
\label{fig:et-xjet-corl} 
\end{figure}

At HERA the amount of energy which is carried by the photon remnant
can be estimated using the variable
$$x_{\gamma}^{jets}=\frac{E_T^{jet1}e^{-\eta^{jet1}}+
  E_T^{jet2}e^{-\eta^{jet2}}}{2E_{\gamma}}$$
where $x_{\gamma}^{jets}$
is the fraction of the photon energy carried by interacting parton,
$E_T^{jet1}$ and $E_T^{jet2}$ are the energies of the two jets with
the highest transverse energies, and $\eta^{jet1}$ and $\eta^{jet2}$
are their pseudorapidities. The energy of the photon, $E_{\gamma}$, is
determined from the energy measured in the electron tagger.
Fig.~\ref{fig:et-xjet-corl}(a) from ref.~\cite{Aid:1995ma} shows the
transverse energy density outside the jets of 2-jet events in the
central rapidity region.  The data decrease as
$x_{\gamma}^{jets}\rightarrow 1$ to the level measured in deep
inelastic $ep$ scattering events, dominated by direct photon
processes. The dashed line in Fig.~\ref{fig:et-xjet-corl} indicates
the energy density measured in minimum bias events (for which
$x_{\gamma}^{jets}$ is not measurable). At small $x_{\gamma}^{jets}$
the energy density increases to the level found in hadron-hadron
collisions ($\approx 0.3$ at the SPS and the Tevatron).  Both PHOJET
and PYTHIA with the MI parameters suitably tuned ($p_t^{mia}$ depending
on choice of the the photon pdf) are able to reproduce the data. This
type of the measurement apparently has no analog in hadron-hadron
collisions.
\par
Energy-energy correlations are sensitive measures of how the energy is
distributed over the available phase space and provide important
information for the modeling of UE.  The rapidity correlation $\Omega$
is defined as
\begin{equation}
  \begin{split}
    &\Omega(\eta^*)\label{eq:Omega}\\
&=\frac{1}{N_{ev}}\sum_{i=1}^{N_ev} \frac{(\langle
      E_{T,\eta^*=0}\rangle-E_{T,\eta^*=0})_i (\langle
      E_{T,\eta^*}\rangle-E_{T,\eta^*})_i} {(E_T^2)_i}
\end{split}
\end{equation}
Here $E_T$ is the total transverse energy measured in the H1
calorimeter and the other terms refer to transverse energies measured
in pseudorapidity bins of size $\Delta\eta=0.22$ in the $\gamma p$
cms. The average values were extracted from all events in the sample.
Fig.~\ref{fig:et-xjet-corl} (b) shows the rapidity correlations from
the high $E_T$ sample. The data show a short range correlation around
the reference bin $\eta^*=0$ and a long range anti-correlation which
results from the hard scattering process. PYTHIA without MI predicts
an anti-correlation which is too strong. Adding MI i.e. the addition
of uncorrelated energy to the event results in a correct description
of the data.  The same conclusion holds for an event sample where
jets are explicitly required (jet sample).
\par
In summary, the underlying event in photoproduction events can be
consistently interpreted as the superposition of a hard scattering
process and interactions between the beam spectators, as modeled by
PYTHIA and PHOJET.  Processes with resolved photons at
$x_{\gamma}\approx 0$ are found to produce 3.5 times the transverse
energy density of minimum bias events comparable with that observed in
hadron-hadron collisions at the SPS (UA1) and the Tevatron (CDF).
Studies of energy-energy correlations demonstrate that the additional
transverse energy in the event is not correlated with the hard
scattering process. Finally, the contribution of higher order
radiation to UE can be studied separately using the kinematic quantity
$x_{\gamma}$ to switch off the beam remnant interactions.

\subsubsection{Explicit Observations of Double Hard Scattering}

The general signature of multiple parton scattering is an increase in
the transverse energy flow of the event. However, in extreme cases,
the transverse energy of a secondary interaction is sufficient to
produce an additional pair of jets. The observation of such events is
highly important for several reasons.  It is sensitive to the
phenomenology of multiple parton interactions and provides direct
information on the structure of the proton in transverse space. It is
also important for estimating backgrounds to processes producing
di-boson ($W^+W^-$, etc.) and boson + jets at the LHC.
\par
Double parton scattering (DP) in the simplest model produces a final
state that mimics a combination of two independent scatterings. It is
customary \cite{Abe:1993rv} to express the cross section for this
process as a product of the cross sections for the individual hard
scatterings divided by a scaling factor, $\sigma_{eff}$:
$$\sigma_{DP}=m\frac{\sigma_A\sigma_B}{2\sigma_{eff}} .$$
The factor $m$
is unity for indistinguishable scatterings and has a value of two when
it is possible to distinguish between A and B.  This formula assumes
that the number of parton-parton interactions follows a Poisson
distribution but can also use other distributions e.g. Poisson
statistics for a given impact parameter \cite{Calucci:1997uw}.  The
parameter $\sigma_{eff}$ describes the spatial distribution of partons
\cite{Sjostrand:1987su} e.g. for a model that assumes a proton with
uniformly distributed partons $\sigma_{eff}=11$~mb.
\par  
Events with four or more high transverse momentum objects (jets,
leptons, prompt photons...)  is an obvious place to look for
signatures of multiple hard parton interactions, although it should be
realized that higher order QCD processes, for which no exact QCD
calculations are available yet, are dominating.  Only few searches for
double parton collisions at the ISR, the SPS and the Tevatron have
been performed and the results are not very consistent
\cite{Akesson:1986iv,Alitti:1991rd,Abe:1993rv}.  Recently CDF
published \cite{Abe:1997xk} a strong signal for double parton
scattering. In this analysis a value of
$\sigma_{eff}=14.5\pm1.7^{+1.7}_{-2.3}$~mb was extracted from data in
a model-independent way by comparing the number of observed double
parton events to the number of events with hard scatterings at the
separate $p\bar{p}$ collisions within the same beam crossing. This
represents a significant improvement over previous measurements and
may be used to constrain models using a parton spatial density.
   
\subsubsection{Multiple interaction component of the underlying event
  at Tevatron and  HERA : summary}

Analyses of hadron-hadron and photon-hadron collisions at the Tevatron
and HERA have firmly established the multiple interaction component of
the underlying event. Only QCD models which include secondary soft or
semi-hard scatterings a la \cite{Sjostrand:1987su} (PYTHIA, HERWIG,
PHOJET) are able to give a reasonable description of the data.  The
energy flow of underlying events as measured outside leading jets was
studied in various phase space regions, applying conditions which help
to disentangle contributions from beam-beam interaction and
initial/final state radiation. At HERA the energy available to the
photon beam remnant was used as an additional constraint. The general
structure of the underlying event is reasonably well described by Monte
Carlo generators like PYTHIA, HERWIG and PHOJET, but a detailed
understanding is still missing. Studies of underlying events at HERA
are not as extensive as those by CDF at the Tevatron and it would
certainly be of great interest to apply the same analysis methods to
high energy $\gamma p$, where $x_{\gamma}$ provides an additional
``degree of freedom''. The effects of the transverse size of hadronic
photon on the underlying event, i.e. the $Q^2$ dependence, has not
been exploited at all so far.  The CDF Collaboration has reported a
firm observation of double hard parton scattering in the $\gamma$ + 3
jets final state and has made an estimation of the effective cross
section for double parton scattering.  This fact is of paramount
importance for the phenomenological understanding of the underlying
event, in constraining the multiple interaction models
\cite{Sjostrand:1987su,Odagiri:2004tt}.


\subsection{Gaps between jets and BFKL}
\label{sec:jetgapjet}
\textit{Main author G.~Ingelman}\\

\begin{figure}
\begin{center}
\epsfig{file=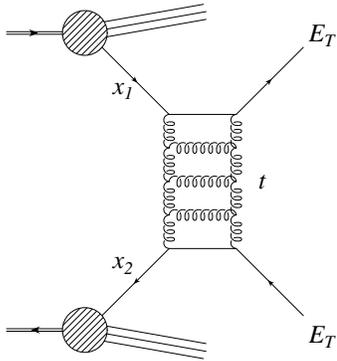,width=0.5\columnwidth}
\end{center}
\caption{Hard color singlet exchange through a BFKL gluon ladder
  giving a rapidity gap between two high-$p_\perp$ jets.}
\label{fig:GI-BFKL-ladder}
\end{figure}

The observation \cite{Abe:1997ie,Abbott:1998jb} at the Tevatron of
events with a rapidity gap between two high transverse-energy ($E_T$)
jets provides strong evidence for BFKL dynamics in terms of color
singlet gluon ladder exchange \cite{Enberg:2001ev}. As illustrated in
Fig.~\ref{fig:GI-BFKL-ladder}, the process can be described by elastic
parton-parton scattering via a hard color singlet gluon ladder. Since
there is no color exchanged, no color fields (strings) will be
formed in between and hence no hadrons produced through hadronization
in the intermediate rapidity region.

In the high energy limit $s/|t| \gg 1$, where the parton cms energy is
much larger than the momentum transfer, the amplitude for this diagram
is dominated by terms $\sim [\alpha_s \, \ln (s/|t|)]^n$ where the
smallness of $\alpha_s$ is compensated by the large logarithm. These
terms are resummed in the BFKL equation, which describes the exchange
of the whole gluon ladder, including virtual corrections and
reggeization of gluons. When solving the equation numerically it was
found that non-leading corrections are very important at the
non-asymptotic energy of the Tevatron
\cite{Enberg:2001ev,Motyka:2001zh}.

\begin{figure}
\begin{center}
\epsfig{file=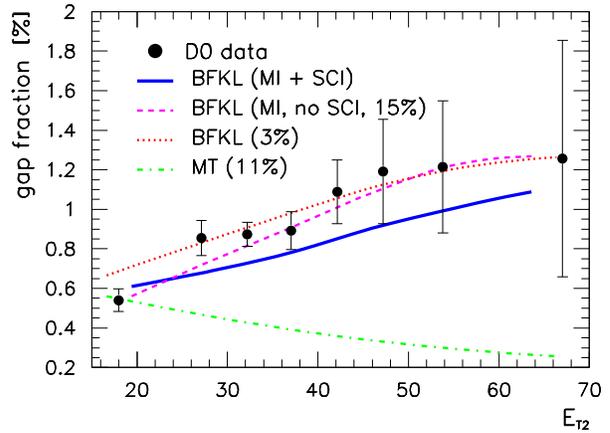,width=0.95\columnwidth}
\end{center}
\caption{Fraction of jet events having a rapidity gap in $|\eta|<1$
  between the jets versus the second-highest jet-$E_T$. D0 data
  compared to the color singlet exchange mechanism
  \protect\cite{Enberg:2001ev} based on the BFKL equation with
  non-leading corrections and with the underlying event treated in
  three ways: simple 3\% gap survival probability, {\sc Pythia}'s
  multiple interactions (MI) and hadronization requiring a 15\% gap
  survival probability, MI plus soft color interactions (SCI) and
  hadronization with no need for an overall renormalization factor.
  Also shown is the Mueller-Tang (MT) asymptotic result with a 11\%
  gap survival probability.}
\label{fig:GI-jet-gap-jet-ET}
\end{figure}

Formulating the results as matrix elements for effective $2\to 2$
parton scattering processes, they could be implemented in the Lund
Monte Carlo {\sc Pythia} such that parton showers and hadronization
could be added to generate complete events. As shown in
Fig.~\ref{fig:GI-jet-gap-jet-ET} these reproduce the data, both in
shape and absolute normalization, which is not at all trivial. The
non-leading corrections are needed since the asymptotic Mueller-Tang
result has the wrong $E_T$ dependence. A free gap survival probability
parameter, which in other models is introduced to get the correct
overall normalization, is not needed in this approach. Amazingly, the
correct gap rate results from the complete model including parton
showers, parton multiple scattering and hadronization through {\sc
  Pythia} together with the soft color interaction model
\cite{Edin:1995gi,Edin:1996mw}. The latter accounts for QCD
rescatterings \cite{Brodsky:2004hi} that are always present and if
these are ignored one needs to introduce an {\it ad hoc} 15\% gap
survival probability factor.

\begin{figure}
\begin{center}
\epsfig{file=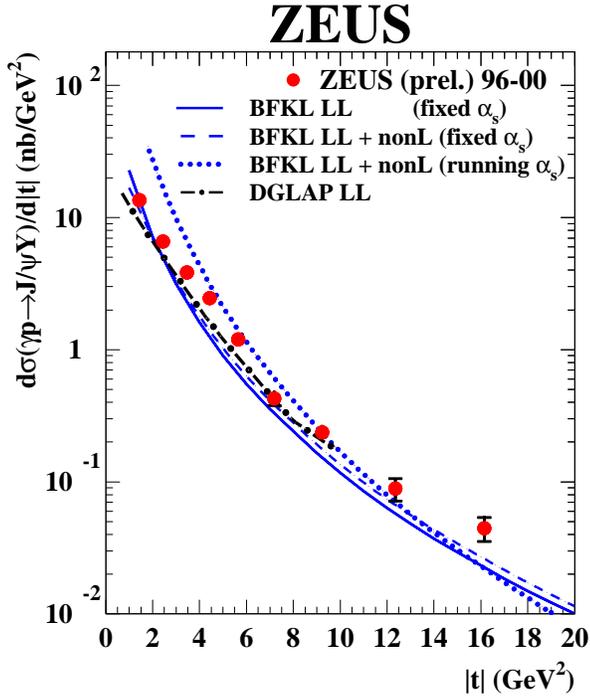,width=0.9\columnwidth}\\
\end{center}
\caption{Differential cross-section $d\sigma/d|t|$ for the process $\gamma +
p\to J/\psi + Y$. ZEUS data compared \protect\cite{ZEUS-diffr-psi-t} to BFKL
model calculations using leading log (LL) with fixed $\alpha_s$, and including
non-leading (non-L) corrections with fixed or running $\alpha_s$ as well as
with a model based on leading log DGLAP.}
\label{fig:GI-ZEUS-diffr-psi-t}
\end{figure}

\begin{figure}
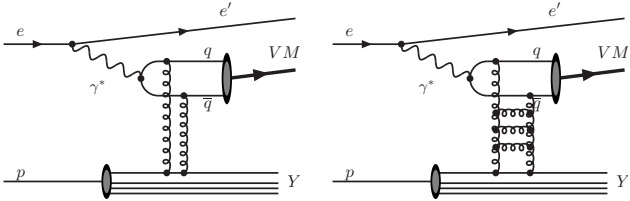

\begin{center}
\epsfig{file=fig/GI-ZEUS-diffr-psi-2gluon.epsi,
bbllx=159pt,bblly=457pt,bburx=333pt,bbury=575pt,clip=,width=0.45\columnwidth}
\hspace*{2mm}
\epsfig{file=fig/GI-ZEUS-diffr-psi-BFKL.epsi, 
bbllx=159pt,bblly=457pt,bburx=333pt,bbury=575pt,clip=,width=0.45\columnwidth}\\
\end{center}
\caption{Diffractive vector meson production at large momentum transfer
  as described by perturbative QCD hard color singlet exchange via
  two gluons and a gluon ladder in the BFKL framework
  \protect\cite{Enberg:2002zy}.}
\label{fig:GI-ZEUS-diffr-psi-BFKL}
\end{figure}

Related to this is the new results from ZEUS \cite{ZEUS-diffr-psi-t}
on the production of $J/\psi$ at large momentum transfer $t$ in
photoproduction at HERA. The data, shown in
Fig.~\ref{fig:GI-ZEUS-diffr-psi-t}, agree well with perturbative QCD
calculations \cite{Enberg:2002zy}, based on the hard scales $t$ and
$m_{c\bar{c}}$, for two-gluon BFKL color singlet exchange. As
illustrated in Fig.~\ref{fig:GI-ZEUS-diffr-psi-BFKL}, not only the
simple two-gluon exchange is included, but also the full gluon ladder
in either leading logarithm approximation or with non-leading
corrections. Using a running $\alpha_s$ does, however, give a somewhat
too steep $t$-dependence compared to the data. The conventional DGLAP
approximation provides a good description in the range
$|t|<m^2_{J/\psi}$ where this model \cite{Gotsman:2001ne} is argued to
be valid due to ordered momenta in the gluon ladder (cf.\ 
Fig.~\ref{fig:GI-ZEUS-diffr-psi-BFKL}). However, the DGLAP model gives
a very weak dependence on the energy $W$, which is in contrast to the
observed increase of the cross-section with energy as also results
from the BFKL-based calculations \cite{ZEUS-diffr-psi-t}. Altogether
this provides another evidence for BFKL dynamics.

\subsection{\boldmath Jets at small-$x$}
\label{sec:xjets}

\textit{Main authors L.~Jönsson and A.~Knutsson}\\

In the region of low $x$-values the interacting parton frequently
produces a cascade of emissions before it interacts with the virtual
photon. Due to the strong ordering in virtuality, the emissions of
the DGLAP evolution are very soft close  to the proton direction,
whereas BFKL emissions can produce large transverse  momenta in this
region. Thus, deviations from the DGLAP parton evolution scheme  are
expected to be most visible in a region close to the direction of
the  proton beam.

HERA has extended the available region in the Bjorken  scaling
variable, $\xbj$, down to values  of $\xbj \simeq 10^{-4} $, for
values of the four momentum transfer squared, $Q^2$, larger than a
few GeV$^2$, where perturbative calculations in  QCD are expected to
be valid. 

A measurement of the forward jet production cross section at small
$\xbj$, as proposed by Mueller and
Navelet~\cite{Mueller:1990er,Mueller:1990gy,Mueller:1986ey}, has 
long been regarded as the most promising test of perturbative parton
dynamics. The idea is to select events with a jet close to the proton
direction having the virtuality of the propagator closest to the
proton approximately equal to the virtuality of the exchanged photon.
This will suppress an evolution with strong ordering in virtuality as
is the case in the DGLAP evolution. The additional requirement that
the forward jet takes a large fraction of the proton momentum,
$x_{jet} = E_{jet}/E_p$, such that $x_{jet} \gg x_{Bj}$ opens up for
an evolution where the propagators are strongly ordered in the
longitudinal momentum fraction like in the BFKL scheme.
Experimentally this is realized by demanding the squared transverse
momentum of the forward jet to be of the same order as $Q^2$ and
$x_{jet}$ to be larger than a preselected value which still gives
reasonable statistics.  More exclusive final states, like those
containing a di-jet system in addition to the forward jet (called
`2+forward jet'), provide an additional handle to control the parton
dynamics.

\paragraph{Production of forward jets in DIS}

The H1 experiment has measured the forward jet cross
section~\cite{Aktas:2005up} 
using data collected in 1997, comprising an
integrated luminosity of 13.7 pb$^{-1}$.
The proton energy is 820~GeV and the positron energy is
27.6~GeV which correspond to a center-of-mass-energy of
$\sqrt{s}\approx$300~GeV. 

DIS events are obtained by applying the cuts $E_{e'} > 10$~GeV, $ 156
\dg < \theta_e < 175\dg$, $0.1 < y < 0.7$ and $5$~GeV$^2 < Q^2 <
85$~GeV$^2$, where $E'_e$ is the energy of the scattered electron,
$\theta_e$ the polar angle, and $y$ is the inelasticity of the
exchanged photon.  Jets are defined using the inclusive $k_t$-jet
algorithm~\cite{Catani:1993hr, Catani:1992zp} applied in the
Breit-frame. A forward jet is defined in the laboratory system as having
$p_{t,jet} > 3.5 $~GeV and being in the angular range $7\dg <
\theta_{jet} < 20\dg$.  In order to enhance BFKL evolution it is
required that $x_{jet} > 0.035$ whereas DGLAP evolution was suppressed
in the single differential cross section measurement by introducing
the requirement $0.5 < p_t^2/Q^2 < 5$.

Another event
sample, called the '2+forward jet' sample, is selected by requiring
that, in addition to the forward jet, at least two more jets are found,
all of them having $p_{t,jet}$ larger than 6~GeV. In
this scenario the $p_t^2/Q^2$-cut is not applied, due to the limited
statistics.

The forward jet cross sections for single and triple differential cross sections
are compared to LO ($\alpha_s$) and NLO
($\alpha_s^2$) calculations of direct photon interactions as obtained
from the DISENT program. Comparisons of the inclusive forward jet
cross sections with the DISENT predictions for a di-jet final state
are adequate, since the forward jet events always contain at least 
one additional jet due to the kinematics. The renormalization scale
$(\mu_r^2)$ is given by the average $p_t^2$ of the di-jets from the
hard scattering process, while the factorization scale $(\mu_f^2)$ is given by 
the average $p_t^2$ of all forward jets in the selected sample. 

In the analysis of events with two jets in addition to the
forward jet, the measured cross sections are compared to the predictions of
NLOJET++. This program provides perturbative calculations of cross
sections for three-jet production in DIS at NLO ($\alpha^3_s$) accuracy. 
In this case the scales $\mu_r=\mu_f$ are set to the
average $p_t^2$ of the three selected jets in the calculated event.

The NLO calculations by DISENT~\cite{Catani:1996jh, Catani:1996vz} and
NLOJET++~\cite{Nagy:2001xb} are performed using the
CTEQ6M~\cite{Pumplin:2002vw} parameterization of the parton distributions in the proton.

\paragraph{Single Differential Cross Section}
\label{sec:inclusive} 

\begin{figure}[htb]
  \begin{center}
    \vspace*{1mm}
    \vspace*{1cm}
    \epsfig{figure=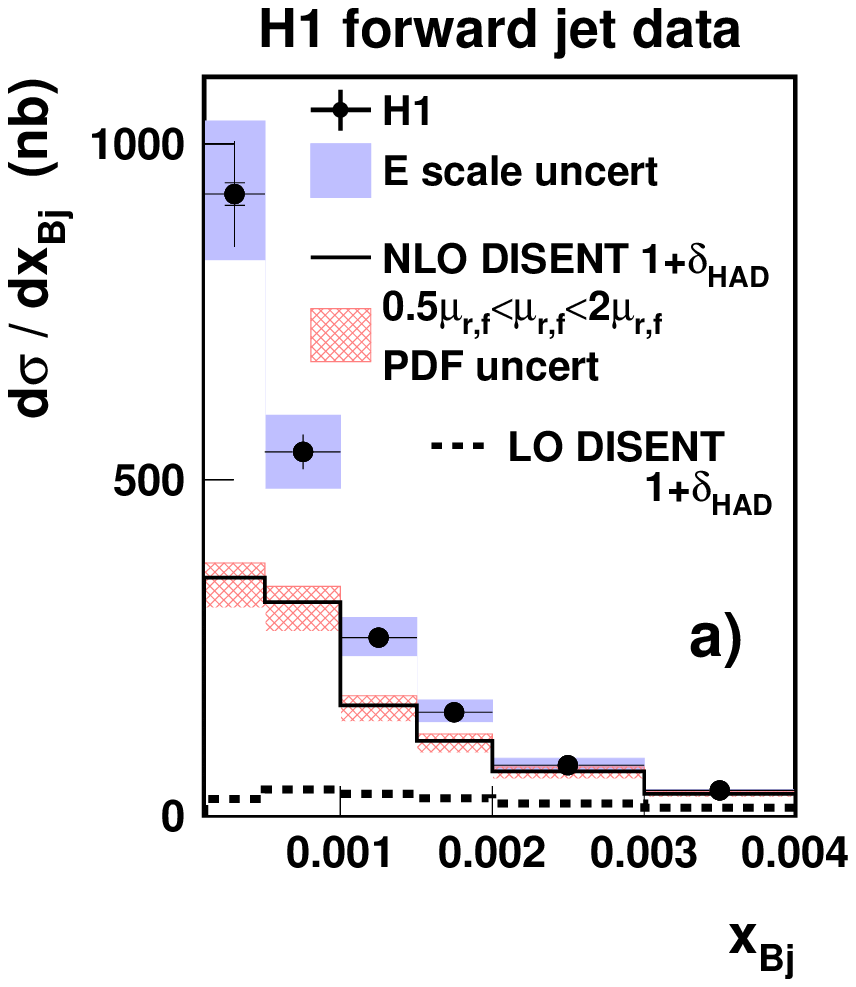,height=0.24\textwidth}
    \epsfig{figure=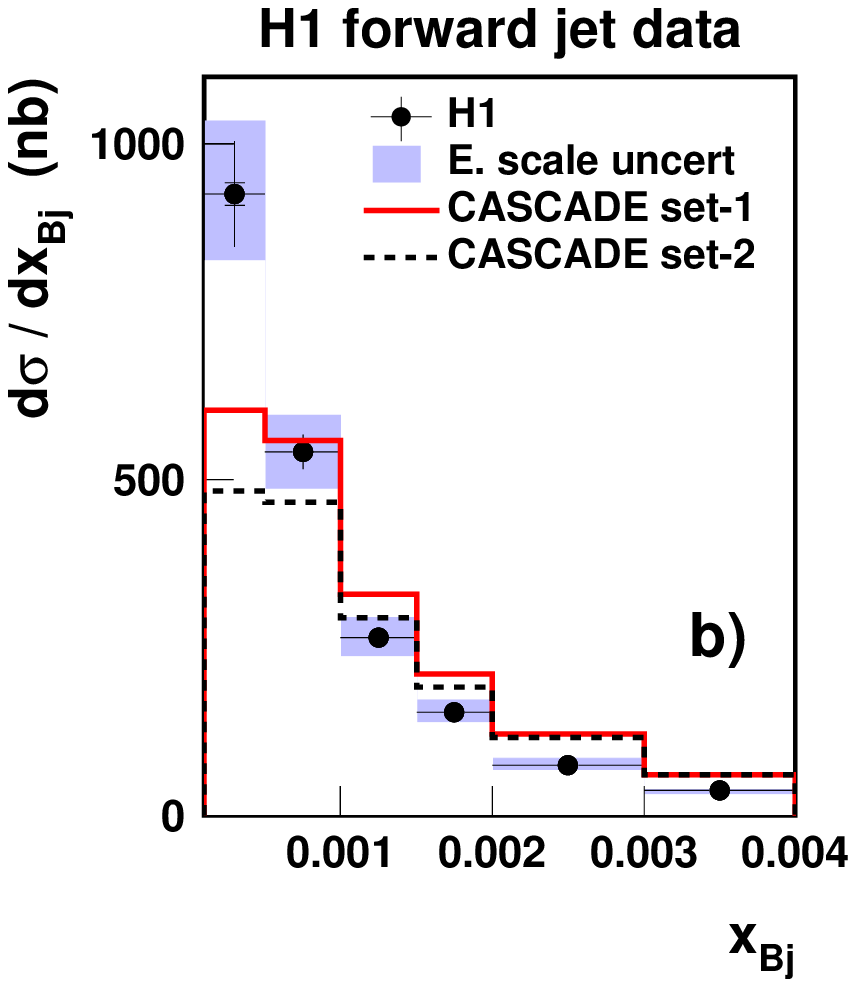,height=0.24\textwidth}
    \epsfig{figure=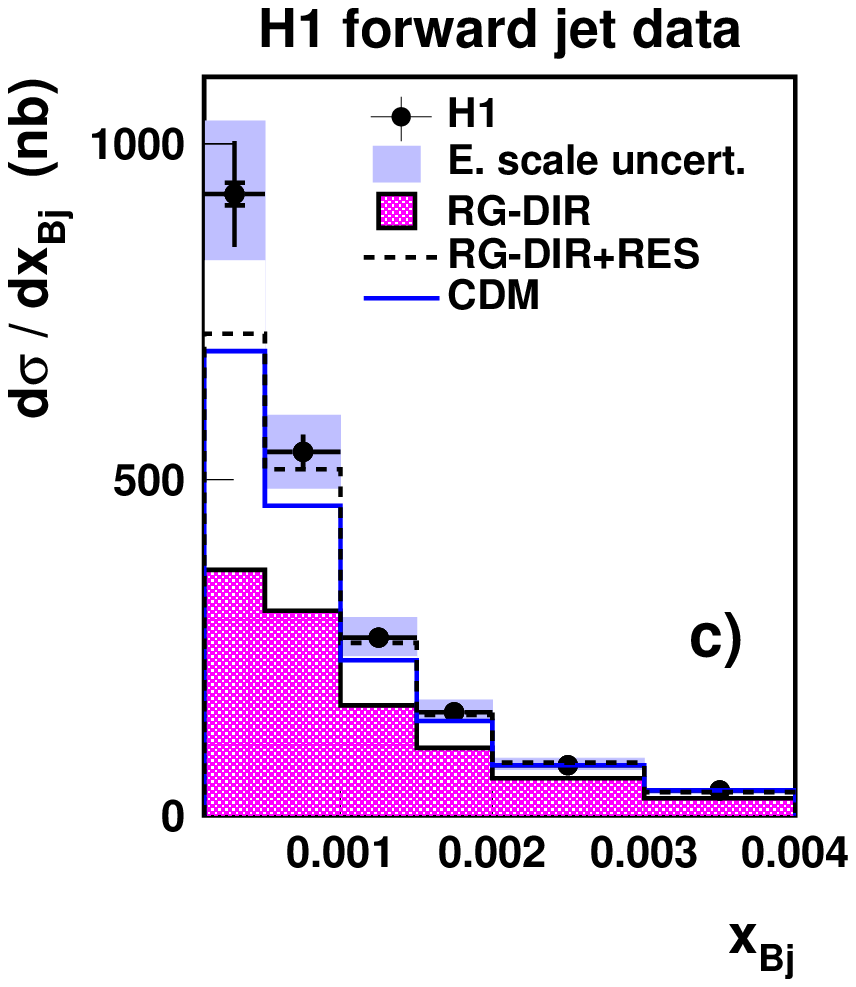,height=0.24\textwidth}
    \caption{
      The hadron level cross section for forward jet production 
      as a
      function of $\xbj$ as measured by H1~\protect\cite{Aktas:2005up}
      compared to NLO predictions from DISENT (a)
      and to QCD Monte Carlo models (b and c). The shaded band around
      the data points shows the error from the uncertainties in the
      energy scales of the liquid argon calorimeter and the SpaCal
      electromagnetic calorimeter. The hatched band around the NLO
      calculations illustrates the theoretical uncertainties in the
      calculations, estimated as described in the text. The dashed
      line in (a) shows the LO contribution.
      \label{xfj_djcorr}}
  \end{center}
\end{figure}

The measured single differential forward jet cross sections on hadron
level are compared with LO ($\alpha_s$) and NLO ($\alpha^2_s$)
calculations from DISENT in Fig.~\ref{xfj_djcorr}a.  In
Fig.~\ref{xfj_djcorr}b and c the data are compared to the various QCD
models.

In Fig.~\ref{xfj_djcorr}a it can be observed that, at small $x_{Bj}$,
the NLO di-jet calculations from DISENT are significantly larger than
the LO contribution. This reflects the fact that the contribution
from forward jets in the LO scenario is suppressed by kinematics. For
small $x_{Bj}$ the NLO contribution is an order of magnitude larger
than the LO contribution. The NLO contribution opens up the phase
space for forward jets and improves the description of the data
considerably. However, the NLO di-jet predictions are still a factor
of 2 below the data at low $\xbj$. The somewhat improved agreement at
higher $x_{Bj}$ can be understood from the fact that  the range in
the longitudinal momentum fraction which is available for higher
order emissions decreases.

From Fig.~\ref{xfj_djcorr}b it is seen that the CCFM model (both set-1
and set-2) predicts a somewhat harder $x_{Bj}$ distribution, which
results in a comparatively poor description of the data.

Fig.~\ref{xfj_djcorr}c shows that the DGLAP model
with direct photon interactions alone (RG-DIR) gives results similar
to the NLO di-jet calculations and falls below the data, particularly
in the low $x_{Bj}$ region. The description of the data by the
DGLAP model is significantly improved if contributions from resolved
virtual photon interactions are included (RG-DIR+RES). However, there
is still a discrepancy in the lowest  $x_{Bj}$-bin, where a possible
BFKL signal would be expected to show up most prominently. The CDM
model, which gives emissions that are non-ordered in transverse
momentum, shows a behavior similar to the RG DIR+RES model.

\paragraph{Events with Reconstructed Di-jets in Addition to the Forward Jet}

By requiring the reconstruction of the two hardest
jets in the event in addition to the forward jet, different
kinematic regions can be investigated by applying cuts on the jet momenta and their
rapidity separation.

\begin{figure}[htb]
  \begin{center}
    \vspace*{1mm}
    \vspace*{1cm}
    \epsfig{figure=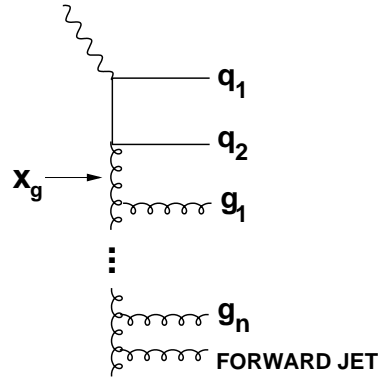,height=5cm}
    \caption{
      A schematic diagram of an event giving a forward jet and two
      additional hard jets. These may stem from the quarks ($q_1$ and
      $q_2$) in the hard scattering vertex or gluons in the parton
      ladder. $x_g$ is the longitudinal momentum fraction carried by
      the gluon, connecting to the hard di-jet system (in this case
      $q_1$ and $q_2$) .
    \label{fwdjet-dijet}}
  \end{center}
\end{figure}

In this scenario it is demanded that all jets have transverse momenta
larger than 6~GeV. By applying the same $p_{t,jet}$ cut to all three
jets, evolution with strong $k_{t}$-ordering is not favored.
The jets are ordered in rapidity according to $\eta_{fwdjet} >
\eta_{jet_2} > \eta_{jet_1} > \eta_e$  with $\eta_e$ being the
rapidity of the scattered electron. The
cross section is  measured by H1~\cite{Aktas:2005up} 
in two intervals of $\Delta \eta_1 = 
\eta_{jet_2} - \eta_{jet_1} $. If 
the di-jet system originates from the quarks $q_1$ and $q_2$ 
(see Fig.~\ref{fwdjet-dijet}), the phase space for evolution in $x$ between 
the di-jet system and the forward jet is increased by requiring that 
$\Delta\eta_1$ is small and that 
$\Delta\eta_2 = \eta_{fwdjet} - \eta_{jet_2}$ is large. $\Delta\eta_1 < 1$ 
favors small invariant masses of the di-jet system and thereby small values 
of $x_g$ (see Fig.~\ref{fwdjet-dijet}).  With $\Delta\eta_2$ large, 
$x_g$ carries only a small fraction 
of the total propagating momentum, leaving the rest for additional radiation.  

 The directions of the other jets are related to the forward 
jet through the $\Delta\eta$ requirements.  When $\Delta\eta_2$ is small, it is 
therefore possible that one or both of the additional jets originate 
from gluon radiation close in rapidity space to the forward jet. 
With $\Delta\eta_1$ large, BFKL-like evolution may then occur between the 
two jets from the di-jet system, or, with both $\Delta\eta_1$ and $\Delta\eta_2$ small, even between 
the di-jet system and the hard scattering vertex. By studying the cross 
section for different $\Delta\eta$ values one can test theory and models 
for event topologies where the $\kt$ ordering is broken at varying 
locations along the evolution chain.

In this investigation the same settings of the QCD models are used as
in sections~\ref{sec:inclusive}, while the NLO three-jet
cross sections are calculated using NLOJET++.

\begin{figure}[htb]
  \begin{center}
    \vspace*{1mm}
    \vspace*{1cm}
    \epsfig{figure=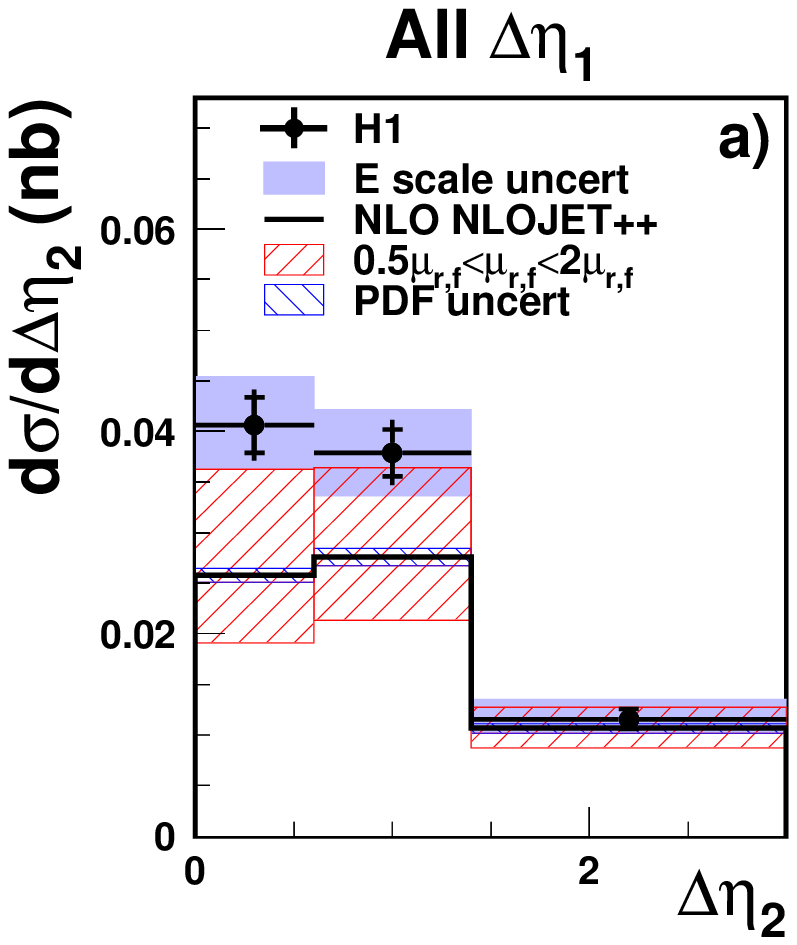,width=0.24\textwidth}
    \epsfig{figure=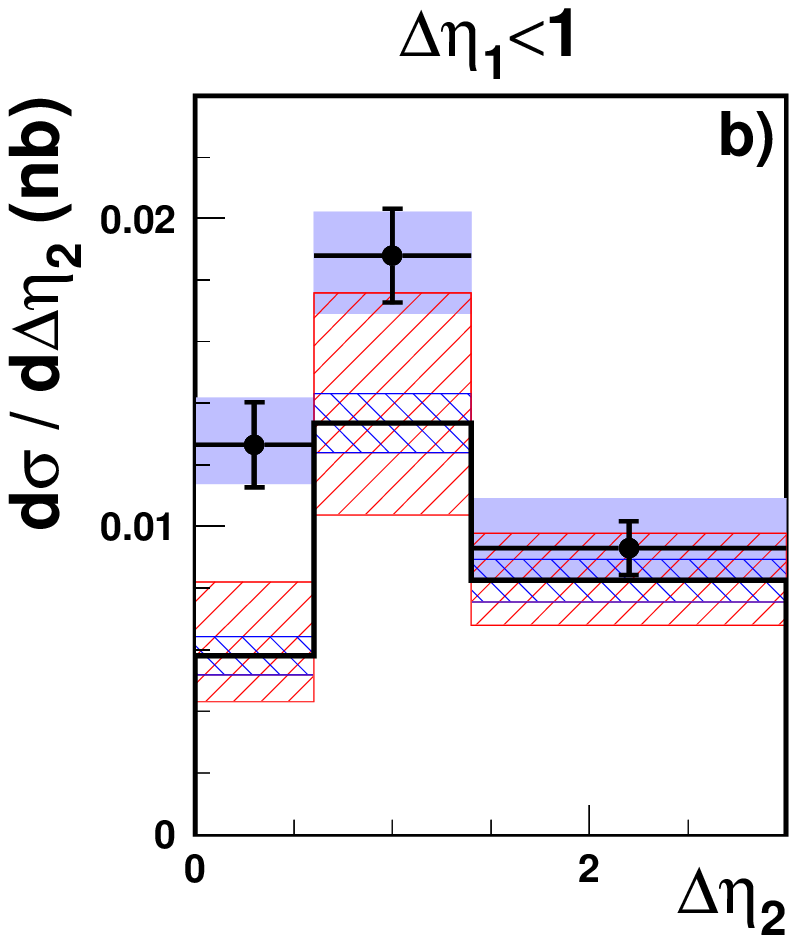,width=0.24\textwidth}
    \epsfig{figure=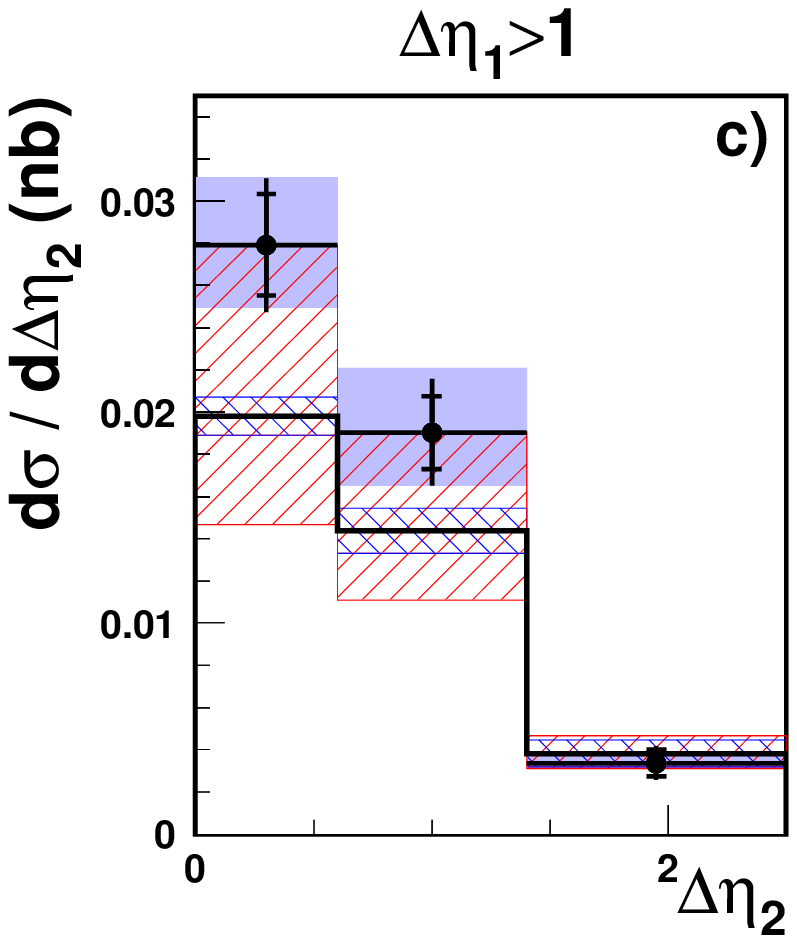,width=0.24\textwidth}
    \caption{
      The cross section for events with a reconstructed high
      transverse momentum di-jet system and a forward jet as a
      function of the rapidity separation between the forward jet and
      the most forward-going additional jet, $\Delta\eta_2$ as measured by 
      H1~\protect\cite{Aktas:2005up}. 
      Results
      are shown for the full sample and for two ranges of the
      separation between the two additional jets, $\Delta\eta_1<1$ and
      $\Delta\eta_1>1$.  The data are compared to the predictions of a
      three-jet NLO calculation from NLOJET++.  The band around the
      data points illustrates the error due to the uncertainties in
      the calorimetric energy scales. The band around the NLO
      calculations illustrates the theoretical uncertainties in the
      calculations.
    \label{2+fwdnlo}}
  \end{center}
\end{figure}

From Fig.~\ref{2+fwdnlo} it is observed that NLO three-jet gives good 
agreement with the data if the two additional hard jets are emitted
in the central region ($\Delta\eta_2$ large). It is  interesting to
note that a fixed order calculation ($\alpha_s^3$),  including the
$\log (1/x)$-term to the first order in $\alpha_s$, is  able to
describe these data well. However, the more the additional  hard jets
are shifted to the forward region ($\Delta\eta_2$ small), the less
well are the data described by NLO three-jet. A possible
explanation is that the more forward the additional jets go, the
higher  the probability is that one of them, or even both, do not
actually originate from quarks but from additional radiated gluons.  
NLO three-jet calculates the NLO contribution to final states  containing
one forward jet and two jets from the di-quarks, i.e. it accounts for
the emission  of one gluon in addition to the three jets. Since the
radiated gluon is predominantly soft it has a small probability to
produce a jet that fulfills the transverse momentum requirement
applied in this analysis. This results in a depletion of the
theoretical cross section in the small $\Delta\eta_2$ region, which
is more pronounced when $\Delta\eta_1$ is also small, i.e. when all
three jets are in the forward region. Consequently a significant
deviation between data and NLOJET++ can be observed for such events
(see the lowest bin in Fig.~\ref{2+fwdnlo}b). Accounting for still
higher orders in $\alpha_s$ might improve the description of the data
in this domain, since an increased number of gluon emissions would
enhance the probability that one of  the radiated gluons produces a
jet which is above the threshold on the transverse momentum.

As explained above, evolution with strong $\kt$-ordering is 
disfavored in this study. Radiation that is non-ordered in $\kt$ 
may occur at different locations along the evolution chain, 
depending on the values of $\Delta\eta_1$ and $\Delta\eta_2$. In a
comparsion to QCD models (these figures are not shown, for details
see~\cite{Aktas:2005up}) the following observations
where made. The colour dipole model gives good agreement in all
cases, whereas the LO DGLAP models give cross sections that are too
low except when both $\Delta\eta_1$ and $\Delta\eta_2$ are large. For
this last topology all models and the NLO calculation agree with the
data, indicating that the available phase space is exhausted and that
little freedom is left for dynamical variations.  

Furthermore it was seen that the `2+forward jet' sample 
differentiates between the CDM and the DGLAP-resolved model, in contrast to  the
more inclusive samples where CDM and RG-DIR+RES give the same 
predictions.   The conclusion is that additional breaking of
the  $\kt$ ordering is needed compared to what is included in the 
resolved photon model (see Ref.~\cite{Aktas:2005up}).

\subsection{Production of neutral strange particles in 
  deep-inelastic scattering at HERA}
\label{sec:strange}
\textit{Main author C.~Risler}\\

In deep-inelastic scattering strange particles can be produced either
if a strange quark is interacting in the hard subprocess, or if
strange quark pairs are produced during the hadronization process.
The production of strange particles is sensitive to soft and hard
parton radiation of the initial and final state partons and is thus a
complementary approach to small $x$ processes.
Other sources of strangeness can be the decays
of charm hadrons or more exotic particles like glueballs, pentaquarks
or instantons.

The production properties of strange particles are not yet fully understood
nor described by the QCD models. 
Since strange particles are also produced
during the hadronization process, 
a measurement of strange particle production
is also a means to test the universality of hadronization in $e^+e^-$, $pp$ or
$ep$ collisions.

The inclusive production cross sections of strange neutral particles,
namely $\kn$-mesons and $\lam$-baryons\footnote{By $\lam$-baryons the
  $\lam$ particle and its antiparticle $\bar\lam$ are referred to.},
in deep-inelastic $ep$-scattering at HERA were measured with the H1
detector~\cite{Risler:2003np}.  The analyzed data were collected in
the years 1996 and 1997 at a center of mass energy of 300 GeV and with an
integrated luminosity of 17.8 ${\rm pb}^{-1}$.  The kinematic region $2
\,\mgevsq < Q^2 < 100 \,\mgevsq$ and $0.1 < y < 0.6$ is investigated,
where $Q^2$ is the squared momentum transfer and y the inelasticity.
This allows for probing very low Bjorken-$x$, $x>10^{-5}$.  $\kn$
mesons and $\lam$ baryons are reconstructed via the decay to
$\pi^-\pi^+$ and $\pi^-p$, respectively.  The production of $\kn$ and
$\lam$ is measured within the visual range, defined by $-1.3 < \eta <
1.3$ and $0.5 \,\mgev < p_T < 3.5 \,\mgev$, where $\eta$ is the
pseudorapidity and
$p_T$ the transverse momentum in the laboratory frame.

Comparisons of the total $\kn$ and $\lam$ cross sections with models using 
the Lund string hadronization \cite{Artru:1974hr,Bowler:1981sb,Andersson:1983ia,Andersson:1983jt,Andersson:1985qr} show that a lower strangeness 
suppression factor of 
$\lambda_s\approx 0.2-0.25$
is preferred to the default value of $\lambda_s=0.3$. 

The differential cross sections in the laboratory and the Breit frame
are compared to different model predictions, namely the MEPS (matrix
element and parton shower) model using the RAPGAP event generator
\cite{Jung:1995gf}, CCFM
\cite{Ciafaloni:1988ur,Catani:1990yc,Catani:1990sg} implemented in the
CASCADE program \cite{Jung:2000hk,Jung:2001hx}, the color dipole model
(CDM)
\cite{Gustafson:1988rq,Andersson:1990dp,Gustafson:1992uh,Andersson:1989gp,Gustafson:1986db,Lonnblad:1992tz}
using DJANGOH \cite{Charchula:1994kf} and to predictions by the HERWIG
\cite{Marchesini:1991ch,Corcella:2000bw} event generator.  The HERWIG prediction for the $\lam$
cross section is normalized to the observed total cross section, since
HERWIG overestimates the
cross section by a factor of 3.

\begin{figure}[btp]
  \includegraphics*[width=0.5\textwidth]{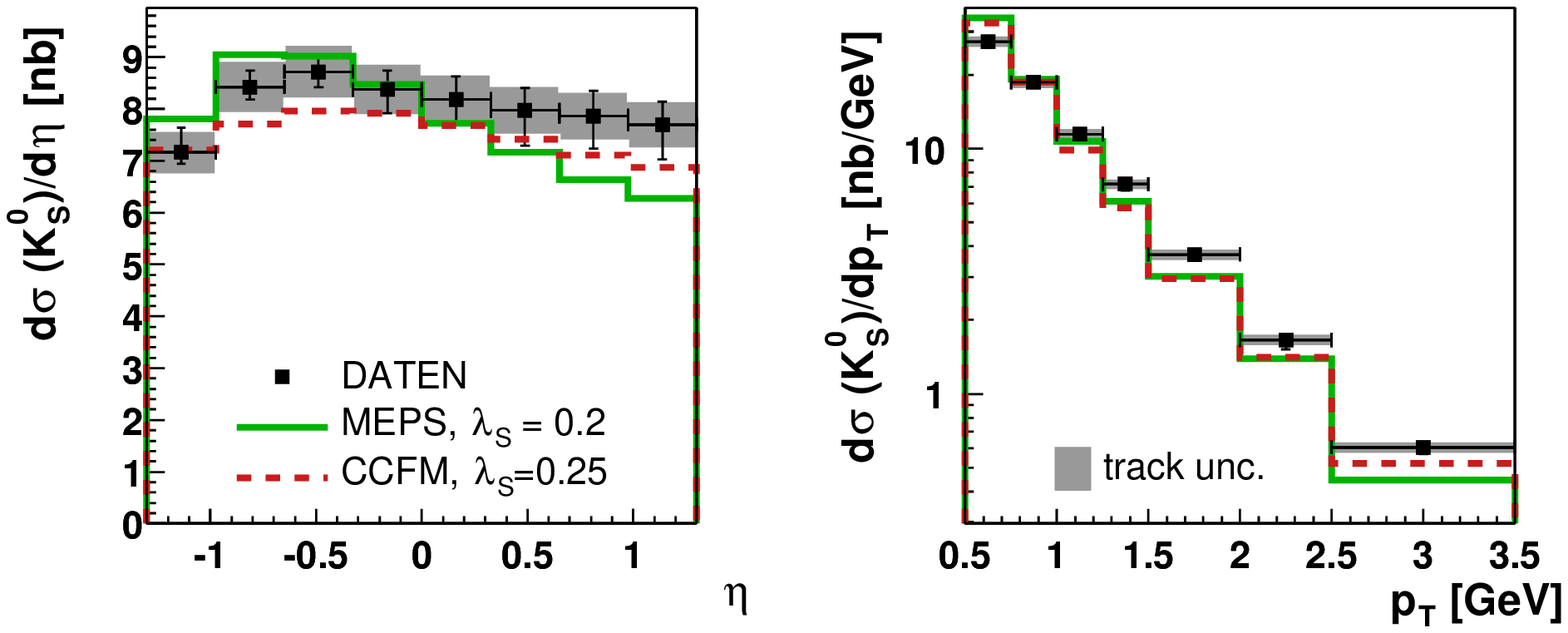}
  \includegraphics*[width=0.5\textwidth]{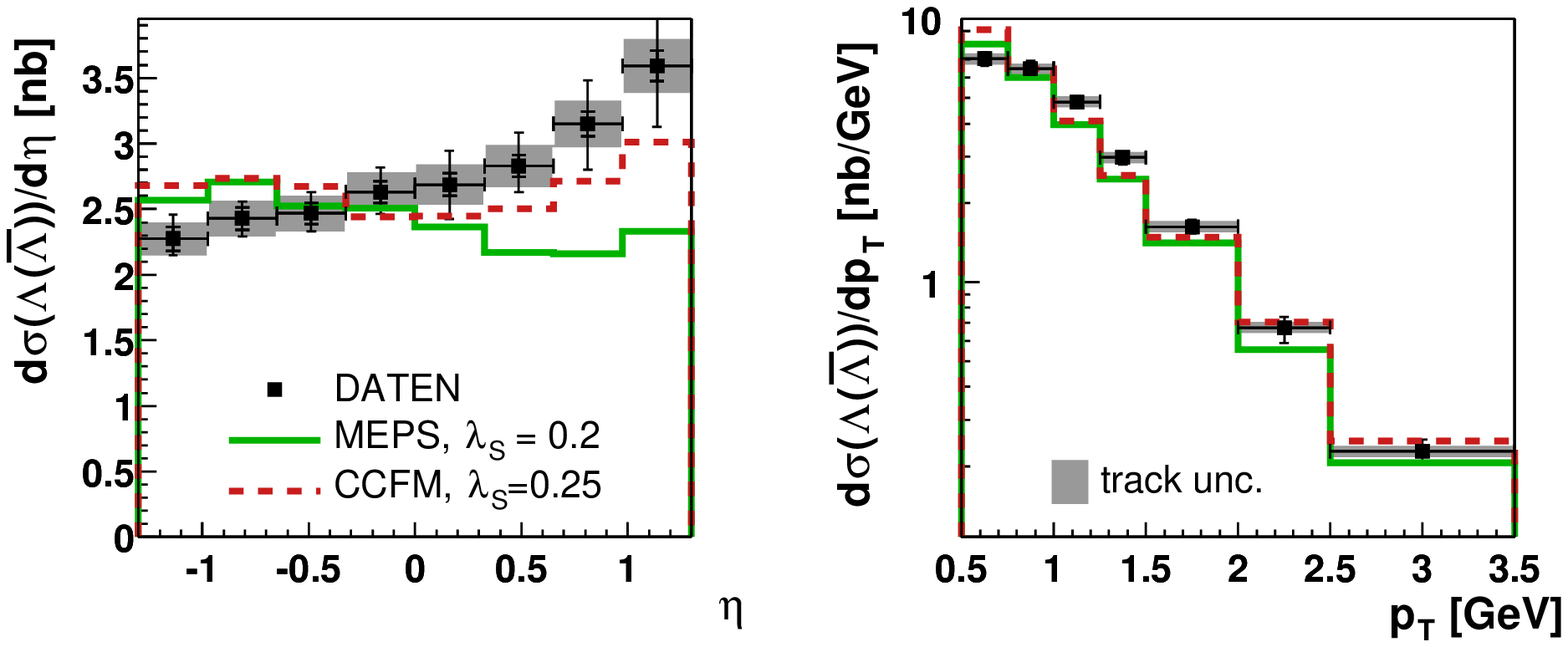}
  \caption{\label{fig:strange1}Differential $\kn$ and $\lam$ production 
    cross sections in the laboratory frame.
    }
\end{figure}

Fig. \ref{fig:strange1} shows the $\eta$ and $p_T$ dependence of the
$\kn$ and $\lam$ cross sections in the laboratory frame compared to
the model predictions by MEPS and CCFM, using a strangeness
suppression factor of $\lambda_S=0.2$ and $\lambda_S=0.25$ in the Lund
string model, respectively.  The preliminary data are shown with statistical
and systematic errors; the systematic uncertainty of the cross section
due to the
uncertainty of the tracking efficiency is separately shown as a grey band.\\
The $\eta$ spectrum of $\kn$ production can not be reproduced by the
MEPS model, while CCFM gives a better description.  In $\lam$
production a rise in the forward direction, defined by the direction
of the outgoing proton beam, is observed, which is not present in any
of the models.
The $p_T$ distribution of the $\kn$ and the $\lam$ cross section
are too soft in  MEPS as well as in CCFM.

The CCFM model yields a better description of the $\eta$ spectra in the data
than the MEPS model. 
In addition,
CCFM, in its implementation in 
the CASCADE program, allows only for gluon induced hard subprocesses.

\begin{figure}[btp]
  \includegraphics*[width=0.25\textwidth]{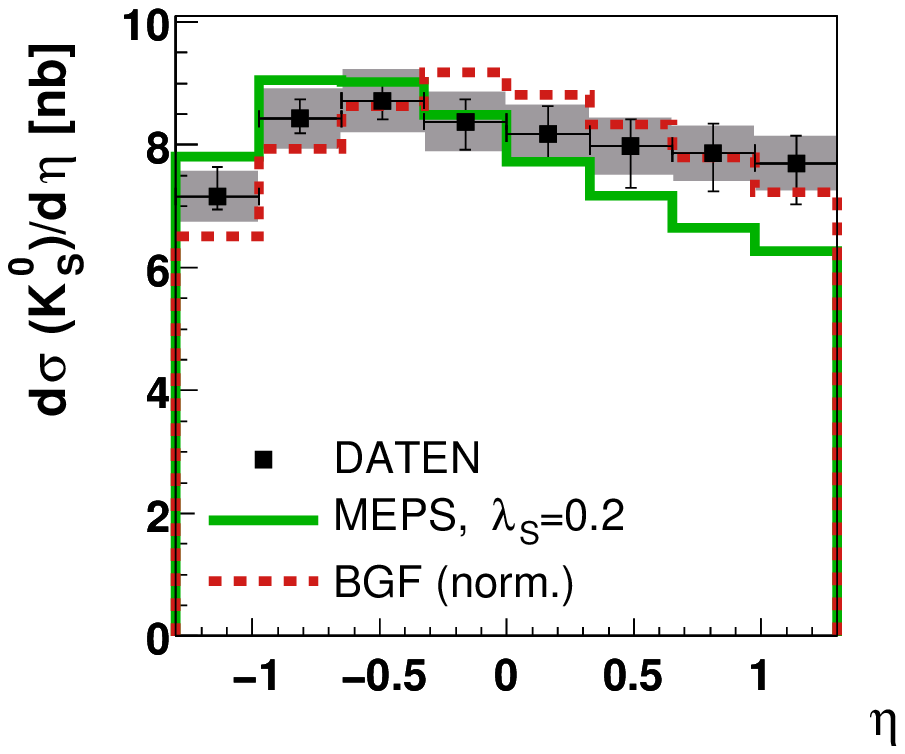}%
  \includegraphics*[width=0.25\textwidth]{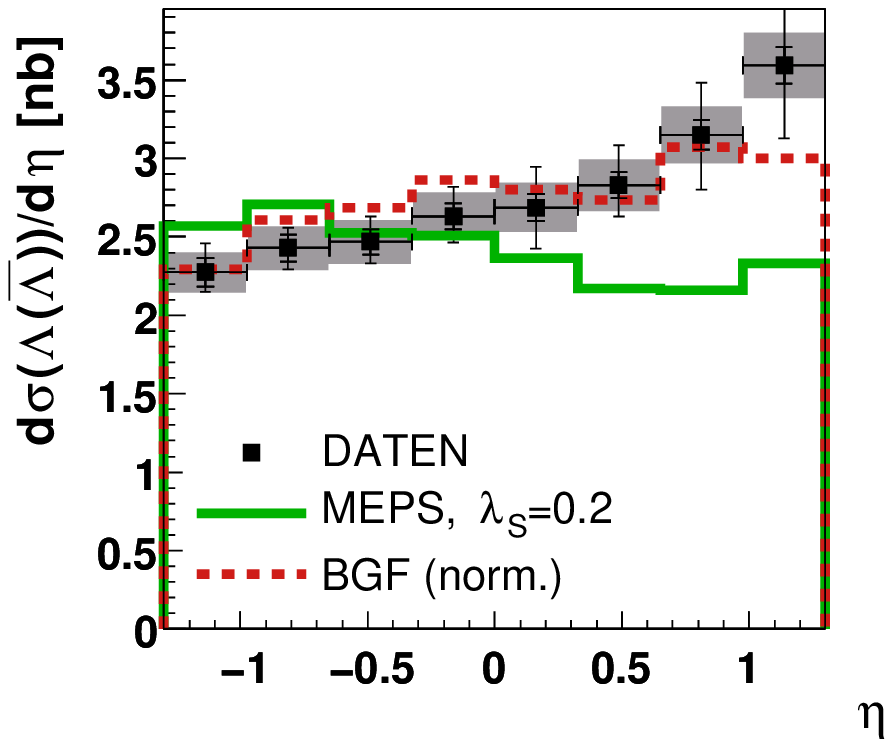}     
  \caption{\label{fig:strange2} $\eta$ dependence of $\kn$ and
    $\lam$ cross sections in the laboratory frame compared to MEPS and
    a modified MEPS model with only BGF hard subprocess.  }
\end{figure}

In Fig.~\ref{fig:strange2} the data are compared to a modified 
MEPS model (BGF), 
where  only boson gluon fusion processes are taken into account, while quark
induced subprocesses are switched off.
This modified MEPS model gives a slightly better description
of the data than the standard MEPS predictions.

The differential $\kn$ and $\lam$ production cross sections are
investigated as functions of $x_p=2 |\vec p|/Q$ and $p_T$
in the Breit frame.

The Breit frame can be divided into the target hemisphere of the
fragmenting proton and the current hemisphere in the direction of the
incoming photon, which is related to
the fragmentation of the current quark.

\begin{figure}[btp]
  \includegraphics*[width=0.5\textwidth]{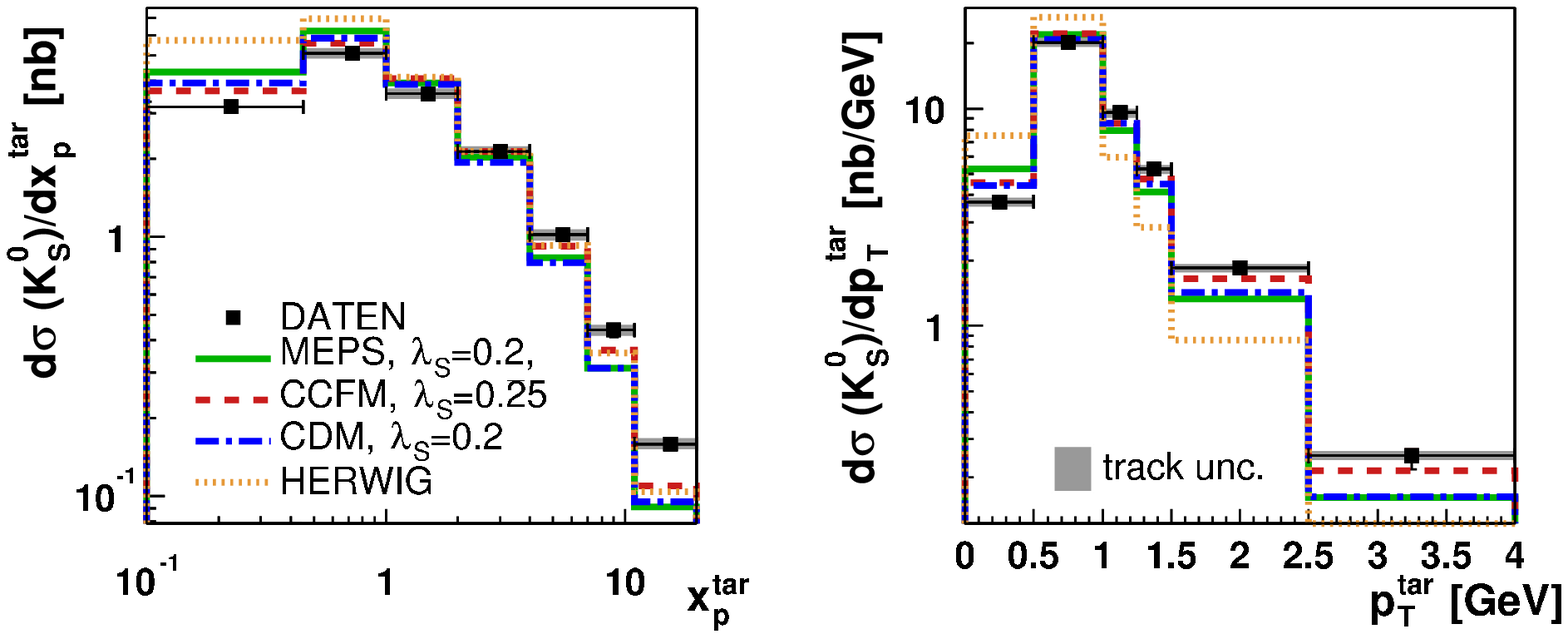}
  \includegraphics*[width=0.5\textwidth]{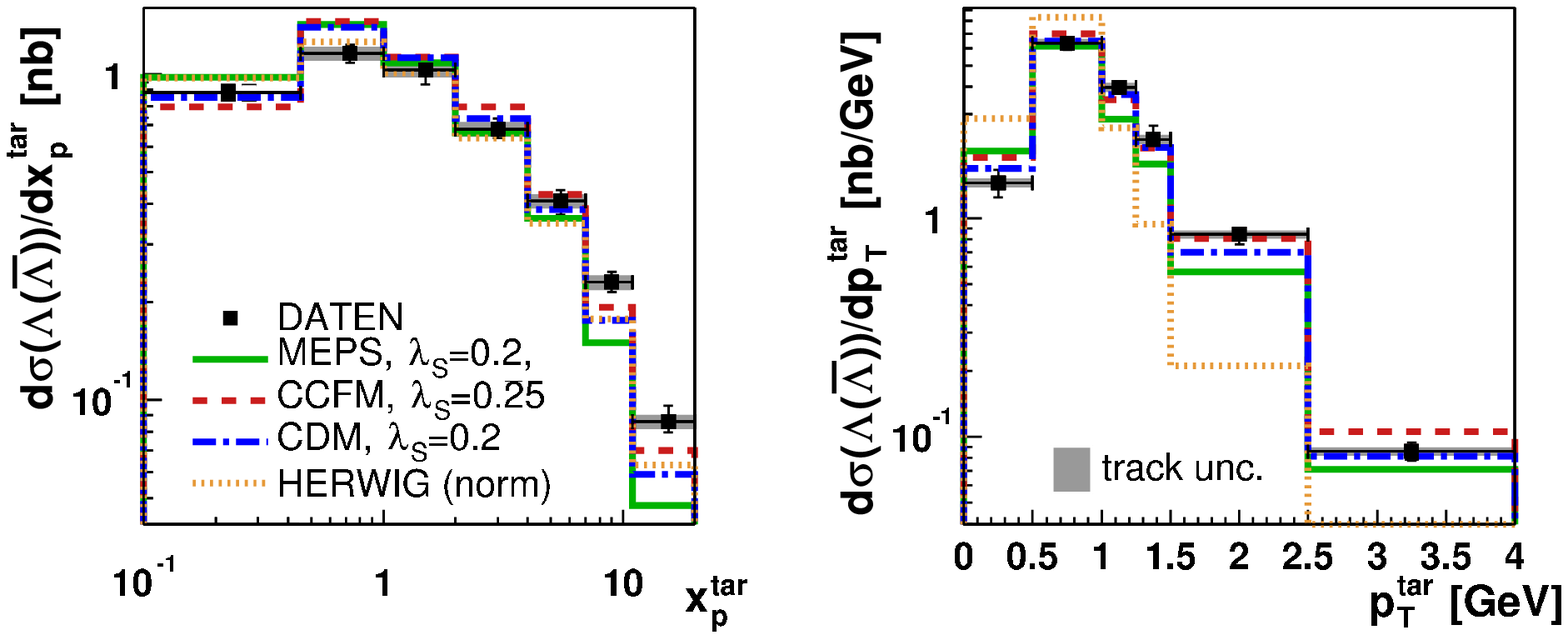}
  \caption{\label{fig:strange3}Target hemisphere of the Breit frame:
    $x_p=2 |\vec p|/Q$ and $p_T$ dependence of $\kn$ and $\lam$ cross
    sections in the target hemisphere of the Breit frame.}
\end{figure}

In the target hemisphere of the Breit frame (fig.~\ref{fig:strange3})
all four models underestimate the $\kn$ and $\lam$ cross section at
large $x_p$ and the $p_T$ spectra are modeled too softly in most of
the models.

\begin{figure}[btp]
  \includegraphics*[width=0.5\textwidth]{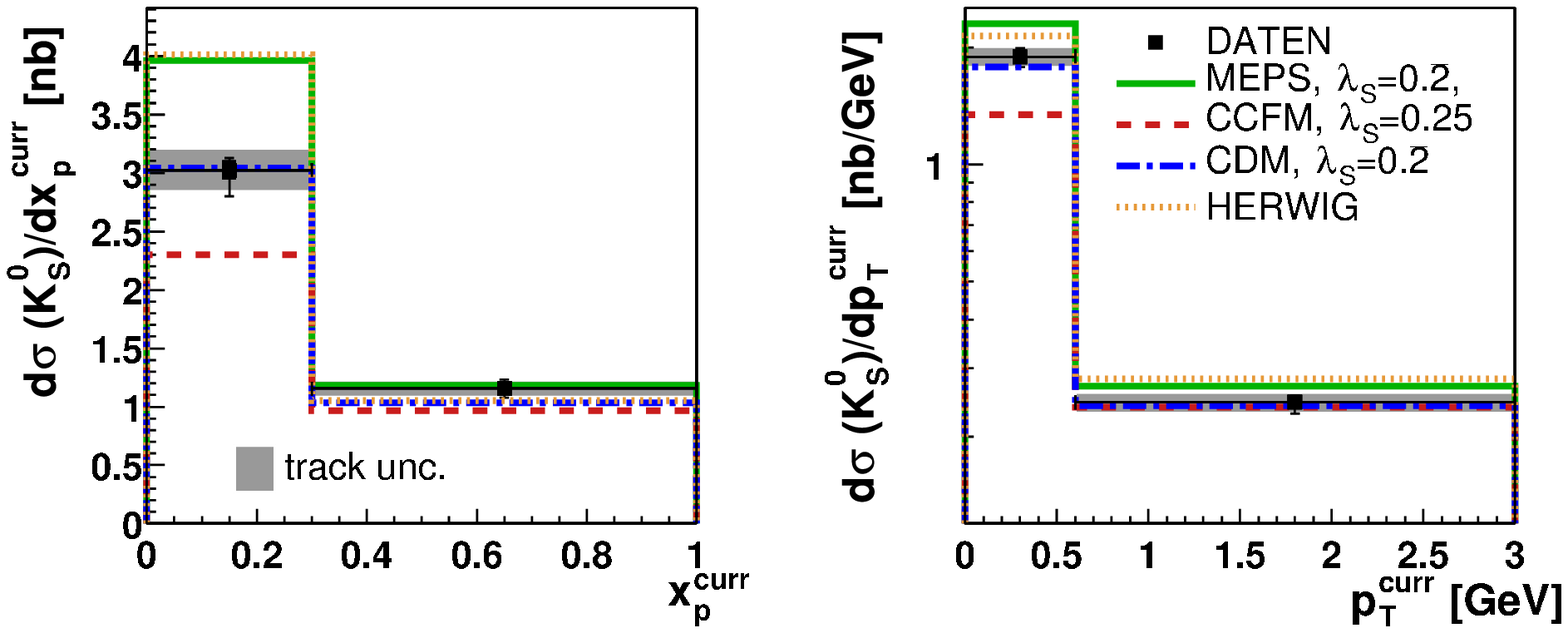}
  \includegraphics*[width=0.5\textwidth]{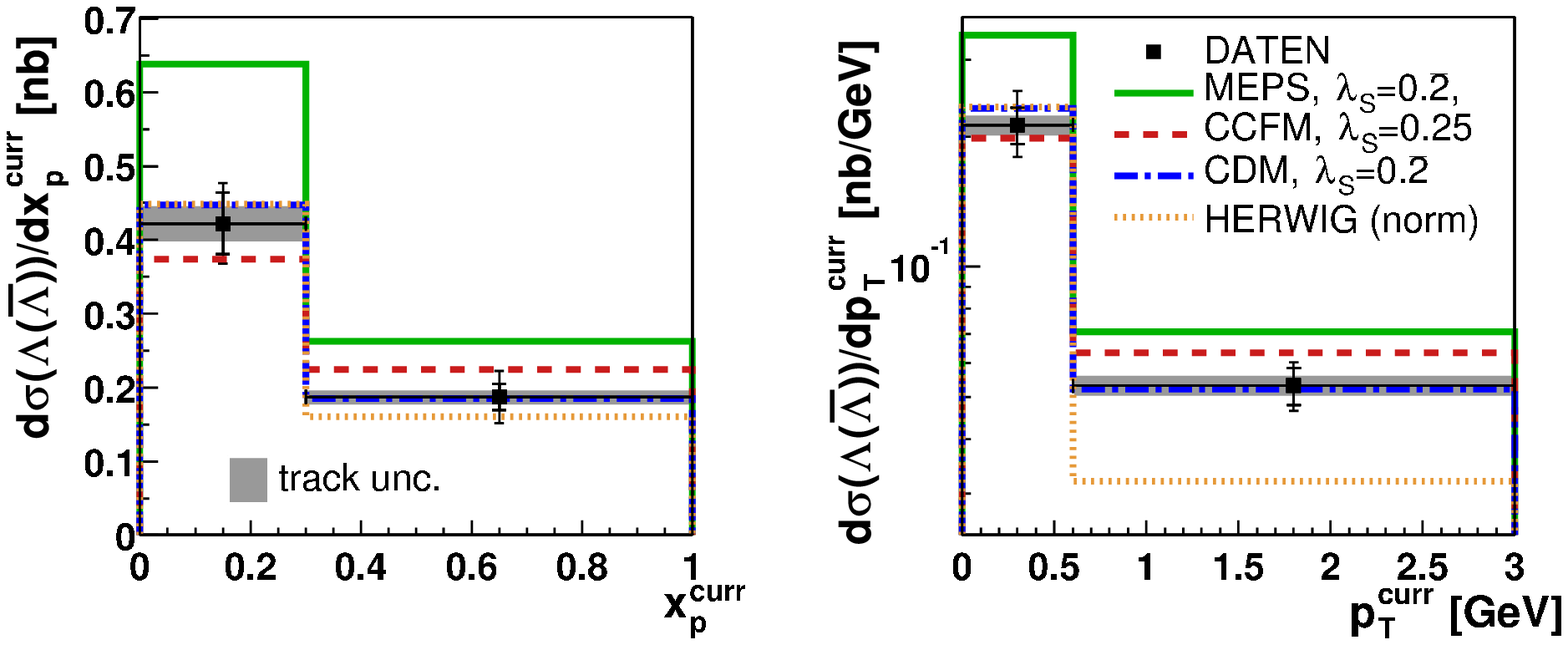}
  \caption{\label{fig:strange4}Current hemisphere of the Breit frame:
    $x_p=2 |\vec p|/Q$ and $p_T$ dependence of 
    $\kn$ and $\lam$ cross sections in the current hemisphere of the Breit frame.
    }
\end{figure}

Only a small fraction of all $\kn$ and $\lam$ decays is found in the
current hemisphere of the Breit frame, leading to large statistical
errors of the differential cross sections shown in
Fig.~\ref{fig:strange4}.  Within these errors CDM gives the best
description of the $x_p$ and $p_T$ dependence of the differential
$\kn$ and $\lam$ cross sections.

Concluding one can say that
none of the models gives a satisfactory description of the 
observed cross sections
of neutral strange particle production.
In particular the simulated transverse momentum spectra are too soft.
A significant increase of
$\lam$-baryon production was observed in the
region $0 < \eta < 1.3$ in the laboratory frame, 
which is not reproduced by any of the models.

The comparison with QCD models using the Lund string hadronization
reveals that a lower strangeness suppression factor than the 
standard LEP-value seems to be preferred at HERA.

The cross sections and comparisons shown here are results of a 
PhD thesis~\cite{Risler:2003np}.


\section{Outlook}
\label{sec:conclusions}

Studies of QCD in high energy $ep$ collisions are interesting in themselves,
as a highly nontrivial theory due to its nonlinear nature with a
nontrivial vaccuum. It is also important in order to fully understand
the background in attempts to find signals for physics beyond the
standard model at the LHC and future high energy colliders.

For the timelike cascades in $e^+ e^-$-annihilation, experimental data
are reproduced to an extent beyond expectations, by a perturbative
parton cascade (if only the first gluon emission is adjusted to matrix
elements) followed by a model for the subsequent non-perturbative
hadronization. To describe the spacelike cascades in $ep$ scattering at
high energies poses a much more difficult challenge.
$k_\perp$-factorization and leading order BFKL evolution offer a
qualitative frame of reference at small $x$, but do not give a
quantitative description of the experimental data. Non-leading
contributions are large, and the separation between perturbative
and non-perturbative effects in the timelike cascades is not realised
in the corresponding spacelike processes.

The non-leading contributions are essential also for the behaviour at
asymptotic energies. They give asymptotically small corrections to the
\emph{evolution equation}, but not to its \emph{solution}. The leading
order equation fixes the solution to the powerlike form $\sim
x^{-\lambda}$ (with logarithmic corrections), but the power $\lambda$
is affected by the non-leading terms, which therefore have a very
large effect. The perturbative--non-perturbative interplay is
important in two regimes. Firstly, the random walk in $\ln k_\perp^2$,
characteristic for the BFKL evolution chain, extends down into the
soft regime. This problem is further enhanced by a running coupling
$\alpha_s$. Secondly, the high gluon densities at small $x$ imply that
unitarity constraints and saturation become essential. This means that
non-perturbative effects are important also at larger $k_\perp$, where
the running coupling is small.

Recent progress, described in this report, includes in particular:
\begin{itemize}
\item Extending the $k_\perp$-factorization formalism introducing
  two-scale unintegrated and doubly unintegrated PDFs and investigation of the
  importance of the correct kinematics even at lowest order.
\item The solution to the BFKL evolution at NLO, and the NLO
  $\gamma^*$ impact factor.
\item BFKL dynamics in other fields, exemplified by
  $Q\bar{Q}$-production and away-from-jet energy flow in $e^+
  e^-$-annihilation.
\item Studies of unitarity corrections and saturation via the
  Balitsky-Kovchegov equation.
\item Going beyond leading order in the BK equation, where in
  particular energy-momentum conservation has a large effect.
\item AGK cutting rules in QCD, multi-pomeron exchange and
  diffraction.
\item Phenomenological applications and comparisons with experimental
  data. Here studies of forward jet and heavy quark production are of
  particular interest.
\end{itemize}

Further work is still needed within all these fields. 
The impact parameter dependence and
correlations, as well as generalisations to $eA$ collisions, need to be studied.
This is particularly important to get a better
understanding of high energy proton-proton collisions. 
 To fully
understand the dynamics of small-$x$ physics we need in the future
also to be able to combine the different routes listed above in a
unifying formalism, which can simultaneously account for the effects
of NLO (and NNLO) contributions and unitarity and saturation effects
including multi-pomeron exchange, pomeron loops and diffraction.

The detailed understanding of small-$x$ physics is essential for the
understanding of the underlying event structure observed at Tevatron and which
is expected to be even more significant at the LHC. Small $x$ physics 
is an important issue on its own right and is important also for the understanding of
the QCD {\it background} for any searches. Small $x$ physics is very complicated
due to the large phase space opened but it offers also the possibility to understand
the transition from a dilute to a dense system in a systematic way and thus
contributes much to the understanding of complicated processes in general.

\section*{Acknowledgements}
We are very grateful to  
the DESY directorate for the hospitality and the 
financial support of this workshop.

The work of J.~R.~Andersen was funded by PPARC (postdoctoral
fellowship PPA/P/S/2003/00281) and that of A.~Sabio~Vera by the
Alexander von Humboldt Foundation.

\bibliographystyle{mysty} 
\raggedright 
\bibliography{ref}

\end{document}